\documentclass[a4paper,11pt]{article}
\pdfoutput=1 
\usepackage{jheppub}
\usepackage[T1]{fontenc} 
\usepackage[all]{xy}
\usepackage{rotating}
\usepackage{float}
\usepackage{tikz}
\usepackage{circuitikz}
\usepackage{tikz-network}
\usepackage{diagbox}
\usepackage{braket}
\usepackage{empheq}
\usepackage{tabularx}
\usepackage{multirow}
\usepackage{slashed}
\usetikzlibrary{decorations.markings}
\usetikzlibrary{arrows.meta}
\tikzset{->-/.style={decoration={
  markings,
  mark=at position #1 with {\arrow{>}}},postaction={decorate}}}

\pgfarrowsdeclare{:}{:}{}{}

\let\a=\alpha \let\b=\beta \let\g=\gamma \let\d=\delta 
\let\z=\zeta  \let\th=\theta  
\let\l=\lambda     \let\r=\rho
\let\s=\sigma \let\t=\tau    
    \let\D=\Delta

\DeclareMathOperator{\Tr}{Tr}

\def\a{\alpha}
\def\b{\beta}

\def\CI{{\cal I}}

\def\CN{{\cal N}}

\def\CP{{\cal P}}

\def\CT{{\cal T}}

\def\Wa{\widetilde{a}}

\def\Wc{\widetilde{c}}

\def\Wm{\widetilde{m}}

\def\Wq{\widetilde{q}}

\def\WN{\widetilde{N}}

\def\WQ{\widetilde{Q}}

\def\beq#1\eeq{\begin{align}#1\end{align}}

\makeatletter
\newcommand*{\rom}[1]{\expandafter\romannumeral #1}

\makeatother

\title{Classification of monopole deformed 3d $\mathcal{N}=2$ Seiberg-like duality with an adjoint matter}

\abstract{We propose a new 3d $\CN=2$ Seiberg--like duality of adjoint SQCD(Kim--Park duality) with linear monopole superpotential terms which encompasses known monopole deformed Kim--Park dualities. Equipped with this, we classify all the monopole deformed Kim--Park dualities up to quadratic powers of monopole deformations, and find all are equivalent either to the original Kim--Park, or to the proposed duality. With the recently developed deconfined perspective, this means all the working monopole deformed Kim--Park dualities up to quadratic terms are assembled by the Aharony and Benini-Benvenuti-Pasquetti dualities.}

\author [a]{Qiang Jia,}
\author[b]{Sungjoon Kim}

\affiliation[a]{Department of Physics, Korea Advanced Institute of Science and Technology, Daejeon 34141, Korea}
\affiliation[b]{Korea Institute for Advanced Study, 85 Hoegiro, Dongdaemun-Gu, Seoul 02455, Korea}

\emailAdd{qjia1993@kaist.ac.kr}
\emailAdd{sungjoon@kias.re.kr}

\begin{document} 
\maketitle
\flushbottom



\section{Introduction}
Duality is one of the most intriguing and mysterious phenomena in physics. Over the past few decades, numerous instances have been discovered in quantum field theories, particularly with supersymmetry, following the pioneering discovery by Seiberg \cite{Seiberg:1994pq}. This is an IR duality between 4d $\CN=1$ UV supersymmetric gauge theories as a generalization of the electric--magnetic duality, and variations to additional matters, or/and to other dimensions have also been intensively examined, enlarging our understanding of the duality and non-perturbative aspects of quantum field theories \cite{Kutasov:1995np,Brodie:1996vx,Giveon:2008zn,Benini:2011mf,Kapustin:2011gh,Hwang:2011qt,Hwang:2011ht,Hanany:2015via,Hwang:2018uyj,Hwang:2020ddr,Bottini:2021vms,Benvenuti:2024mpn,Benvenuti:2024seb,Benvenuti:2025huk,Benvenuti:2025qnq,Hayashi:2025guk,Okazaki:2021pnc,Okazaki:2021gkk,Cho:2024civ,Kang:2024inx,Amariti:2025zgj}.
Especially, when the Seiberg duality is compactified to 3-dimension, non--trivial monopole operator superpotential plays a crucial role \cite{Aharony:2013dha} to correctly achieve the anticipated 3d $\CN=2$ IR duality found by Aharony \cite{Aharony:1997gp}. The monopole operator is a disorder operator defined by summing over all possible gauge field configurations of a unit magnetic flux on a two-sphere $S^2$ around its insertion point \cite{Borokhov:2002ib,Borokhov:2002cg,Kapustin:2005py}. Recently, the importance of such monopole operator superpotential becomes more aware of, controlling diverse aspects of the IR dynamics in 3d supersymmetric gauge theories, such as construction of the rank-0 theories \cite{Gang:2018huc,Gang:2021hrd,Gang:2023rei,Gang:2024loa,Creutzig:2024ljv,Ferrari:2023fez,Gang:2022kpe,Gang:2023ggt,Baek:2024tuo,Gang:2024tlp,Baek:2025uev,Jeong:2025xid}, twisted 3d reduction of 4d $\CN=2$ Argyes--Douglas theories \cite{Dedushenko:2023cvd,Gaiotto:2024ioj,Go:2025ixu,Kim:2024dxu,ArabiArdehali:2024ysy,ArabiArdehali:2024vli,Kim:2025rog} that provides a bridge for the 4d/2d correspondence, and the s-confinement phenomena \cite{Csaki:1996sm,Amariti:2025lem,Amariti:2024gco,Bajeot:2022kwt,Benvenuti:2021nwt,Hwang:2021xyw,Bajeot:2023gyl,Hwang:2024hhy,Benvenuti:2024glr,Okazaki:2023kpq}.
One of the most interesting roles of the monopole operator is that they navigate a way to new 3d $\CN=2$ IR dualities. For example, the work \cite{Benini:2017dud} considers deformations by linear monopole superpotential terms on the Aharony duality and obtains a new duality whose dual gauge rank is reduced by one or two compared to that of the original Aharony duality, lifting the Coulomb branch. This monopole deformed Aharony duality is called the Benini-Benvenuti-Pasquetti (BBP) duality in literature. Likewise, a series of works \cite{Kim:2013cma,Amariti:2018wht,Amariti:2019rhc,Hwang:2022jjs} have explored similar 3d $\mathcal{N}=2$ Seiberg–like dualities for gauge theories with unitary groups and adjoint tensor matter, commonly referred to as the Kim–Park (KP) duality, and have shown that deformations by monopole superpotentials can lead to new IR dualities.

In this work, we complete the classification of such monopole deformed KP dualities up to quadratic powers of monopole superpotential terms, of which power covers all the known monopole deformed KP dualities. The classification is based on the monopole deformed KP duality with linear monopole superpotential terms which is proposed in the present work. For this new duality, we confirm the perfect matching of the superconformal indices of the dual theories and check the relevance of the monopole deformations for many cases by performing the F-maximization computations with particular RG flows. Interestingly, this new duality holds even if we extend the conventional upper bound of the dressing number of the monopoles and allows us to unify all the known monopole deformed KP dualities by interpreting them in terms of our new duality. Such extension is crucial for the classification.
We classify the monopole deformed KP dualities by simply assuming consistent duality maps for the chiral ring elements between candidate dual theories. We scan all possible generic monopole superpotential deformed KP dualities up to quadratic powers and find all compatible dualities are always equivalent to the original KP, or our new duality. Combined with a recently developed deconfined picture of adjoint SQCD \cite{Benvenuti:2024glr,Hwang:2024hhy}, this observation suggests evidence that two basic dualities, Aharony and BBP, are the sufficient building blocks for deriving all the other KP-type dualities, including all possible monopole deformed ones. Similar evidences were also discussed in the context of $SL(2,\mathbb{Z})$ duals of 3d the $\CN=4$ quiver theories as well \cite{Benvenuti:2023qtv,Hwang:2020wpd,Hwang:2021ulb,Bottini:2022vpy,Comi:2022aqo,Giacomelli:2023zkk,Giacomelli:2024laq}. To summarize, our classification of monopole deformed KP dualities indicates a duality-decomposition structure where all the KP-type dualities arise from the two fundamental dualities.

This paper is organized as follows. In section \ref{sec: review}, we review 3d $\CN=2$ Seiberg--like IR dualities that are relevant to our discussion. In section \ref{sec: linear}, we propose a new 3d $\CN=2$ monopole deformed KP dualities with linear superpotential terms of monopole operators $\hat V_\a^\pm$ and show it works even if we extend the conventional upper bound of the dressing number $\a$. In section \ref{sec: bootstrap}, we classify a landscape of monopole deformed KP dualities by assuming a consistent chiral ring map of candidate duality, together with a restriction by generic monopole superpotential deformations. In section \ref{sec: conclusion}, we end with a summary and possible future directions. We also provide an appendix \ref{app: index f-max}, where we summarize our conventions for the supersymmetric observables used in this work with the superconformal index results for duality check as well as the F-maximization results for relevance check of the monopole superpotential deformations.

\section{Review of 3d \texorpdfstring{$\CN=2$}{N=2} IR dualities} \label{sec: review}
In this section, we review some 3d $\CN=2$ IR dualities, in particular, between SQCDs with unitary gauge groups that are relevant to our duality proposal in section \ref{sec: linear}. Let us begin with the most pioneering supersymmetric IR duality in 4d $\CN=1$ SQCD as a generalization of the electric--magnetic duality, named {\it Seiberg duality} \cite{Seiberg:1994pq}:
\begin{equation}
\begin{tikzpicture}
  \tikzset{vertex/.style={circle,fill=white!25,minimum size=12pt,inner sep=2pt}}
  \tikzset{every loop/.style={}}
    \node[vertex] (eNc) at (-1,0) [shape=circle,dotted,draw=black,minimum size=2em] {$N$};
    \node[vertex] (eNf1) at (1,0.7) [shape=rectangle,draw=black,minimum height=2em, minimum width=2em] {$F$};
    \node[vertex] (eNf2) at (1,-0.7) [shape=rectangle,draw=black,minimum height=2em, minimum width=2em] {$F$};

    \draw[->-=.5] ([yshift= 2pt] eNc.east) to ([yshift= 0pt] eNf1.west);
    \draw[->-=.5] ([yshift= 0pt] eNf2.west) to ([yshift= -2pt] eNc.east);

    \node at (0,0.7) {\tiny$Q$};
    \node at (0,-0.7) {\tiny$\WQ$};
    \node at (0,-1.5) {$W_A=  0$};

    \node at (3.5,0.2) {$\overset{\text{Dual}}{\longleftrightarrow}$};

    \node[vertex] (mNc) at (6,0) [shape=circle,dotted,draw=black,minimum size=2em] {$\WN$};
    \node[vertex] (mNf1) at (8,0.7) [shape=rectangle,draw=black,minimum height=2em, minimum width=2em] {$F$};
    \node[vertex] (mNf2) at (8,-0.7) [shape=rectangle,draw=black,minimum height=2em, minimum width=2em] {$F$};

    \draw[->-=.5] ([yshift= 0pt] mNf1.west) to ([yshift= 2pt] mNc.east);
    \draw[->-=.5] ([yshift= -2pt] mNc.east) to ([yshift= 0pt] mNf2.west);
    \draw[->-=.5] ( mNf2.north) to ( mNf1.south);

    \node at (7,0.7) {\tiny$\Wq$};
    \node at (7,-0.7) {\tiny$q$};
    \node at (8.5,0) {\tiny$M$};
    \node at (7,-1.5) {$W_B = M\, \Wq q$};

\end{tikzpicture}
\label{eq: Seiberg quiver}
\end{equation}
Through the paper, we use quivers to conveniently depict various dualities. The (dotted) circular and the square nodes represent ($SU(N)$) $U(N)$ gauge and $SU(F)$ flavor symmetry groups, respectively. An arrow represents a chiral multiplet in the bi--fundamental representation, i.e., (anti)--fundamental representation under the (ending) beginning node. So, the LHS theory of \eqref{eq: Seiberg quiver} is a 4d $\CN=1$ $SU(N)$ SQCD with $F$ flavors $(Q,\WQ)$ and trivial superpotential, while the RHS theory is a $SU(\WN)$ SQCD ($\WN=F-N$) with $F$ flavors $(q,\Wq)$ and the meson--singlet coupling superpotential $W_B$ where the gauge and flavor indices are contracted. The chiral ring generators, $\WQ Q$ on the LHS are mapped to the singlets $M$ on the RHS. Such meson to singlet map is the universal feature for the other Seiberg--like dualities.

\paragraph{Aharony duality} The 3d version of the Seiberg duality was suggested by Aharony \cite{Aharony:1997gp} and named as the {\it Aharony duality}:
\begin{equation}
\begin{tikzpicture}
  \tikzset{vertex/.style={circle,fill=white!25,minimum size=12pt,inner sep=2pt}}
  \tikzset{every loop/.style={}}
    \node[vertex] (eNc) at (-1,0) [shape=circle,draw=black,minimum size=2em] {$N$};
    \node[vertex] (eNf1) at (1,0.7) [shape=rectangle,draw=black,minimum height=2em, minimum width=2em] {$F$};
    \node[vertex] (eNf2) at (1,-0.7) [shape=rectangle,draw=black,minimum height=2em, minimum width=2em] {$F$};

    \draw[->-=.5] ([yshift= 2pt] eNc.east) to ([yshift= 0pt] eNf1.west);
    \draw[->-=.5] ([yshift= 0pt] eNf2.west) to ([yshift= -2pt] eNc.east);

    \node at (0,0.7) {\tiny$Q$};
    \node at (0,-0.7) {\tiny$\WQ$};
    \node at (0,-1.5) {$W_A=  0$};

    \node at (3.5,0.2) {$\overset{\text{Dual}}{\longleftrightarrow}$};

    \node[vertex] (mNc) at (6,0) [shape=circle,draw=black,minimum size=2em] {$\WN$};
    \node[vertex] (mNf1) at (8,0.7) [shape=rectangle,draw=black,minimum height=2em, minimum width=2em] {$F$};
    \node[vertex] (mNf2) at (8,-0.7) [shape=rectangle,draw=black,minimum height=2em, minimum width=2em] {$F$};

    \draw[->-=.5] ([yshift= 0pt] mNf1.west) to ([yshift= 2pt] mNc.east);
    \draw[->-=.5] ([yshift= -2pt] mNc.east) to ([yshift= 0pt] mNf2.west);
    \draw[->-=.5] ( mNf2.north) to ( mNf1.south);

    \node at (7,0.7) {\tiny$\Wq$};
    \node at (7,-0.7) {\tiny$q$};
    \node at (8.5,0) {\tiny$M$};
    \node at (7,-1.5) {$W_B = M\, \Wq q + V^+ \hat{v}^+ + V^- \hat{v}^-$};

\end{tikzpicture}
\label{eq: Aharony quiver}
\end{equation}
where the gauge groups are now unitary, $U(N)$ and $U(\WN)$ respectively ($\WN = F-N$), also there are a pair of extra singlets, $V^\pm$, on the dual side that correspond to the monopole operators $\hat V^\pm$ on the original side. Those singlets are coupled to the dual monopole operators $\hat v^\pm$ respectively as given in the superpotential $W_B$.\footnote{The monopole operators here are charged under the $U(1)_T$ topological symmetry from the diagonal of the unitary gauge group, whose charges are denoted on the upper indices of the monopole operators. We set the generators of both $U(1)_T$ topological symmetries of the dual theories, say $T$ and $T'$, such that they are mapped as $T \leftrightarrow -T'$ under the duality so that the terms $V^{\pm} \hat v^\pm$ in $W_B$ are indeed $U(1)_T$ singlets.} Although it is similar to the Seiberg duality, it took a while to figure out a correct derivation by dimensional reduction due to a non-trivial monopole superpotential effect \cite{Aharony:2013dha}. Such monopole operators play very interesting and non-trivial roles and provide rich physics in 3d supersymmetric gauge theories.

\paragraph{Benini--Benvenuti--Pasquetti (BBP) duality}
One such interesting application with the monopole operator is to consider a monopole superpotential deformed version of the Aharony duality as studied in \cite{Benini:2017dud}, and named as the {\it Benini--Benvenutti--Pasquetti(BBP) duality}:
\begin{equation}
\begin{tikzpicture}
  \tikzset{vertex/.style={circle,fill=white!25,minimum size=12pt,inner sep=2pt}}
  \tikzset{every loop/.style={}}
    \node[vertex] (eNc) at (-1,0) [shape=circle,draw=black,minimum size=2em] {$N$};
    \node[vertex] (eNf1) at (1,0.7) [shape=rectangle,draw=black,minimum height=2em, minimum width=2em] {$F$};
    \node[vertex] (eNf2) at (1,-0.7) [shape=rectangle,draw=black,minimum height=2em, minimum width=2em] {$F$};

    \draw[->-=.5] ([yshift= 2pt] eNc.east) to ([yshift= 0pt] eNf1.west);
    \draw[->-=.5] ([yshift= 0pt] eNf2.west) to ([yshift= -2pt] eNc.east);

    \node at (0,0.7) {\tiny$Q$};
    \node at (0,-0.7) {\tiny$\WQ$};
    \node at (0,-1.5) {$W_A=  \hat V^\pm $};

    \node at (3.5,0.2) {$\overset{\text{Dual}}{\longleftrightarrow}$};

    \node[vertex] (mNc) at (6,0) [shape=circle,draw=black,minimum size=2em] {$\WN$};
    \node[vertex] (mNf1) at (8,0.7) [shape=rectangle,draw=black,minimum height=2em, minimum width=2em] {$F$};
    \node[vertex] (mNf2) at (8,-0.7) [shape=rectangle,draw=black,minimum height=2em, minimum width=2em] {$F$};

    \draw[->-=.5] ([yshift= 0pt] mNf1.west) to ([yshift= 2pt] mNc.east);
    \draw[->-=.5] ([yshift= -2pt] mNc.east) to ([yshift= 0pt] mNf2.west);
    \draw[->-=.5] ( mNf2.north) to ( mNf1.south);

    \node at (7,0.7) {\tiny$\Wq$};
    \node at (7,-0.7) {\tiny$q$};
    \node at (8.5,0) {\tiny$M$};
    \node at (7,-1.5) {$W_B = \hat v^\pm + M\, \Wq q + V^\mp \hat{v}^\mp$};

\end{tikzpicture}
\label{eq: BBP quiver}
\end{equation}
where the rank of the dual gauge group is reduced by one, $\WN = F-N-1$, compared to the Aharony duality. Let us denote this duality as $\text{BBP}^\pm$. If we consider $\text{BBP}^+$, for instance, the monopole operator $\hat V^-$ is mapped to a singlet $V^-$ that couples to the dual monopole operator $\hat v^-$ in the RHS theory, while the monopole operator $\hat V^+$ is directly mapped to the dual monopole operator $\hat v^+$.

Let us consider a further deformation at the LHS theory by the other monopole operator, i.e., $W_A = \hat V^+ + \hat V^-$, then one can check that a successive dualization as in \eqref{eq: BBP2 quiver} along the counterclockwise direction achieves another BBP duality:
\begin{equation}
\centering
\begin{tikzpicture}
  \tikzset{vertex/.style={circle,fill=white!25,minimum size=12pt,inner sep=2pt}}
  \tikzset{every loop/.style={}}
    \node[vertex] (eNc) at (-1,0) [shape=circle,draw=black,minimum size=2em] {$N$};
    \node[vertex] (eNf1) at (1,0.7) [shape=rectangle,draw=black,minimum height=2em, minimum width=2em] {$F$};
    \node[vertex] (eNf2) at (1,-0.7) [shape=rectangle,draw=black,minimum height=2em, minimum width=2em] {$F$};

    \draw[->-=.5] ([yshift= 2pt] eNc.east) to ([yshift= 0pt] eNf1.west);
    \draw[->-=.5] ([yshift= 0pt] eNf2.west) to ([yshift= -2pt] eNc.east);

    \node at (0,0.7) {\tiny$Q$};
    \node at (0,-0.7) {\tiny$\WQ$};
    \node at (0,-1.5) {\scriptsize$W_A=  \hat V^+ + \hat V^- $};

    \node at (3.5,0.2) {$\overset{\text{Dual}}{\longleftrightarrow}$};

    \node[vertex] (mNc) at (6,0) [shape=circle,draw=black,minimum size=2em] {$\WN$};
    \node[vertex] (mNf1) at (8,0.7) [shape=rectangle,draw=black,minimum height=2em, minimum width=2em] {$F$};
    \node[vertex] (mNf2) at (8,-0.7) [shape=rectangle,draw=black,minimum height=2em, minimum width=2em] {$F$};

    \draw[->-=.5] ([yshift= 0pt] mNf1.west) to ([yshift= 2pt] mNc.east);
    \draw[->-=.5] ([yshift= -2pt] mNc.east) to ([yshift= 0pt] mNf2.west);
    \draw[->-=.5] ( mNf2.north) to ( mNf1.south);

    \node at (7,0.7) {\tiny$\Wq$};
    \node at (7,-0.7) {\tiny$q$};
    \node at (8.5,0) {\tiny$M$};
    \node at (7,-1.5) {\scriptsize$W_B = \hat v^+ + \hat v^- + M\, \Wq q$};

    \node[vertex] (eeNc) at (-1,-4) [shape=circle,draw=black,minimum size=2em] {$N'$};
    \node[vertex] (eeNf1) at (1,-4+0.7) [shape=rectangle,draw=black,minimum height=2em, minimum width=2em] {$F$};
    \node[vertex] (eeNf2) at (1,-4-0.7) [shape=rectangle,draw=black,minimum height=2em, minimum width=2em] {$F$};

    \draw[->-=.5]  ([yshift= 0pt] eeNf1.west) to ([yshift= 2pt] eeNc.east);
    \draw[->-=.5]  ([yshift= -2pt] eeNc.east) to ([yshift= 0pt] eeNf2.west);
    \draw[->-=.5] ( eeNf2.north) to ( eeNf1.south);

    \node at (0,-4+0.7) {\tiny$q'$};
    \node at (0,-4-0.7) {\tiny$\Wq'$};
    \node at (1.5,-4) {\tiny$M'$};
    \node at (0.5,-4-1.5) {\scriptsize$W_{A'}=  \hat U^+ + V^- + V^-\hat U^- + M'\Wq'q'$};

    \node at (3.5,-4+0.2) {$\overset{\text{Aharony}}{\longleftrightarrow}$};

    \node[vertex] (mmNc) at (6,-4) [shape=circle,draw=black,minimum size=2em] {$\WN'$};
    \node[vertex] (mmNf1) at (8,-4+0.7) [shape=rectangle,draw=black,minimum height=2em, minimum width=2em] {$F$};
    \node[vertex] (mmNf2) at (8,-4-0.7) [shape=rectangle,draw=black,minimum height=2em, minimum width=2em] {$F$};

    \draw[->-=.5] ([yshift= 2pt] mmNc.east) to ([yshift= 0pt] mmNf1.west);
    \draw[->-=.5] ([yshift= 0pt] mmNf2.west) to ([yshift= -2pt] mmNc.east);

    \node at (7,-4+0.7) {\tiny$\WQ'$};
    \node at (7,-4-0.7) {\tiny$Q'$};
    \node at (7,-4-1.5) {\scriptsize$W_{B'} = U^+ + \hat u^- + U^+ \hat u^+$};

    \node at (-0.7,-2.5) {\scriptsize$\text{BBP}^+$};
    \node at (0,-2.5){$\updownarrow$};

    \node at (7+0.7,-2.5) {\scriptsize$\text{BBP}^-$};
    \node at (7,-2.5){$\updownarrow$};
    
\end{tikzpicture}
\label{eq: BBP2 quiver}
\end{equation}
where the dual gauge rank is now reduced by two compared to the Aharony duality,
\begin{align}
    \WN &= F - \WN'-1
    \nonumber\\
    & = F - (F-N') - 1
    \nonumber\\
    & = F - \big( F - (F - N - 1) \big) - 1
    \nonumber\\
    & = F - N - 2 \,,
\end{align}
and the monopole operators $\hat V^\pm$ are successively mapped along the process to $\hat v^\pm$:
\begin{align}
    &\hat V^+ 
    \;\;\rightarrow\;\;
    \hat U^+
    \;\;\rightarrow\;\;
    U^+
    \;\;\rightarrow\;\;
    \hat v^+
    \nonumber\\
    &\hat V^-
    \;\;\rightarrow\;\;
    V^-
    \;\;\rightarrow\;\;
    \hat u^-
    \;\;\rightarrow\;\;
    \hat v^-
\end{align}
This was also proposed in the same paper which we will denote as $\text{BBP}_2$. The duality chain \eqref{eq: BBP2 quiver} tells us that $\text{BBP}_2$ can be understood from the Aharony and $\text{BBP}^\pm$ dualities. Conversely, the Aharony and $\text{BBP}^{\pm}$ can be obtained from real mass deformations on $\text{BBP}_2$ \cite{Hwang:2020wpd}. The point is that the Aharony and BBP dualities are not independent but closely related to each other.

\paragraph{Kim--Park(KP) duality}
Another way to generalize the Aharony duality is to add tensorial matters. One plausible attempt would be adding a chiral multiplet in the adjoint representation with a monomial superpotential, which was studied in \cite{Kim:2013cma} and named, the {\it Kim--Park duality}:
\begin{equation}
\begin{tikzpicture}
  \tikzset{vertex/.style={circle,fill=white!25,minimum size=12pt,inner sep=2pt}}
  \tikzset{every loop/.style={}}
    \node[vertex] (eNc) at (-1,0) [shape=circle,draw=black,minimum size=2em] {$N$};
    \node[vertex] (eNf1) at (1,0.7) [shape=rectangle,draw=black,minimum height=2em, minimum width=2em] {$F$};
    \node[vertex] (eNf2) at (1,-0.7) [shape=rectangle,draw=black,minimum height=2em, minimum width=2em] {$F$};

    \draw[-to, min distance=1cm]  (eNc) edge [out=150, in=210] node {} (eNc);
    \draw[->-=.5] ([yshift= 2pt] eNc.east) to ([yshift= 0pt] eNf1.west);
    \draw[->-=.5] ([yshift= 0pt] eNf2.west) to ([yshift= -2pt] eNc.east);

    \node at (-1.6,0.6) {\tiny$X$};
    \node at (0,0.7) {\tiny$Q$};
    \node at (0,-0.7) {\tiny$\WQ$};
    \node at (0,-1.5) {$W_A=  \Tr X^{p+1}$};

    \node at (3.5,0.2) {$\overset{\text{Dual}}{\longleftrightarrow}$};

    \node[vertex] (mNc) at (7,0) [shape=circle,draw=black,minimum size=2em] {$\WN$};
    \node[vertex] (mNf1) at (9,0.7) [shape=rectangle,draw=black,minimum height=2em, minimum width=2em] {$F$};
    \node[vertex] (mNf2) at (9,-0.7) [shape=rectangle,draw=black,minimum height=2em, minimum width=2em] {$F$};

    \draw[-to, min distance=1cm]  (mNc) edge [out=150, in=210] node {} (mNc);
    \draw[->-=.5] ([yshift= 0pt] mNf1.west) to ([yshift= 2pt] mNc.east);
    \draw[->-=.5] ([yshift= -2pt] mNc.east) to ([yshift= 0pt] mNf2.west);
    \draw[->-=.5] ( mNf2.north) to ( mNf1.south);

    \node at (6.4,0.6) {\scriptsize$x$};
    \node at (8,0.7) {\tiny$\Wq$};
    \node at (8,-0.7) {\tiny$q$};
    \node at (9.5,0) {\tiny$M_i$};
    \node at (8,-1.5) {$W_B = \Tr x^{p+1} \!+\! M_i \,\Wq x^{p\texttt{-}i\texttt{-}1} q + V_{p\texttt{-}i\texttt{-}1}^+ \hat v_{i}^+ + V_{p\texttt{-}i\texttt{-}1}^- \hat v_{i}^-$};

\end{tikzpicture}
\label{eq: KP quiver}
\end{equation}
with dual gauge rank $\WN = pF-N$, $X$ and $x$ are the adjoint chiral multiplets, that give rise to $p$ pairs of monopole operators by dressing powers of them to the bare monopole operators:
\begin{align}
    \hat V_i^\pm \equiv \Tr_{U(N-1)}' X^i \hat V_0^\pm
    \quad,\quad
    \hat v_i^\pm \equiv \Tr_{U(\WN-1)}' x^i \hat v_0^\pm
    \label{eq: dressed monopole}
\end{align}
for $i=0,\cdots,p-1$. Here the traces are taken over the unbroken $U(N-1)$ and $U(\WN-1)$ sectors of the background magnetic fluxes $m = (\pm 1,0,\cdots,0)$ by the bare monopole operators $\hat V_0^\pm$ and $\hat v_0^\pm$ respectively. We use $\Tr' X^n$ to conveniently denote a possible linear combination of products of traces $\Tr X^{i_1} \Tr X^{i_2}\cdots \Tr X^{i_l}$ with $i_1+i_2+\cdots+i_l=n$ such that each factor is free from the trace relation by the gauge rank in consideration\footnote{See \cite{Hwang:2018uyj,Hwang:2022jjs} for more rigorous explanations from the operator-state correspondence with examples.}. The truncation of the dressing number is due to the F-term equation $X^p = 0$. We explicitly leave the indices $i$ in the superpotential $W_B$ for the summations over $i=0,\cdots,p-1$. The monopole operators $\hat V_i^\pm$ in the LHS are mapped to the singlets $V_i^\pm$ in the RHS that couple to the dual monopole operators $\hat v_{p\texttt{-}i\texttt{-}1}^\pm$. Likewise, the mesons, $\WQ X^i Q$, $i=0,\cdots,p-1$ in the LHS are mapped to the singlets, $M_i$ in the RHS. Note that when $p=1$, the adjoints in both sides become massive, exactly reproducing the Aharony duality. In this sense, the KP duality is a natural generalization of the Aharony duality.

\paragraph{Monopole deformed Kim--Park duality}
\begin{table}[tbp]
\centering
\renewcommand{\arraystretch}{1.2}
\begin{tabular}{|c|c|c|c|}
\hline
Duality & $\D W_A$ & $\D W_B$ & $\WN$ \\ 
\hline
\hline
Kim--Park \cite{Kim:2013cma} & $0$ & $0$ & $pF-N$ \\
\hline
\multirow{2}{*}{Amariti--Cassia \cite{Amariti:2018wht}} & $\hat V_0^+$ & $\hat v_0^+ + V_i^-\,\hat v_{p\texttt{-}i\texttt{-}1}^-$ & $pF-N-p$ \\
\cline{2-4}
 & $\hat V_0^+ + \hat V_0^-$ & $\hat v_0^+ + \hat v_0^-$ & $pF-N-2p$ \\
\hline
\multirow{3}{11em}{Amariti--Cassia--Garozzo--Mekareeya \cite{Amariti:2019rhc}} \!\!\!\!\!\!\!\!\!& $(\hat V_\a^+)^2 + (\hat V_{p\texttt{-}\a\texttt{-}1}^-)^2$ & $(\hat v_\a^+)^2 + (\hat v_{p\texttt{-}\a\texttt{-}1}^-)^2$ & $pF-N$ \\
\cline{2-4} & $\big(\hat V_{\frac{p\texttt{-}1}{2}}^+\big)^2 + \hat V_0^-$ & $\big(\hat v_{\frac{p\texttt{-}1}{2}}^+\big)^2 + \hat v_0^-$ & $pF - N-p$ \\
\cline{2-4} & $\big( \hat V_{\frac{p\texttt{-}1}{2}}^+ \big)^2$ & $\big( \hat v_{\frac{p\texttt{-}1}{2}}^+ \big)^2 + V_i^- \, \hat v_{p\texttt{-}i\texttt{-}1}^-$ & $pF-N$ \\
\hline
Hwang--Kim--Park \cite{Hwang:2022jjs} & $\hat V_\a^+ + \hat V_\a^-$ & $\hat v_\a^+ + \hat v_\a^-$ & $pF-N-2p+2\a$ \\
 \hline
\end{tabular}
\caption{\label{tab: monopole KP} Proposed 3d $\CN=2$ IR dualities between two SQCDs of gauge group $U(N)$ and $U(\WN)$ with an adjoint matter $X$ and $x$ having superpotential $W_A = \Tr X^{p+1} + \D W_A$ and $W_B = \Tr x^{p+1} + M_i\, \Wq x^{p\texttt{-}i\texttt{-}1} q + \D W_B$ respectively.}
\end{table}
Like the BBP duality, one can also think of monopole superpotential deformed version of the KP duality \cite{Amariti:2018wht,Amariti:2019rhc,Hwang:2022jjs} which are organized in table \ref{tab: monopole KP}. These are, to the best of current knowledge, the only known monopole deformed KP dualities. Among them, let us review the one discussed in \cite{Hwang:2022jjs}:
\begin{equation}
\begin{tikzpicture}
  \tikzset{vertex/.style={circle,fill=white!25,minimum size=12pt,inner sep=2pt}}
  \tikzset{every loop/.style={}}
    \node[vertex] (eNc) at (-1,0) [shape=circle,draw=black,minimum size=2em] {$N$};
    \node[vertex] (eNf1) at (1,0.7) [shape=rectangle,draw=black,minimum height=2em, minimum width=2em] {$F$};
    \node[vertex] (eNf2) at (1,-0.7) [shape=rectangle,draw=black,minimum height=2em, minimum width=2em] {$F$};

    \draw[-to, min distance=1cm]  (eNc) edge [out=150, in=210] node {} (eNc);
    \draw[->-=.5] ([yshift= 2pt] eNc.east) to ([yshift= 0pt] eNf1.west);
    \draw[->-=.5] ([yshift= 0pt] eNf2.west) to ([yshift= -2pt] eNc.east);

    \node at (-1.6,0.6) {\tiny$X$};
    \node at (0,0.7) {\tiny$Q$};
    \node at (0,-0.7) {\tiny$\WQ$};
    \node at (0,-1.5) {$W_A= \hat V_\a^+ + \hat V_\a^- +  \Tr X^{p+1}$};

    \node at (3.5,0.2) {$\overset{\text{Dual}}{\longleftrightarrow}$};

    \node[vertex] (mNc) at (7,0) [shape=circle,draw=black,minimum size=2em] {$\WN$};
    \node[vertex] (mNf1) at (9,0.7) [shape=rectangle,draw=black,minimum height=2em, minimum width=2em] {$F$};
    \node[vertex] (mNf2) at (9,-0.7) [shape=rectangle,draw=black,minimum height=2em, minimum width=2em] {$F$};

    \draw[-to, min distance=1cm]  (mNc) edge [out=150, in=210] node {} (mNc);
    \draw[->-=.5] ([yshift= 0pt] mNf1.west) to ([yshift= 2pt] mNc.east);
    \draw[->-=.5] ([yshift= -2pt] mNc.east) to ([yshift= 0pt] mNf2.west);
    \draw[->-=.5] ( mNf2.north) to ( mNf1.south);

    \node at (6.4,0.6) {\scriptsize$x$};
    \node at (8,0.7) {\tiny$\Wq$};
    \node at (8,-0.7) {\tiny$q$};
    \node at (9.5,0) {\tiny$M_i$};
    \node at (8,-1.5) {$W_B = \hat v_\a^+ + \hat v_\a^- + \Tr x^{p+1} \!+\! M_i \,\Wq x^{p\texttt{-}i\texttt{-}1} q$};

\end{tikzpicture}
\label{eq: monopole KP quiver}
\end{equation}
where the dual gauge rank is $\WN = pF-N-2p+2\a$, and the monopole operators $\hat V_i^\pm$ are mapped directly to the monopole operators $\hat v_i^\pm$ in the dual theory. Note that when $p=1$, only the bare monopole operators exist and this duality reproduces the $\text{BBP}_2$ duality. One of the main contents of the present work is to generalize this duality to arbitrary choices of the monopoles in the superpotential which is discussed in the next section.

\paragraph{$D^{3d}_p[SU(N)]$ coupled to $U(N)$ SQCD}
The $D_p[SU(N)]$ theory is a class of 4d $\CN=2$ Argyres--Douglas theory with at least $SU(N)$ flavor symmetry \cite{Cecotti:2012jx,Cecotti:2013lda}. One remarkable observation is that when $\text{gcd}(p,N)=1$, the $D_p[SU(N)]$ theory behaves as an adjoint matter with non-trivial superpotential if it is coupled to a $SU(N)$ gauge group via a 4d $\CN=1$ vector multiplet with a proper Coulomb branch operator deformation \cite{Maruyoshi:2023mnv}. This is interesting, given that the $D_p[SU(N)]$ theory does not have a UV Lagrangian description for $\text{gcd}(p,N)=1$. However, when this is reduced to 3d $\CN=4$, it becomes a linear quiver gauge theory denoted as $D_p^{3d}[SU(N)]$ apparently having a Lagrangian description \cite{Closset:2020afy,Giacomelli:2020ryy}:
\begin{equation}
\begin{tikzpicture}
  \tikzset{vertex/.style={circle,fill=white!25,minimum size=12pt,inner sep=2pt}}
  \tikzset{every loop/.style={}}
    \node at (-0.5,0) {$D^{3d}_p[SU(N)]\quad:\qquad$ };    
    \node[vertex] (m1) at (1.5,0) [shape=circle,draw=black] {$m_1$};
    \node[vertex] (m2) at (3,0) [shape=circle,draw=black] {$m_2$};
    \node[draw=none,fill=none] (ndots) at (4.5,0)  {$\cdots$};
    \node[vertex] (mp1) at (6,0) [shape=circle,draw=black] {\scriptsize $\!m_{p\texttt{-}1}\!$};
    \node[vertex] (Np) at (7.5,0) [shape=rectangle,draw=black,minimum height=2em, minimum width=2em] {$N$};

      \draw[-to, min distance=1cm]  (m1) edge [out=120, in=60] node {} (m1);
      \draw[-to, min distance=1cm]  (m2) edge [out=120, in=60] node {} (m2);
      \draw[-to, min distance=1cm]  (mp1) edge [out=120, in=60] node {} (mp1);
      
      \draw[->-=.5] ([yshift= 2pt] m1.east) to ([yshift= 2pt] m2.west);
      \draw[->-=.5] ([yshift= -2pt] m2.west) to ([yshift= -2pt] m1.east);
      \draw[->-=.5] ([yshift= 2pt] m2.east) to ([yshift= 2pt] ndots.west);
      \draw[->-=.5] ([yshift= -2pt] ndots.west) to ([yshift= -2pt] m2.east);
      \draw[->-=.5] ([yshift= 2pt] ndots.east) to ([yshift= 2pt] mp1.west);
      \draw[->-=.5] ([yshift= -2pt] mp1.west) to ([yshift= -2pt] ndots.east);
      \draw[->-=.5] ([yshift= 2pt] mp1.east) to ([yshift= 2pt] Np.west);
      \draw[->-=.5] ([yshift= -2pt] Np.west) to ([yshift= -2pt] mp1.east);
      
\end{tikzpicture}
\label{eq: 3d DpG}
\end{equation}
The rank of each $U(m_j)$ node is $m_j = \lfloor\, j\,N /p \rfloor$. Upon proper monopole superpotential deformation, $D_p^{3d}[SU(N)]$ theory is IR dual to a Wess--Zumino theory of a chiral multiplet $X$ in the adjoint representation of the flavor symmetry with monomial superpotential $\Tr X^{p+1}$ \cite{Benvenuti:2024glr,Hwang:2024hhy}:
\begin{equation}
    \begin{tikzpicture}
      \tikzset{vertex/.style={circle,fill=white!25,minimum size=12pt,inner sep=2pt}}
      \tikzset{every loop/.style={}}
        \node[vertex] (N) at (0,0) [shape=rectangle,draw=black,minimum height=2em, minimum width=2em] {$N$};
    
        \draw[-to, min distance=1cm]  (N) edge [out=120, in=60] node {} (N);
    
        \node at (-4,0) {$D_p^{3d}[SU(N)]$};
        \node at (-1.8,0.2) {$\xrightarrow{ W_{\text{monopole}}}$};
        
        \node at (0,1.2) {\tiny $X$};
        \node at (2,0) {$\;,\quad W=\Tr X^{p+1}$};
    \end{tikzpicture} \,.
    \label{eq: Dp s-conf}
\end{equation}
which is proven by only assuming Aharony and $\text{BBP}^\pm$ dualities. This s--confinement picture provides a useful setup for understanding the KP duality. More precisely, if we gauge the flavor node of the LHS theory in \eqref{eq: Dp s-conf} with $U(N)$ SQCD of $F$ flavors and apply sequential Aharony dualities from the right to the left gauge nodes, one can precisely obtain the dual theory of the KP duality with $U(pF-N)$ gauge group. Namely, the KP duality can be proven by only assuming Aharony and the s--confinement \eqref{eq: Dp s-conf}. Furthermore, it is also partially checked that the monopole deformed KP can also be understood with this deconfined picture, and will be discussed further in future work \cite{future}. This deconfined picture serves as a substantial clue for pursuing a full generalization of the monopole deformed KP duality that will be proposed in the next section.

\section{Duality proposal and extension \label{sec: linear}}
In this section, we propose a couple of new dualities. We also extend the conventional upper bound of the dressing number of the monopole operators to extend the applicable domain of the duality. By doing so, we unify the known monopole deformed KP dualities in our proposed framework.

\subsection{Duality proposal: Linear superpotential with a dressed monopole}
\begin{table}[tbp]
\centering
\begin{tabular}{|c|c|c|c|c|}
\hline
 & $U(1)_R$ & $SU(F)_t$ & $SU(F)_b$ & $U(1)_D$ \\
\hline 
$Q$ & $\D_Q$ & $\overline{\mathbf{F}}$ & $\mathbf 1$ & $1$ \\
$\WQ$ & $\D_Q$ & $\mathbf 1$ & $\mathbf{F}$ & $1$ \\
$X$ & $2/(p+1)$ & $\mathbf 1$ & $\mathbf 1$ & $0$ \\
$\hat{V}_i^-$ & $2 F (1-\Delta_Q)-\frac{2}{p+1}(2 N + p - \a - 1 - i)$ & $\mathbf 1$ & $\mathbf 1$ & $-2 F$ \\
\hline
$q$ & $\frac{2}{p+1}-\D_Q$ & $\mathbf 1$ & $\overline{\mathbf{F}}$ & $-1$ \\
$\Wq$ & $\frac{2}{p+1}-\D_Q$ & $\mathbf{F}$ & $\mathbf 1$ & $-1$ \\
$x$ & $2/(p+1)$ & $\mathbf 1$ & $\mathbf 1$ & $0$ \\
$M_i$ & $2 \D_Q+2 i/(p+1)$ & $\overline{\mathbf{F}}$ & $\mathbf{F}$ & $2$ \\
$V_i^-$ & $2 F (1-\Delta_Q)-\frac{2}{p+1}(2 N + p - \a - 1 - i)$ & $\mathbf 1$ & $\mathbf 1$ & $-2 F$ \\
\hline
\end{tabular}
\caption{\label{tab: one monopole KP charge} The charges of the elementary matter fields of the theories in the single linear monopole deformed KP duality. The subscript $i$ runs from $0$ to $p-1$. $U(1)_D$ is the unbroken diagonal of $U(1)_A\times U(1)_T$, where $U(1)_A$ and $U(1)_T$ are the axial and topological symmetries respectively.}
\end{table}
We propose a monopole deformed Kim--Park duality with a single linear term, $\hat V_\a^+ \leftrightarrow \hat v_\a^+$:
\begin{equation}
\begin{tikzpicture}
  \tikzset{vertex/.style={circle,fill=white!25,minimum size=12pt,inner sep=2pt}}
  \tikzset{every loop/.style={}}
    \node[vertex] (eNc) at (-1,0) [shape=circle,draw=black,minimum size=2em] {$N$};
    \node[vertex] (eNf1) at (1,0.7) [shape=rectangle,draw=black,minimum height=2em, minimum width=2em] {$F$};
    \node[vertex] (eNf2) at (1,-0.7) [shape=rectangle,draw=black,minimum height=2em, minimum width=2em] {$F$};

    \draw[-to, min distance=1cm]  (eNc) edge [out=150, in=210] node {} (eNc);
    \draw[->-=.5] ([yshift= 2pt] eNc.east) to ([yshift= 0pt] eNf1.west);
    \draw[->-=.5] ([yshift= 0pt] eNf2.west) to ([yshift= -2pt] eNc.east);

    \node at (-1.6,0.6) {\tiny$X$};
    \node at (0,0.7) {\tiny$Q$};
    \node at (0,-0.7) {\tiny$\WQ$};
    \node at (0,-1.5) {$W_A= \hat V_\a^+ +  \Tr X^{p+1}$};

    \node at (3.5,0.4) {\scriptsize $\text{KP}_\a^+$};
    \node at (3.5,0) {$\longleftrightarrow$};

    \node[vertex] (mNc) at (7,0) [shape=circle,draw=black,minimum size=2em] {$\WN$};
    \node[vertex] (mNf1) at (9,0.7) [shape=rectangle,draw=black,minimum height=2em, minimum width=2em] {$F$};
    \node[vertex] (mNf2) at (9,-0.7) [shape=rectangle,draw=black,minimum height=2em, minimum width=2em] {$F$};

    \draw[-to, min distance=1cm]  (mNc) edge [out=150, in=210] node {} (mNc);
    \draw[->-=.5] ([yshift= 0pt] mNf1.west) to ([yshift= 2pt] mNc.east);
    \draw[->-=.5] ([yshift= -2pt] mNc.east) to ([yshift= 0pt] mNf2.west);
    \draw[->-=.5] ( mNf2.north) to ( mNf1.south);

    \node at (6.4,0.6) {\scriptsize$x$};
    \node at (8,0.7) {\tiny$\Wq$};
    \node at (8,-0.7) {\tiny$q$};
    \node at (9.5,0) {\tiny$M_i$};
    \node at (8,-1.5) {$W_B = \hat v_\a^+ + \Tr x^{p+1} \!+\! M_i \,\Wq x^{p\texttt{-}i\texttt{-}1} q + V_i^- \hat v_{p\texttt{-}i\texttt{-}1}^-$};

\end{tikzpicture}
\label{eq: linaer KP quiver}
\end{equation}
where the symmetry charges of the matter fields are given in table \ref{tab: one monopole KP charge}. The dual gauge rank is $\WN = pF-N-p+\a$, for $\a = 0,\cdots,p-1$ and the indices $i$ in the superpotential $W_B$ are summed over $i=0,\cdots,p-1$. The chiral rings on both sides are mapped as
\begin{align}
    \Tr X^i
    \qquad&\longleftrightarrow\qquad
    \Tr x^i
    \nonumber\\
    \WQ X^i Q
    \qquad&\longleftrightarrow\qquad
    M_i
    \nonumber\\
    \hat V_{i}^+
    \qquad&\longleftrightarrow\qquad
    \hat v_{i}^+
    \nonumber\\
    \hat V_i^-
    \qquad&\longleftrightarrow\qquad
    V_i^-
\end{align}
and the Coulomb branch parametrized by the monopole operators $\hat V_j^+$ and $\hat v_j^+$ for $j\geq \a$ are lifted by the same way as discussed in \cite{Hwang:2022jjs}. Note that just a simple sign flipping of the monopole operators/singlets gives an equivalent duality deformed by the oppositely charged monopole operators, $\hat V_\a^- \leftrightarrow \hat v_\a^-$. Also, $\a=0$ reproduces the Amariti--Cassia duality \cite{Amariti:2018wht}. For shorthand notation, let us denote this duality as $\text{KP}_\a^\pm$.

\paragraph{Duality checks} As a consistency check, let us first investigate the quantum numbers of the chiral ring elements. We call the LHS theory in \eqref{eq: linaer KP quiver} as the original theory and the RHS theory as the dual theory. Before we turn on the monopole operator superpotential, the global symmetries of the 3d $\CN=2$ adjoint SQCD with $F$ flavors are $U(1)_T\times U(1)_A\times U(1)_R\times SU(F)_t \times SU(F)_b$, where $U(1)_T$ is the topological symmetry, $U(1)_A$ is the axial symmetry, $U(1)_R$ is the R-symmetry, and $SU(F)_t \times SU(F)_b$ are the flavor symmetries. Since the monopole operators $\hat V_\a^+$ and $\hat v_\a^+$ are non--trivially charged under $U(1)_T\times U(1)_A\times U(1)_R$, only a combination of $U(1)_T\times U(1)_A$ remains unbroken if we turn on them and let us denote it as $U(1)_D$. At the same time, we have to demand the $U(1)_R$ charges of the monopole operators to be 2, so, we redefine the R-charge by mixing it with $A$ and $T$, the charges of the $U(1)_T$ and $U(1)_A$ respectively, and denote the charge of $U(1)_D$ as $D$:
\begin{align}
    R\to R + \Wa\, A + \t\, T
    \quad&\longleftrightarrow\quad
    R\to R + \Wa' \, A' + \t' \, T'
    \nonumber\\
    D = a\,A + t\,T
    \qquad&\longleftrightarrow\qquad
    D = a'\,A' + t'\,T'
\end{align}
where $\Wa,\t,\Wa'$ and $\t'$ are some mixing parameters. For rather comprehensive setup, we introduce $A'$ and $T'$ as the charges of $U(1)_{A'}$ and $U(1)_{T'}$ of the dual theory in the usual sense, i.e., $U(1)_{A'}$ and $U(1)_{T'}$ rotate flavors $(q,\Wq)$ and positively charged monopole operators $\hat v_i^+$ with positive unit charge. The R-charge 2 and the flavor singlet condition from the monopole superpotential terms, $\hat V_\a^+ \leftrightarrow \hat v_\a^+$, restrict the coefficients as
\begin{align}
    R[\hat{V}_\a^+] &= F(1-\D_Q-\Wa) + \frac{2}{p+1}(1-N+\a)+\t =2
    \nonumber\\
    R[\hat{v}_\a^+] &= F\Big(1-\big(\frac{2}{p+1}-\D_Q\big) - \Wa' \Big) + \frac{2}{p+1}(1-\WN+\a)+\t' =2
    \nonumber\\
    D[\hat{V}_\a^+] &= -a\,F + t\, = 0
    \nonumber\\
    D[\hat{v}_\a^+] &= -a'\,F + t'\, = 0
    \,.
    \label{eq: one monopole cond1}
\end{align}
Also from the usual meson--singlet coupled terms in the dual side, $M_i\,\Wq x^{p-i-1} q$,  we have constraints:
\begin{align}
    R[ M_i ] + R[\Wq x^{p-i-1} q] &=
    2\D_Q + \frac{2i}{p+1} + 2\Wa + 2\Big( \frac{2}{p+1} - \D_Q \Big) + \frac{2}{p+1}(p-i-1) + 2\Wa'
    =2
    \nonumber\\
    D[ M_i ] + D[\Wq x^{p-i-1} q] &= 2a + 2a'
    =0
\end{align}
which give rise to $\Wa + \Wa' = a + a' = 0$. Furthermore, the monopole--singlet terms, $V_\a^-\,\hat v_{p-i-1}^-$, constrain,
\begin{align}
    R[V_i^-] + R[\hat v_{p-i-1}^-] &= F\Big( 2-\frac{2}{p+1}-\Wa-\Wa' \Big) + \frac{2}{p+1}(2-N-\WN + p - 1) - \t - \t' =2
    \nonumber\\
    D[V_i^-] + D[\hat v_{p-i-1}^-] &= -F(a+a') - t-t' =0
    \label{eq: one monopole cond3}
\end{align}
which sets $t+t'=0$ so that we get $D = -a\, A' - t\, T'$ in the dual theory, implying $T' = - T$ and $A' = - A$. By adding $R[\hat V_\a^+]$ and $R[\hat v_\a^+]$ in \eqref{eq: one monopole cond1} together with \eqref{eq: one monopole cond3}, one can solve,
\begin{align}
    p\,F - N - \WN - p + \a = 0
\end{align}
which reads the correct dual gauge rank $\WN = pF-N-p+\a$ as a simple consistency check.

We also confirm the perfect matching of the superconformal indices of the theories in $\text{KP}_\a^+$ for various cases as a non--trivial duality check. We provide the results in appendix \ref{app: index result}. Furthermore, we could exactly match the squashed three-sphere partition function for $p=2,3$, with arbitrary $F,N,\a$ using the integral identities from Aharony and $\text{BBP}^\pm$ dualities together with the $D^{3d}_p[SU(N)]$ deconfined picture. More investigations in this direction will be discussed in future work \cite{future}.

\paragraph{Two monopole terms}
Given the $\text{KP}_\a^\pm$ duality, we can derive another duality with two monopole terms, i.e., $\D W_A = \hat V_\a^+ + \hat V_\b^-$, by assuming the KP duality as in the following duality chain:
\begin{equation}
\centering
\begin{tikzpicture}
  \tikzset{vertex/.style={circle,fill=white!25,minimum size=12pt,inner sep=2pt}}
  \tikzset{every loop/.style={}}
    \node[vertex] (eNc) at (-1,0) [shape=circle,draw=black,minimum size=2em] {$N$};
    \node[vertex] (eNf1) at (1,0.7) [shape=rectangle,draw=black,minimum height=2em, minimum width=2em] {$F$};
    \node[vertex] (eNf2) at (1,-0.7) [shape=rectangle,draw=black,minimum height=2em, minimum width=2em] {$F$};

     \draw[-to, min distance=1cm]  (eNc) edge [out=150, in=210] node {} (eNc);
    \draw[->-=.5] ([yshift= 2pt] eNc.east) to ([yshift= 0pt] eNf1.west);
    \draw[->-=.5] ([yshift= 0pt] eNf2.west) to ([yshift= -2pt] eNc.east);

    \node at (-1.7,0.6) {\tiny$X$};
    \node at (0,0.7) {\tiny$Q$};
    \node at (0,-0.7) {\tiny$\WQ$};
    \node at (0,-1.5) {\scriptsize$W_A=  \hat V_\a^+ + \hat V_\b^- + \Tr X^{p+1}$};

    \node at (3.5,0.4) {\scriptsize$\text{KP}_{\a,\b}$};
    \node at (3.5,0) {$\longleftrightarrow$};
    
    \node[vertex] (mNc) at (6,0) [shape=circle,draw=black,minimum size=2em] {$\WN$};
    \node[vertex] (mNf1) at (8,0.7) [shape=rectangle,draw=black,minimum height=2em, minimum width=2em] {$F$};
    \node[vertex] (mNf2) at (8,-0.7) [shape=rectangle,draw=black,minimum height=2em, minimum width=2em] {$F$};

     \draw[-to, min distance=1cm]  (mNc) edge [out=150, in=210] node {} (mNc);
    \draw[->-=.5] ([yshift= 0pt] mNf1.west) to ([yshift= 2pt] mNc.east);
    \draw[->-=.5] ([yshift= -2pt] mNc.east) to ([yshift= 0pt] mNf2.west);
    \draw[->-=.5] ( mNf2.north) to ( mNf1.south);

    \node at (7-1.7,0.6) {\tiny$x$};
    \node at (7,0.7) {\tiny$\Wq$};
    \node at (7,-0.7) {\tiny$q$};
    \node at (8.5,0) {\tiny$M_i$};
    \node at (7,-1.5) {\scriptsize$W_B = \hat v_\a^+ + \hat v_\b^- + \Tr x^{p+1} + M_i\, \Wq x^{p\texttt{-}i\texttt{-}1} q$};

    \node[vertex] (eeNc) at (-1,-4) [shape=circle,draw=black,minimum size=2em] {$N'$};
    \node[vertex] (eeNf1) at (1,-4+0.7) [shape=rectangle,draw=black,minimum height=2em, minimum width=2em] {$F$};
    \node[vertex] (eeNf2) at (1,-4-0.7) [shape=rectangle,draw=black,minimum height=2em, minimum width=2em] {$F$};

     \draw[-to, min distance=1cm]  (eeNc) edge [out=150, in=210] node {} (eeNc);
    \draw[->-=.5]  ([yshift= 0pt] eeNf1.west) to ([yshift= 2pt] eeNc.east);
    \draw[->-=.5]  ([yshift= -2pt] eeNc.east) to ([yshift= 0pt] eeNf2.west);
    \draw[->-=.5] ( eeNf2.north) to ( eeNf1.south);

    \node at (-1.7,-4+0.6) {\tiny$x'$};
    \node at (0,-4+0.7) {\tiny$q'$};
    \node at (0,-4-0.7) {\tiny$\Wq'$};
    \node at (1.5,-4) {\tiny$M'_i$};
    \node at (0,-4-1.5) {\scriptsize$W_{A'}=  \hat U_\a^+ + V_\b^- + V_i^-\,\hat U_{p\texttt{-}i\texttt{-}1}^- $};
    \node at (0.5,-4-2) {\scriptsize$+ \Tr x'^{p+1} + M_i'\Wq'x'^{p\texttt{-}i\texttt{-}1}q'$};

    \node at (3.5,-4+0.2) {$\overset{\text{KP}}{\longleftrightarrow}$};

    \node[vertex] (mmNc) at (6,-4) [shape=circle,draw=black,minimum size=2em] {$\WN'$};
    \node[vertex] (mmNf1) at (8,-4+0.7) [shape=rectangle,draw=black,minimum height=2em, minimum width=2em] {$F$};
    \node[vertex] (mmNf2) at (8,-4-0.7) [shape=rectangle,draw=black,minimum height=2em, minimum width=2em] {$F$};

     \draw[-to, min distance=1cm]  (mmNc) edge [out=150, in=210] node {} (mmNc);
    \draw[->-=.5] ([yshift= 2pt] mmNc.east) to ([yshift= 0pt] mmNf1.west);
    \draw[->-=.5] ([yshift= 0pt] mmNf2.west) to ([yshift= -2pt] mmNc.east);

    \node at (7-1.7,-4+0.6) {\tiny$X'$};
    \node at (7,-4+0.7) {\tiny$\WQ'$};
    \node at (7,-4-0.7) {\tiny$Q'$};
    \node at (7,-4-1.5) {\scriptsize$W_{B'} = U_\a^+ + \hat u_\b^- + \Tr X'^{p+1} + U_i^+ \hat u_{p\texttt{-}i\texttt{-}1}^+$};

    \node at (-0.7,-2.5) {\scriptsize$\text{KP}_\a^+$};
    \node at (0,-2.5){$\updownarrow$};

    \node at (7+0.7,-2.5) {\scriptsize$\text{KP}_\b^-$};
    \node at (7,-2.5){$\updownarrow$};
    
\end{tikzpicture}
\label{eq: KPab quiver}
\end{equation}
which is analogous to the process for deriving the $\text{BBP}_2$ duality in \eqref{eq: BBP2 quiver}. We will denote this two monopole deformed duality as $\text{KP}_{\a,\b}$, whose dual gauge rank $\WN$ can be traced back as,
\begin{align}
    \WN & = pF - \WN' - p + \b
    \nonumber\\
    & = pF - (pF-N') - p + \b
    \nonumber\\
    & = pF - \big(pF- (pF - N - p + \a) \big) - p + \b
    \nonumber\\
    & = p F - N - 2 p + \a + \b \,.
\end{align}
\begin{table}[tbp]
\centering
\begin{tabular}{|c|c|c|c|c|c|}
\hline
$A$ & $A'$ & $B'$ & $B$ & $SU(F)_t\times SU(F)_b$ & $U(1)_R$ \\
\hline
$\Tr X^i$ & $\Tr x^i$ & $\Tr X'^i$ & $\Tr x^i$ & $\bf{1} \times \bf{1}$ & $\frac{2i}{p+1}$ \\
$\WQ X^i Q$ & $M'_i$ & $\WQ' X'^i Q'$ & $M_i$ & $\overline{\bf{F}}\times \bf{F}$ & $2\D_Q + \frac{2i}{p+1}$ \\
$\hat V_i^+$ & $\hat U_i^+$ & $U_i^+$ & $\hat v_i^+$ & $\bf{1} \times \bf{1}$ & $2+\frac{2(i-\a)}{p+1}$ \\
$\hat V_i^-$ & $V_i^-$ & $\hat u_i^-$ & $\hat v_i^-$ & $\bf{1} \times \bf{1}$ & $2+\frac{2(i-\b)}{p+1}$ \\
\hline
\end{tabular}
\caption{\label{tab: two monopole KP chiral ring map} Chiral ring generators in each frame for deriving the $\text{KP}_{\a,\b}$ duality and their charges under the global symmetries. $U(1)_A \times U(1)_T$ are completely broken by the monopole superpotential.}
\end{table}
Note that $\b=\a$ reproduces the duality discussed in \cite{Hwang:2022jjs}. Let us explain the derivation in \eqref{eq: KPab quiver}. Consider $U(N)$ SQCD with $F$ flavors $(Q,\WQ)$ and adjoint $X$ with superpotential,
\begin{align}
    W_A = \hat V_\a^+ + \hat V_\b^- + \Tr X^{p+1}
\end{align}
and since there is a linear term of the $\a$-th dressed monopole operator $\hat V_\a^+$, we can apply $\text{KP}_\a^+$ to get, $U(N')$ SQCD ($N'=pF-N-p+\a$) with $F$ flavors $(q',\Wq')$, adjoint $x'$, and superpotential,
\begin{align}
    W_{A'} & = \hat U_\a^+ + V_\b^- + \Tr x'^{p+1} + V_i^-\,\hat U_{p\texttt{-}i\texttt{-}1}^- + M_i'\,\Wq' x'^{p-i-1} q'
\end{align}
where the monopole operators of positive charge are mapped to the monopole operators as, $\hat V_i^+ \to \hat U_i^+$, while the negatively charged ones are mapped to singlets $\hat V_i^- \to V_i^-$ that couple to dual monopole operators $\hat U_{p-i-1}^-$. The mesons are mapped to singlets, $\WQ X^i Q \to M_i'$ coupled to dual mesons $\Wq' x^{p-i-1} q'$. Next, we apply the KP duality to get $U(\WN')$ SQCD $(\WN' = pF - N')$ with $F$ flavors $(Q',\WQ')$, adjoint $X'$, and superpotential,
\begin{align}
    W_{B'} & = U_\a^+ + V_\b^- + \Tr X'^{p+1} + V_i^- \, U_{p\texttt{-}i\texttt{-}1}^- + M_i' \, m_{p\texttt{-}i\texttt{-}1}
    \nonumber\\
    &\qquad + m_{p\texttt{-}i\texttt{-}1}\, \WQ' X'^i Q
    + U_{p\texttt{-}i\texttt{-}1}^+ \hat u_{i}^+ + U_{p\texttt{-}i\texttt{-}1}^- \hat u_{i}^-
\end{align}
where the monopole operators are mapped to singlets, $\hat U_i^\pm \to U_i^\pm$, coupled to dual monopoles $\hat u_{p\texttt{-}i\texttt{-}1}^\pm$, and the mesons are mapped to singlets, $\Wq' x'^i q' \to m_i$, coupled to dual mesons $\WQ' X'^i Q'$. By integrating out the massive singlets $V_i^-$, $U_{i}^-$, $M_i'$, and $m_i$, we get,
\begin{align}
    W_{B'} = U_\a^+ + \hat u_\b^- + \Tr X'^{p+1} + U_{p-i-1}^+\,\hat u_i^+
    \,.
\end{align}
Thanks to the linear monopole term $\hat u_\b^-$, we can apply $\text{KP}_\b^-$ to obtain $U(\WN)$ SQCD ($\WN = pF - \WN' - p + \b$) with $F$ flavors $(q,\Wq)$, adjoint $x$, and superpotential,
\begin{align}
    W_B = \hat v_\a^+ + \hat v_\b^- + \Tr x^{p+1} + M_i\,\Wq x^{p\texttt{-}i\texttt{-}1}q
\end{align}
where the mesons are mapped to singlets, $\WQ' X'^i Q' \to M_i$, coupled to dual mesons, $\Wq x^{p\texttt{-}i\texttt{-}1} q$, and the negatively charged monopoles are mapped to the dual monopoles of negative charge, $\hat u_i^- \to \hat v_i^-$, while the positively charged ones are mapped to singlets, $\hat u_i^+ \to u_i^+$, which are massive and integrated out, generating the linear term $\hat v_\b^-$. This completes the derivation.
The symmetry charges of the chiral rings and their map along the derivation are summarized in Table \ref{tab: two monopole KP chiral ring map} with the resulting chiral ring map,
\begin{align}
    \Tr X^i \qquad&\longleftrightarrow\qquad \Tr x^i
    \nonumber\\
    \WQ X^i Q \qquad&\longleftrightarrow\qquad M_i
    \nonumber\\
    \hat V_i^\pm \qquad&\longleftrightarrow\qquad \hat v_i^\pm
\end{align}
where the Coulomb branch parametrized by $\hat V_j^+$, $\hat v_j^+$ for $j\geq\a$, and $\hat V_l^-$, $\hat v_l^-$ for $l\geq\b$ are lifted \cite{Hwang:2022jjs}. As a duality check, we could confirm the perfect matching of the superconformal indices of the dual theories in $\text{KP}_{\a,\b}$ which is rather guaranteed by the KP and $\text{KP}_\a^\pm$ dualities, see appendix \ref{app: index result}.
\paragraph{F--maximization and relevance check} 
\begin{table}[tbp]
\renewcommand{\arraystretch}{1.1}
\centering
\begin{tabular}{|c|c|c|c|c|}
    \hline
    $N$ & $p$ & $F$ & $\D W_1 = \hat V_\a^+$ & $\D W_2 = \hat V_\b^-$ \\
    \hline
    \multirow{6}{*}{1} & \multirow{2}{*}{2} & 2 & $\a = 0,1$ & $(0,\b)_{\b\leq 1}$,$(1,\b)_{\b\leq 1}$ \\
    & & 3 & $\a= 0$ & (0,0) \\
    \cline{2-5}
    & \multirow{2}{*}{3} & 2 & $\a=0,1$ & $(0,\b)_{\b\leq 2}$,$(1,\b)_{\b\leq 1}$ \\
    & & 3 & $\a=0$ & (0,0) \\
    \cline{2-5}
    & 4 & 2 & $\a=0,1,2$ & $(0,\b)_{\b\leq 3}$,$(1,\b)_{\b\leq 2}$,$(2,\b)_{\b\leq 2}$ \\
    \cline{2-5}
    & 5 & 2 & $\a=0,1,2$ & $(0,\b)_{\b\leq 3}$,$(1,\b)_{\b\leq 3}$,$(2,\b)_{\b\leq 2}$ \\
    \hline
    \multirow{8}{*}{2} & \multirow{3}{*}{2} & 2 & $\a=0,1$ & $(1,\b)_{\b\leq 2}$ \\
    & & 3 & $\a=0,1$ & $(0,\b)_{\b\leq 1}$,$(1,\b)_{\b\leq 1}$ \\
    & & 4 & $\a=0$ & $(0,0)$ \\
    \cline{2-5}
    & \multirow{2}{*}{3} & 2 & $\a=0,1,2$ & $(1,\b)_{\b\leq 2}$,$(2,\b)_{\b\leq 2}$\\
    & & 3 & $\a=0,1$ & $(0,\b)_{\b\leq 1}$,$(1,\b)_{\b\leq 1}$\\
    \cline{2-5}
    & \multirow{2}{*}{4} & 2 & $\a=0,1,2$ & $(0,\b)_{\b\leq 3}$,$(1,\b)_{\b\leq 3}$,$(2,\b)_{\b\leq 3}$\\
    & & 3 & $\a=0,1$ & $(0,\b)_{\b\leq 2}$,$(1,\b)_{\b\leq 1}$\\
    \cline{2-5}
    & 5 & 2 & $\a=0,1,2,3$ & $(0,\b)_{\b\leq 4}$,$(1,\b)_{\b\leq 4}$,$(2,\b)_{\b\leq 3}$,$(3,\b)_{\b\leq 3}$ \\
    \hline
    \multirow{8}{*}{3} & \multirow{2}{*}{2} & 3 & $\a=0,1$ & $(0,\b)_{\b\leq 2}$,$(1,\b)_{\b\leq 2}$\\
    & & 4 & $\a=0,1$ & $(0,\b)_{\b\leq 1}$,$(1,\b)_{\b\leq 1}$ \\
    \cline{2-5}
    & \multirow{2}{*}{3} & 2 & $\a=0,1,2$ & $(2,\b)_{\b\leq 2}$\\
    & & 3 & $\a=0,1$ & $(0,\b)_{\b\leq 2}$,$(1,\b)_{\b\leq 2}$ \\
    \cline{2-5}
    & \multirow{2}{*}{4} & 2 & $\a=0,1,2,3$ & $(2,\b)_{\b\leq 3}$,$(3,\b)_{\b\leq 3}$\\
    & & 3 & $\a=0,1,2$ & $(0,\b)_{\b\leq 3}$,$(1,\b)_{\b\leq 2}$,$(2,\b)_{\b\leq 2}$ \\
    \cline{2-5}
    & \multirow{2}{*}{5} & 2 & $\a=0,1,2,3$ & $(1,\b)_{\b\leq 4}$,$(2,\b)_{\b\leq 4}$,$(3,\b)_{\b\leq 4}$\\
    & & 3 & $\a=0,1,2$ & $(0,\b)_{\b\leq 3}$,$(1,\b)_{\b\leq 2}$,$(2,\b)_{\b\leq 2}$ \\
    \hline
\end{tabular}
\caption{\label{tab: relevance} Relevant monopole operators from the F-maximization computations. We start with $U(N)$ gauge theory with $F$ flavors and an adjoint $X$ with superpotential $W = \Tr X^{p+1}$. Then the first monopole superpotential deformation $\D W_1 = \hat V_\a^+$ is relevant for $\a$ as in the table and triggers RG to a distinct fixed point of the IR duality $\text{KP}_\a^+$, at which, the second monopole superpotential deformations $\D W_2 = \hat V_\b^-$ are relevant for $(\a,\b)$ as in the last column, triggering another RG flows to the fixed points of the $\text{KP}_{\a,\b}$ duality. Detailed procedures are discussed in appendix \ref{app: f-max result}.}
\end{table}
If we think of $\text{KP}_\a^\pm$ and $\text{KP}_{\a,\b}$ as obtained by the monopole superpotential deformations from the adjoint SQCD, it is crucial to check the relevance of the deformation which can be done by the F--maximization \cite{Jafferis:2010un}. We could confirm the relevance of the deformations for many cases as in table \ref{tab: relevance}, but not for all the possible cases, see appendix \ref{app: f-max result} for detailed procedure. Since the RG flow depends on the ordering of a sequence of deformations, irrelevant monopole operators in one frame might be turned on in the IR theory via other sequences of RG flows. The irrelevant monopole operators in our RG still have such a possibility. On the other hand, one may find other UV completions as similarly found in \cite{Benini:2017dud} for the BBP duality, where the UV completion of the monopole deformed 3d $\CN=2$ SQCD is given by coupling copies of auxiliary Ising-SCFTs.
The RG that we adopt starts by assuming the IR fixed point with the adjoint superpotential term $\Tr X^{p+1}$, say $\CT$, then check whether $\hat V_\a^+$ is relevant from the F--maximization. If so, we deform the theory by turning on $\hat V_\a^+$ to trigger an RG flow to a new fixed point with the monopole operator superpotential term, say $\CT_\a$. Here, we again check the relevance of $\hat V_\b^-$, and the relevant ones will trigger further RG flows to new fixed points with two linear monopole terms, say $\CT_{\a,\b}$. It is worth mentioning that, for some cases, $\CT_{\a,\b}$ is not allowed, but it is for $\CT_{\b,\a}$, revealing the importance of the ordering of RG flows, see appendix \ref{app: f-max result} for details.

\subsection{Extension of the dressing number}
Interestingly, the proposed duality $\text{KP}_\a^\pm$ still works even if we extend the dressing number of the monopole operator in the superpotential term:
\begin{align}
    0 \leq \a < p
    \quad \longrightarrow \quad
    0 \leq \a \leq 2p
    \,.
\end{align}
The dressed monopole operators $\hat V_\a^\pm$ are not well-defined with the conventional definition of \eqref{eq: dressed monopole} for $\a \geq p$ due to the F-term condition, $X^p = 0$. However, let us continue to use the same notation $\hat V_\a^\pm$ and $\hat v_\a^\pm$ to conveniently denote the composite operators for $p\leq \a \leq 2p$,
\begin{align}
    \hat V_\a^\pm \equiv 
    \sum_{i=0}^{p-1} c_i\,(\Tr' X^{\a-p+1-i} )\, 
    \hat V_{i}^\pm
    \;\;\;\;\;,\qquad
    \hat v_\a^\pm \equiv 
    \sum_{i=0}^{p-1} \Wc_i\,
    (\Tr' x^{\a-p+1-i} )\, \hat v_{i}^\pm
    \,,
    \label{eq: extended dressed monopoles}
\end{align}
with some coefficients $c_i$ and $\Wc_i$. Again, $\Tr' X^{\a-p+1-i}$ represents a possible linear combination of products of traces whose sum of powers is $\a-p+1-i$, similarly for $x$ as well. Namely, \eqref{eq: extended dressed monopoles} means that we properly attach traces of adjoint operators to dressed monopole operators such that the total stack of adjoints in $\hat V_\a^\pm$ and $\hat v_\a^\pm$ to be $\a$. With this convention, let us keep using the same notations $\text{KP}_\a^\pm$ for $p\leq \a \leq 2p$ cases also\footnote{There is a subtlety for large $p$ that some of the traces of $X$ and $x$ with low powers decouple by the unitarity bound. Nevertheless, we could check for various $p$ that there always exists a possible partitioning of the powers of adjoints such that composite operators $\hat V_\a^\pm$ and $\hat v_\a^\pm$ for $p\leq \a \leq 2p$.}.

\paragraph{Duality checks} 
\begin{table}[tbp]
\centering
\begin{tabular}{|c|c|c|}
    \hline
    $\a$ & $N$ & Dualization $(A') \to (B')$ \\
    \hline
    \hline
    $2$ & odd,even & $\underset{2}{\text{A}} \to \underset{1}{\text{A}}$ \\
    \hline
    \multirow{2}{*}{$3$} & odd & $\underset{2}{\text{A}} \to \underset{2}{\text{B}^+}\to \underset{2}{\text{A}} \to \underset{1}{\text{A}}$ \\
    \cline{2-3}
     & even & $\underset{2}{\text{A}} \to \underset{1}{\text{A}}\to \underset{1}{\text{B}^+} \to \underset{1}{\text{A}}$ \\
    \hline
    \multirow{2}{*}{$4$} & odd & $\underset{2}{\text{A}} \to \underset{1}{\text{A}}\to \underset{1}{\text{B}^+} \to \underset{2}{\text{B}^+}\to\underset{2}{\text{A}} \to \underset{1}{\text{A}} $  \\
    \cline{2-3}
     & even & $\underset{2}{\text{A}} \to \underset{2}{\text{B}^+}\to \underset{2}{\text{A}} \to \underset{1}{\text{A}}\to\underset{1}{\text{B}^+} \to \underset{1}{\text{A}} $ \\
    \hline
\end{tabular}
\caption{\label{tab: dualization} Sequences of the dualization from $(A')$ to $(B')$ in \eqref{eq: p=2 KPa quiver} to derive $\text{KP}_\a^+$ duality for $p=2$ with $\a \geq 2$ cases. The procedures $\underset{j}{\text{A}}$ and $\underset{j}{\text{B}^+}$ denote Aharony and $\text{BBP}^+$ dualities at the $j$-th gauge node respectively.}
\end{table}
We could confirm the perfect matching of the superconformal indices of the $\text{KP}_\a^\pm$ dual theories with $p\leq \a\leq 2p$ for various cases whose results are organized in appendix \ref{app: index result}. Since we did not assume any bound on $\a$ in the previous subsection, all the quantum number analysis work as well. To illustrate why this extension works, let us provide another explicit duality check by matching the squashed three-sphere partition function from the deconfined picture for $p=2$ case ($\D = 1/3$):
\begin{equation}
        \begin{tikzpicture}[thick,scale=0.9, every node/.style={scale=0.9}]
  \tikzset{vertex/.style={circle,fill=white!25,minimum size=12pt,inner sep=2pt}}
  \tikzset{every loop/.style={}}
    
    
    \node[vertex] (mp-1) at (4.5,0) [shape=circle,draw=black,minimum size=2em] {\tiny$\lfloor \frac{N}{2} \rfloor$};
    \node[vertex] (Nc) at (6,0) [shape=circle,draw=black,minimum size=2em] {$N$};
    \node[vertex] (eNf1) at (8,0.8) [shape=rectangle,draw=black,minimum height=2em, minimum width=2em] {$F$};
    \node[vertex] (eNf2) at (8,-0.8) [shape=rectangle,draw=black,minimum height=2em, minimum width=2em] {$F$};

    \draw[-to, min distance=1cm]  (mp-1) edge [out=120, in=60] node {} (mp-1);
    \draw[->-=.5] ([yshift= 2pt] mp-1.east) to ([yshift= 2pt] Nc.west);
    \draw[->-=.5] ([yshift= -2pt] Nc.west) to ([yshift= -2pt] mp-1.east);
    \draw[->-=.5] ([yshift= 2pt] Nc.east) to ([yshift= 0pt] eNf1.west);
    \draw[->-=.5] ([yshift= 0pt] eNf2.west) to ([yshift= -2pt] Nc.east);


    \node at (4,2) {($A'$)};

    \node at (4.5,1.3) {$2\texttt{-}2\D$};
    \node at (5.3,0.4) {$\D$};
    \node at (6.8,0.8) {$\D_Q$};

    \node at (4.5,-0.8) {$\t_1$};
    \node at (6,-0.8) {$\t_2$};

    \node at (10.5,0.7) {Sequential};
    \node at (10.5,0.3) {dualities};
    \node at (10.5,-0.2) {\Large $\Longrightarrow$};
    \node at (10.5,-0.7) {as table \ref{tab: dualization}};
    
    
    \node[vertex] (dmp-1) at (13.5,0) [shape=circle,draw=black,minimum size=2em] {\tiny$\lfloor \frac{\WN}{2} \rfloor$};
    \node[vertex] (dNc) at (15,0) [shape=circle,draw=black,minimum size=2em] {$\WN$};
    \node[vertex] (dNf1) at (17,0.8) [shape=rectangle,draw=black,minimum height=2em, minimum width=2em] {$F$};
    \node[vertex] (dNf2) at (17,-0.8) [shape=rectangle,draw=black,minimum height=2em, minimum width=2em] {$F$};

    \draw[-to, min distance=1cm]  (dmp-1) edge [out=120, in=60] node {} (dmp-1);
    \draw[->-=.5] ([yshift= 2pt] dNc.west) to ([yshift= 2pt] dmp-1.east);
    \draw[->-=.5] ([yshift= -2pt] dmp-1.east) to ([yshift= -2pt] dNc.west);
    \draw[->-=.5] ([yshift= 0pt] dNf1.west) to ([yshift= 2pt] dNc.east);
    \draw[->-=.5] ([yshift= -2pt] dNc.east) to ([yshift= -2pt] dNf2.west);
    \draw[->-=.7] ([xshift= -5pt] dNf2.north) to ([xshift= -5pt] dNf1.south);
    \draw[->-=.7] ([xshift= 5pt] dNf2.north) to ([xshift= 5pt] dNf1.south);
    

    \node at (12.5,2) {($B'$)};

    \node at (13.5,1.3) {$2\texttt{-}2\D$};
    \node at (14.3,0.4) {$\D$};
    \node at (15.4,0.8) {$2\D\texttt{-}\D_Q$};
    \node at (18,0.2) {$2\D_Q$};
    \node at (18.3,-0.2) {$2\D_Q\texttt{+}2\D$};

    \node at (13.5,-0.8) {$\widetilde{\t}_1$};
    \node at (15,-0.8) {$\widetilde{\t}_2$};

    \node at (6.3-1,+3.5) 
    {\begin{tikzpicture}
      \node [rotate=90] {\LARGE\textcolor{black}{$\Longleftarrow$}};    
    \end{tikzpicture}
    };
    \node at (6.3-2.3,+3.5) {Deconfine};

    \node at (15.3,+3.5) 
    {\begin{tikzpicture}
      \node [rotate=-90] {\LARGE\textcolor{black}{$\Longleftarrow$}};    
    \end{tikzpicture}
    };
    \node at (15.3+1.3,+3.5) {Confine};
    
    \node[vertex] (kNc) at (5,+6) [shape=circle,draw=black,minimum size=2em] {$N$};
    \node[vertex] (keNf1) at (7,0.8+6) [shape=rectangle,draw=black,minimum height=2em, minimum width=2em] {$F$};
    \node[vertex] (keNf2) at (7,-0.8+6) [shape=rectangle,draw=black,minimum height=2em, minimum width=2em] {$F$};

    \draw[-to, min distance=1cm]  (kNc) edge [out=120, in=60] node {} (kNc);
    \draw[->-=.5] ([yshift= 2pt] kNc.east) to ([yshift= 0pt] keNf1.west);
    \draw[->-=.5] ([yshift= 0pt] keNf2.west) to ([yshift= -2pt] kNc.east);


    \node at (4,1.5+6) {($A$)};
    \node at (5,1.3+6) {$2\D$};
    \node at (5,-0.8+6) {$\t$};
    \node at (6,1+6) {$\D_Q$};

    \node[vertex] (kmNc) at (14,+6) [shape=circle,draw=black,minimum size=2em] {$\WN$};
    \node[vertex] (kmNf1) at (16,0.8+6) [shape=rectangle,draw=black,minimum height=2em, minimum width=2em] {$F$};
    \node[vertex] (kmNf2) at (16,-0.8+6) [shape=rectangle,draw=black,minimum height=2em, minimum width=2em] {$F$};

    \draw[-to, min distance=1cm]  (kmNc) edge [out=120, in=60] node {} (kmNc);
    \draw[->-=.5] ([yshift= 0pt] kmNf1.west) to ([yshift= 2pt] kmNc.east);
    \draw[->-=.5] ([yshift= -2pt] kmNc.east) to ([yshift= 0pt] kmNf2.west);
    \draw[->-=.7] ([xshift= -5pt] kmNf2.north) to ([xshift= -5pt] kmNf1.south);
    \draw[->-=.7] ([xshift= 5pt] kmNf2.north) to ([xshift= 5pt] kmNf1.south);

    
    \node at (13,1.5+6) {($B$)};
    \node at (14,1.3+6) {$2\D$};
    \node at (15,1+6) {\scriptsize $2\D\texttt{-}\D_Q$};
    \node at (14,-0.8+6) {$\widetilde{\t}$};

    \node at (17,6+0.2) {$2\D_Q$};
    \node at (17.3,6-0.2) {$2\D_Q \texttt{+} 2\D$};

    \node at (10.5,0.3+6) {$\text{KP}_\a^+$};
    \node at (10.5,-0.2+6) {\Large $\Longleftrightarrow$};

\end{tikzpicture}
\label{eq: p=2 KPa quiver}
\end{equation}
where the quivers here represent the integral expression of the squashed three-sphere partition function, see appendix \ref{app: index f-max} for the recipe. In those quivers, we write the real mass parameters near the corresponding matters, i.e., arrows\footnote{We omit to write the real masses of the gauge and $SU(F)_t\times SU(F)_b$ symmetries since these are obvious from the diagram so that writing them is redundant.} and the real mass for the $U(1)_T$ topological symmetry, i.e., FI parameter, is written below each gauge node. Then the R--charge 2 condition from monopole superpotential term $\hat V_\a^+$ in the theory ($A$) for $\a \geq 2$ gives,
\begin{align}
    R[\hat V_\a^+] = F(1-\D_Q) - 2\D(N-1-\a) + \t = 2
\end{align}
which restricts the FI parameter $\t$. By deconfining the adjoint matter, we get theory ($A'$) and the sequential dualizations in table \ref{tab: dualization} give rise to the theory ($B'$) with $\WN = 2F-N-2+\a$ which is again the deconfined form, thus, upon confining the quiver tail, we get dual adjoint SQCD, theory ($B$) with the expected dual rank and FI parameter,
\begin{align}
    \widetilde{\t} = -\t - 2\D (2-\a)
\end{align}
which correctly implies the monopole superpotential term $\hat v_\a^+$ in the dual theory ($B$):
\begin{align}
    R[\hat v_\a^+] = F(1-2\D+\D_Q) - 2\D(\WN - 1 - \a) + \widetilde{\t} = 2.
\end{align}
The singlet contributions $V_i^-$ that couple to the dual monopole operators can also be perfectly traced along the derivation. To summarize, we derived the partition function matching of the $\text{KP}_\a^+$ duality for $p=2$ with $2\leq \a\leq 4$ by using the deconfined picture, which is strong evidence for the duality.
Although it seems natural to explore the region $\a > 2p$, we did not find any example in which the superconformal index matches once the bound $\a = 2p$ is exceeded. Namely, $\text{KP}_\a^\pm$ does not work for $\a > 2p$. We find consistency of the bound from the deconfined picture. Note that the dual gauge rank $\WN = pF-N-p+\a$ is bigger than that of the original $\text{KP}$ duality for $\a > p$. To achieve such cases from successive dualizations in the deconfined picture, we need to first start by applying Aharony duality from the right--most to the left gauge nodes, possibly mixing with $\text{BBP}^\pm$ duality. Also, we should end the dualization at the left--most gauge node with Aharony duality. Along the procedure, each node undergoes $\text{BBP}^\pm$ dualization at most once. Consequently, there are at most $p$ of $\text{BBP}^\pm$ dualizations through the whole procedure, whose applied number is the same as the amount of reduced dual gauge rank compared to the original $\text{KP}$ duality, i.e., $\a-p$. Thus, we have a condition,
\begin{align}
    \a-p \leq p
\end{align}
which gives rise to the observed upper bound, $\a \leq 2p$.
\paragraph{Two monopoles} We can also consider turning on two dressed monopole operators $\D W_A = \hat V_\a + \hat V_\b \leftrightarrow \D W_B = \hat v_\a + \hat v_\b$ with the extended dressing numbers $0\leq \a,\b \leq 2p$. Note that the extended definition of the dressed monopole operators in \eqref{eq: extended dressed monopoles} is just composites of the usual monopole operators and traces of adjoint. So, we can follow the same duality chain derivation in \eqref{eq: KPab quiver} without any additional cost by just separately mapping the composite factors with the chiral ring map in table \ref{tab: two monopole KP chiral ring map}, achieving $\text{KP}_{\a,\b}$ with extended range $0\leq \a,\b \leq 2p$. We could also confirm the perfect matching of the superconformal indices for the dual theories in $\text{KP}_{\a,\b}$ with $0\leq \a,\b \leq 2p$ for various cases as well, see appendix \ref{app: index result} for the results.
\paragraph{Amariti--Cassia--Garozzo--Mekareeya duality} Let us revisit Amariti--Cassia--Garozzo--Mekareeya(ACGM) dualities \cite{Amariti:2019rhc} in terms of our $\text{KP}_\a^\pm$ and $\text{KP}_{\a,\b}$ dualities.
\begin{itemize}
    \item $\D W_A = \big( \hat V_{\frac{p-1}{2}}^+ \big)^2\;\;\longleftrightarrow \;\; \D W_B = \big( \hat v_{\frac{p-1}{2}}^+ \big)^2 +  V_i^-\,\hat v_{p\texttt{-}i\texttt{-}1}^-$
    \\
    \\
    Let us first consider a single quadratic term case with odd $p$ whose dual gauge rank is $\WN = pF-N$. The R--charge 2 conditions from the superpotential term are,
    \begin{align}
        &R\big[\hat V_{\frac{p-1}{2}}^+\big] = F(1-\D_Q) - \frac{2}{p+1}\big(N-1-\frac{p-1}{2}\big) + \t =1
        \nonumber\\
        &R\big[\hat v_{\frac{p-1}{2}}^+\big] = F\Big(1-(\frac{2}{p+1}-\D_Q)\Big) - \frac{2}{p+1}\big(\WN-1-\frac{p-1}{2}\big) + \t' =1
    \end{align}
    where $\t$ and $\t'$ are the mixing parameters for the $U(1)_T$ topological symmetry. One can check that these conditions imply,
    \begin{align}
        R[\hat V_{p}^+] = R[\hat v_{p}^+] = 2
    \end{align}
    which means that at the IR fixed point by the quadratic monopole superpotential deformation, the $p$-th dressed linear monopole operator is exactly marginal and parametrizes a non-trivial conformal manifold. Note that the dual gauge rank and the matter contents of the quadratic deformed ACGM and $\text{KP}_p^+$ are exactly the same. Thus, for the two dualities, the supersymmetric observables such as the superconformal index and squashed three-sphere partition function are the same. 
    
    More generally, let $\mathfrak{D}$ be an IR duality of two distinct UV theories, $\CT_A$ and $\CT_B$ that flow to the same IR fixed point by some relevant deformation $\D W_A$ and $\D W_B$ respectively, where the fixed point lies on possibly some conformal manifold. On the other hand, let $\mathfrak{D}'$ be another IR duality of the same pair of the UV theories by another deformation $\D W_A'$ and $\D W_B'$. If the IR fixed point of $\mathfrak{D}'$ lies on the same conformal manifold, we would say the dualities $\mathfrak{D}$ and $\mathfrak{D}'$ are {\it equivalent} up to conformal manifold, or just simply equivalent in the sense that all the BPS observables of the theories in the two dualities are the same. Thus, the duality $\text{KP}_p^+$ is equivalent to the ACGM duality with quadratic monopole deformation. An interesting point is that $p$ can naturally be an even integer with $\text{KP}_p^+$ perspective, while it is not with the ACGM perspective.

    \item $\D W_A = \big( \hat V_{\frac{p-1}{2}}^+ \big)^2 + \hat V_0^-\;\;\longleftrightarrow \;\; \D W_B = \big( \hat v_{\frac{p-1}{2}}^+ \big)^2 + \hat v_0^-$
    \\
    \\
    Similarly, the quadratic and linear monopole superpotential deformation implies an equivalent duality with deformation,
    \begin{align}
        \D W_A' = \hat V_p^+ + \hat V_0^-
        \quad \longleftrightarrow \quad 
        \D W_B' =
        \hat v_p^+ + \hat v_0^-
    \end{align}
    which gives rise to $\text{KP}_{p,0}$ whose dual gauge rank and the matter contents are the same as those of the ACGM duality with quadratic and bare monopole deformation. So, we would say the two dualities are equivalent.

    \item $\D W_A = \big( \hat V_{\a}^+ \big)^2 + 
    \big( \hat V_{p\texttt{-}\a\texttt{-}1}^+ \big)^2
    \;\;\longleftrightarrow \;\; 
    \D W_B = \big( \hat v_{\a}^+ \big)^2 + 
    \big( \hat v_{p\texttt{-}\a\texttt{-}1}^+ \big)^2$
    \\
    \\
    As a last case, consider two quadratic monopole superpotential deformations whose dual gauge rank is $\WN = pF-N$. The R--charge 2 conditions imply,
    \begin{align}
        R\big[\hat V_{\a\texttt{+}\frac{p+1}{2}}^+\big] = R\big[\hat V_{p\texttt{-}\a\texttt{+}\frac{p-1}{2}}^+\big] = 2
        \nonumber\\
        R\big[\hat v_{\a\texttt{+}\frac{p+1}{2}}^+\big] = R\big[\hat v_{p\texttt{-}\a\texttt{+}\frac{p-1}{2}}^+\big] = 2
    \end{align}
    which tell us an equivalence to the $\text{KP}_{\a\texttt{+}\frac{p+1}{2},p\texttt{-}\a\texttt{+}\frac{p-1}{2}}$ duality for odd $p$, whose dual gauge rank and the matter contents are also the same with those of two quadratic monopole deformed AGCM duality. For even $p$, we don't have $\text{KP}_{a,b}$ interpretation, however, we can understand it in terms of the original KP duality. By the duality map $\hat V_i^\pm \leftrightarrow \hat v_i^\pm$, one can check that the singlet contributions $V_i^\pm$ of the dual theory in KP exactly cancel in the partition function as well as in the superconformal index. Thus, the identities from the KP duality guarantee the matching of the BPS observables for the two quadratic monopole deformed duality. Namely, we would say the match of the BPS observables in this duality is automatic from the KP duality.
    \\
\end{itemize}
In conclusion, our $\text{KP}_\a^\pm$ and $\text{KP}_{\a,\b}$ with an extended range of $\a$ and $\b$, together with the original KP, seem to provide a unified framework for the known monopole deformed dualities. This observation motivates us to bootstrap the landscape of monopole deformed KP dualities toward generic monopole superpotential deformations as discussed in the next section.
Before we move on, let us make some comment. Since operators of the conformal dimension larger than two are irrelevant, our monopole operators with extended dressings are less likely to be relevant deformations and, in fact, except for a few cases, we failed to check the relevance of the monopoles $\hat V_\a^\pm$ with $\a \geq p$ by starting the RG flow from the adjoint SQCD. As mentioned previously, a similar issue in the BBP duality was resolved by considering a UV completion with auxiliary Ising-SCFTs coupled to SQCD \cite{Benini:2017dud}. It would be interesting to identify a similar resolution in our setup.
\paragraph{Summary}
Let us finish this section with a summary. In this section, we proposed new 3d $\CN=2$ IR dualities:
\begin{equation}
\begin{tikzpicture}
  \tikzset{vertex/.style={circle,fill=white!25,minimum size=12pt,inner sep=2pt}}
  \tikzset{every loop/.style={}}
    \node[vertex] (eNc) at (-1,0) [shape=circle,draw=black,minimum size=2em] {$N$};
    \node[vertex] (eNf1) at (1,0.7) [shape=rectangle,draw=black,minimum height=2em, minimum width=2em] {$F$};
    \node[vertex] (eNf2) at (1,-0.7) [shape=rectangle,draw=black,minimum height=2em, minimum width=2em] {$F$};

    \draw[-to, min distance=1cm]  (eNc) edge [out=150, in=210] node {} (eNc);
    \draw[->-=.5] ([yshift= 2pt] eNc.east) to ([yshift= 0pt] eNf1.west);
    \draw[->-=.5] ([yshift= 0pt] eNf2.west) to ([yshift= -2pt] eNc.east);

    \node at (-1.6,0.6) {\tiny$X$};
    \node at (0,0.7) {\tiny$Q$};
    \node at (0,-0.7) {\tiny$\WQ$};
    \node at (0,-1.5) {$W_A= \Tr X^{p+1} + \D W_A$};

    \node at (3.5,0.4) {\scriptsize $\text{KP}_\a^\pm$ or $\text{KP}_{\a,\b}$};
    \node at (3.5,0) {$\longleftrightarrow$};

    \node[vertex] (mNc) at (7,0) [shape=circle,draw=black,minimum size=2em] {$\WN$};
    \node[vertex] (mNf1) at (9,0.7) [shape=rectangle,draw=black,minimum height=2em, minimum width=2em] {$F$};
    \node[vertex] (mNf2) at (9,-0.7) [shape=rectangle,draw=black,minimum height=2em, minimum width=2em] {$F$};

    \draw[-to, min distance=1cm]  (mNc) edge [out=150, in=210] node {} (mNc);
    \draw[->-=.5] ([yshift= 0pt] mNf1.west) to ([yshift= 2pt] mNc.east);
    \draw[->-=.5] ([yshift= -2pt] mNc.east) to ([yshift= 0pt] mNf2.west);
    \draw[->-=.5] ( mNf2.north) to ( mNf1.south);

    \node at (6.4,0.6) {\scriptsize$x$};
    \node at (8,0.7) {\tiny$\Wq$};
    \node at (8,-0.7) {\tiny$q$};
    \node at (9.5,0) {\tiny$M_i$};
    \node at (8,-1.5) {$W_B = \Tr x^{p+1} \!+\! \sum_{i=0}^{p-1} M_i \,\Wq x^{p\texttt{-}i\texttt{-}1} q + \D W_B$};

\end{tikzpicture}
\nonumber
\end{equation}
with
\begin{align}
    \text{KP}_\a^\pm
    \; &: \;\;
    \WN = pF-N-p+\a \;\;,\;\;\;\;\;\;
    \D W_A = \hat V_\a^\pm 
    \;\leftrightarrow \;
    \D W_B = \hat v_\a^\pm + \sum_{i=0}^{p-1} V_i^\mp\,\hat v_{p\texttt{-}i\texttt{-}1}^\mp
    \nonumber\\
    \text{KP}_{\a,\b}
     &: \;\;
    \WN = pF-N-2p+\a+\b \;\;,\;\;
    \D W_A = \hat V_\a^+ + \hat V_\b^-
    \;\leftrightarrow \;
    \D W_B = \hat v_\a^+ + \hat v_\b^+
    \label{eq: duality proposal}
\end{align}
for $0\leq \a,\b \leq 2p$ with symmetry charges as in table \ref{tab: one monopole KP charge} and \ref{tab: two monopole KP chiral ring map}. We could confirm the perfect matching of the superconformal index and the squashed three-sphere partition function for various cases. We could also check the relevance of the monopole operator deformations for various cases. We interpret the ACGM dualities with the proposed dualities together with the original KP duality. We emphasize that $\text{KP}_{\a,\b}$ follows from $\text{KP}_\a^\pm$ and $\text{KP}$, where the latter two follow from the Aharony and $\text{BBP}^\pm$ dualities via the deconfined picture.

\section{Generic monopole deformations} \label{sec: bootstrap}
In this section, we classify all possible monopole deformed Kim--Park dualities by considering generic monopole superpotential deformations. Let us begin with an investigation of the chiral ring generators and their possible maps under the Seiberg--like dualities of the 3d $\CN=2$ adjoint SQCD of $U(N)$ gauge group, $F$ flavors $(Q,\WQ)$, and superpotential $W=\Tr X^{p+1}$ of the adjoint $X$. There are three kinds of gauge invariant chiral ring generators:
\begin{align}
    \Tr X^i
    \;,\quad
    \WQ X^i Q
    \;,\quad
    \hat V_i^\pm\;,
\end{align}
for $i = 0,\cdots,p-1$, with the truncation by the F-term equation, $X^p=0$. Let us call them traces of adjoint, mesons, and dressed monopole operators respectively. On the other hand, we can think of a candidate Seiberg--like dual adjoint SQCD of gauge group $U(\WN)$, adjoint $x$ with superpotential term $\Tr x^{p+1}$, and $F$ flavors $(q,\Wq)$. We assume the universal meson--singlet coupling superpotential term in the dual theory, $\sum_{i=0}^{p-1} M_i\,\Wq x^{p-i-1}q$, where the singlets $M_i$ are mapped from the mesons of the original theory. By assuming the Higgs branch of the original theory is mapped to the Higgs branch of the dual theory, the map for the traces of adjoints and mesons is uniquely determined to respect the global symmetry charges:
\begin{align}
    \Tr X^i \qquad&\longleftrightarrow\qquad \Tr x^i
    \nonumber\\
    \WQ X^i Q \qquad&\longleftrightarrow\qquad M_i
\end{align}
By setting the R-charge of the flavors $(Q,\WQ)$ and $(q,\Wq)$ as $\D_Q$ and $\D_q$ respectively\footnote{The value of $\D_Q$, so as $\D_q$, can be controlled by mixing $U(1)_A$ symmetry to the $U(1)_R$ R-symmetry.}, the R-cherge 2 condition from the meson--singlet coupling term gives ($\D \equiv \frac{1}{p+1}$ for convenience),
\begin{align}
    \D_Q + \D_q= 2\D\,,
\end{align}
from which we can write the R-charge of the dressed monopole operators by counting the fermion zero modes as,
\begin{align}
    R[\hat V_i^\pm] = R_0 + 2i\,\D \pm \t
    \;\;,\qquad
    R[\hat v_i^\pm] = r_0 + 2i\,\D \pm \widetilde{\t}
\end{align}
with the R-charge of the bare monopole operators,
\begin{align}
    R_0 &\equiv F(1-\D_Q) - 2\D (N-1)
    \nonumber\\
    r_0 &\equiv F\big( 1 - (2\D - \D_Q) \big) - 2\D (\WN - 1)
\end{align}
where $\t$ and $\widetilde{\t}$ are mixing parameters of the respective $U(1)_T$ topological symmetries from the $U(N)$ and $U(\WN)$ gauge group. Note that if we set $\WN = p\,F-N-f$ for some integer $f$ which measures a difference of the dual gauge rank compared to that of the KP, we find
\begin{align}
    R_0 + r_0 = 2(f+2)\D\,.
    \label{eq: f}
\end{align}
Since R-charge spectrum of the dressed monopole operators increases along the dressing number $i$, each family of positively and negatively charged dressed monopole operators, say $\hat V_i^\pm$, can be mapped either to the dual monopole operators $\hat v_i^\pm$, or to singlets $V_i^\pm$ which couple to the dual monopole operators $\hat v_{p-i-1}^\pm$ for respective $i$, whose R-charge decreases with $i$. With the coupling terms in the latter case, the degrees of freedom of the dual monopole operators are replaced by the singlets from the F-term condition. Here, we don't redundantly consider mapping the monopole operators to oppositely charged dual monopoles or singlets because such cases can always be interpreted as our cases by taking the opposite sign convention for the dual $U(1)_T$ topological symmetry. There may be other maps for matching the Coulomb branch spectrum in an exotic way other than ours, however, we stick to rather natural our options. Then, there are three options for mapping the Coulomb branch, each of which is discussed below:
\begin{itemize}
    \item Type-1 : $(\hat V_i^+,\hat V_i^-) \quad \leftrightarrow \quad (\hat v_i^+,\hat v_i^-)$
    \\
    \\
    Let us denote Type-1 for the map if all the monopole operators are mapped to the dual monopole operators. In this case, the condition from the R-charge matching gives,
    \begin{align}
        &R[\hat V_i^\pm] = R[\hat v_i^\pm]
        \;\;\;\;\Rightarrow \;\; R_0 = r_0 \;,\;\; \t = \widetilde{\t}
        \label{eq: case1-1}
    \end{align}
    and combined with the condition \eqref{eq: f}, we find
    \begin{align}
        R_0 = (f+2)\D
        \label{eq: case1-2}
    \end{align}
    which means the R--charge spectrum of the monopole operators is simply determined by $f$ and $\t$: 
    \begin{align}
        R[\hat V_i^\pm] = R[\hat v_i^\pm] = (f+2i+2)\D \pm \t \,.
    \end{align}

    \item Type-2 :  $(\hat V_i^+,\hat V_i^-) \quad \leftrightarrow \quad (\hat v_i^+,V_i^-)$
    \\
    \\
    We denote Type-2 for the map that positively charged monopoles are mapped to the dual monopoles of positive charge, while the negatively charged ones are mapped to singlets coupled to the negatively charged dual monopoles. In this case, we have conditions from the R-charge spectrum matching and the singlet-monopole coupling superpotential terms as,
    \begin{align}
        &R[\hat V_i^+] = R[\hat v_i^+]
        \;\;,\;\;\;
        R[ V_i^-] + R[\hat v_{p-i-1}^-] = 2
        \nonumber\\
        &\qquad\Rightarrow \;\;
        R_0 = \widetilde{\t} + 2\D \;,\;\; r_0 = \t + 2\D
        \label{eq: case2-1}
    \end{align}
    and from $\eqref{eq: f}$, we have
    \begin{align}
       2\D\,f = \t + \widetilde{\t}\,.
       \label{eq: case2-2}
    \end{align}
    So, one can check,
    \begin{align}
        &R\big[ \hat V_{p-f}^+ \big] = R_0 + 2\D(p-f) + \t = 2
        \nonumber\\
        &R\big[ \hat v_{p-f}^+ \big] = r_0 + 2\D(p-f) + \widetilde{\t} = 2\,,
    \end{align}
    which is the condition for the $\text{KP}_{p-f}^+$ duality. Indeed, if we set the dressing number $\a = p-f$, the dual gauge rank $\WN = pF - N - p + (p-f) = pF-N-p+\a$ exactly reproduces the expected one. Hence, a duality with Type-2 map always can be interpreted as $\text{KP}_{p-f}^+$ if $0\leq p-f \leq 2p$, or $|f|\leq p$. According to the examination in the previous section, the duality fails to work outside of the range.

    \item Type-3 : $(\hat V_i^+,\hat V_i^-) \quad \leftrightarrow \quad (V_i^+,V_i^-)$
    \\
    \\
    We denote Type-3 if all the monopole operators are mapped to singlets coupled to the dual monopole operators. In this case, we have the R-charge 2 conditions,
    \begin{align}
        & R[ V_i^\pm] + R[\hat v_{p-i-1}^\pm] = 2
        \;\;\Rightarrow\;\;\;
        R_0 + r_0 = 4\D \;\,\;\; 
        \t+ \widetilde{\t} = 0
    \end{align}
    which implies $f=0$ from \eqref{eq: f}. Thus, with a Type-3 map, there is no possibility of getting a dual gauge rank other than that of the KP duality. In fact, this is exactly the KP duality.
\end{itemize}
As one can see, the possible 3d $\CN=2$ Seiberg--like dualities of the adjoint SQCD are rather restricted from the chiral ring map. In particular, for the Type-3, the gauge rank is uniquely fixed as $\WN = pF-N$, leaving $\t = - \widetilde{\t}$ as a free parameter, reproducing the original KP duality. Also, for the Type-2, whenever $f$ is an integer in a range $|f|\leq p$, it just becomes $\text{KP}_{p-f}^+$. In the remaining of this section, we investigate to find new dualities inequivalent to the known ones: KP, $\text{KP}_\a^\pm$, $\text{KP}_{\a,\b}$, by considering a generic form of monopole superpotential deformations certainly with the Type-1 map.
Before we start the search, let us make a few remarks. As shortly argued, if IR fixed points of the two IR dualities $\mathfrak{D}$ and $\mathfrak{D}'$ of the same pair of UV theories lie on the same conformal manifold, we would say the two dualities are equivalent up to conformal manifold and we will not distinguish them further since our classification will be based on the superconformal index which is blind of the conformal manifold.

Also, if the matching of two BPS observables in a duality is guaranteed by a simple parameter tuning of a known BPS observable identity from another duality, we refer to the former as being {\it automatic} by the latter. An example is provided by the ACGM dualities with one or two quadratic monopole deformations. Their BPS observables automatically match due to the BPS observable identities of the KP duality. Although the details of two such dualities may be distinct, we will treat the former as derived one in the sense that its BPS observable matching is ensured by a known duality.

The last remark is that, for KP duality, we can introduce at most two non-collinear terms of products of monopoles. Here, non-collinear means that, the $U(1)_T \times U(1)_A \times U(1)_R$ symmetry charges of the two deformation terms are not aligned. Turning on more than two non-collinear terms breaks the $U(1)_R$ R-symmetry, thereby destroying the superconformal symmetry.

\subsection{Single-term deformation}
Let us search for the possible monopole deformed KP dualities with a single term of generic monopole superpotential:
\begin{align}
    \D W_A = \Tr X^\a \big(\hat V_0^+ \big)^m \big(\hat V_0^- \big)^n
    \qquad\longleftrightarrow\qquad
    \D W_B = \Tr x^\a \big(\hat v_0^+ \big)^m \big(\hat v_0^- \big)^n
    \label{eq: single generic monopole}
\end{align}
where $\a,m,n \geq 0$ are non-negative integers. We take the deformation shape on both sides to be the same as analogous to the known ones. All the possible deformations consisting of dressed monopole operators can be thought of equivalently as a choice of non--negative integer parameters ($\a,m,n$) \footnote{For example, a superpotential deformation $\D W_A = (\hat V_3^+)^2(\hat V_2^-)$ is equivalent to \eqref{eq: single generic monopole} with $(\a,m,n)=(5,2,1)$ in the quantum number sense.}, in the sense that we are looking for additional constraints on quantum numbers from the deformation to determine the possible IR dualities. Namely, the R--charge 2 conditions from \eqref{eq: single generic monopole} give rise to constraints,
\begin{align}
    &2\D \a + (m+n)R_0 + (m-n)\t = 2
    \nonumber\\
    &2\D \a + (m+n)r_0 + (m-n)\widetilde{\t} = 2\,.
    \label{eq: single monopole condition}
\end{align}
Let us examine whether these conditions pinpoint down to a new duality that is {\it inequivalent} to known ones. In the remaining, we will only focus on the map of Type-1 since Type-2 and Type-3 always reproduce $\text{KP}_\a^\pm$ and KP respectively.

\paragraph{Type-1} If $m\neq 0$ and $n\neq 0$, there is no way but to adopt the Type-1 map since both positively and negatively charged monopole operators should be mapped to each of the dual ones respectively to be consistent with the deformation \eqref{eq: single generic monopole}. So, from the constraints of the map, \eqref{eq: case1-1} and \eqref{eq: case1-2}, we find the R-charge 2 condition \eqref{eq: single monopole condition} boils down to,
\begin{align}
    (m-n)\t = \D\big(2p+2-2\a-(m+n)(f+2)\big)
    \label{eq: case1 single term}
\end{align}
which means that for $m\neq n$, the FI parameter $\t$ is fixed once the dual gauge rank is chosen by $f$. Note that if $m=n$, $\t$ becomes a free parameter and the dual gauge rank is fixed as $f = \frac{p+1-\a}{m}-2$. Let us discuss those possibilities soon.

\paragraph{Type-2} Although we know the Type-2 map always gives $\text{KP}_\a^+$ interpretation, let us illustrate when it happens. Without loss of generality, if $n=0$, the negatively charged monopoles do not enter the deformation and there arises a loophole for adopting the Type-2 map by mapping the negative charge monopoles to singlets. So, combined with the constraints \eqref{eq: case2-1} and \eqref{eq: case2-2} of the Type-2 map, the R-charge two conditions \eqref{eq: single monopole condition} fix the dual gauge rank measure $f$ as,
\begin{align}
    f = \frac{1}{m}(p+1-\a) -1
    \label{eq: single f case 2}
\end{align}
so that whenever $f$ is an integer in a range $|f|\leq p$, we always have $\text{KP}_{p-f}^+$ duality interpretation. In other words, when $n=0$ and $(\a,m)$ are chosen such that $f$ is an integer in a range $|f|\leq p$, the resulting monopole deformed duality with generic deformation \eqref{eq: single generic monopole} is equivalent to $\text{KP}_{p-f}^+$. Outside of the range, we could not find any instance of the duality.

\paragraph{Searching for new duality}
To sum up, a possibility of new monopole deformed KP duality that is inequivalent to the known ones can only arise with the Type-1 map as we consider the generic single term deformation \eqref{eq: single generic monopole}. In this work, we restrict the searching up to quadratic powers of monopoles for simplicity, i.e., $m+n\leq 2$, however, the procedure is straightforward to be extended to higher-power term deformations, $m+n > 2$. Without loss of generality, only three kinds of monopole deformations are possible with $m+n\leq 2$:
\begin{align}
    &(m,n) = (1,0) 
    \;\; : \;\;
    \D W_A = \Tr X^\a \,\hat V_0^+ = \hat V_\a^+
    \nonumber\\
    &(m,n) = (2,0) 
    \;\; : \;\;
    \D W_A = \Tr X^\a \,\big( \hat V_0^+\big)^2
    \nonumber\\
    &(m,n) = (1,1) 
    \;\; : \;\;
    \D W_A = \Tr X^\a \,\hat V_0^+\,\hat V_0^-\,.
\end{align}
Obviously, the case $(m,n)=(1,0)$ is exactly the $\text{KP}_\a^+$ duality with Type-2 map, correctly reproducing the dual gauge rank measure $f=p-\a$ from the formula \eqref{eq: single f case 2} with a range $|f|\leq p$. The case $(m,n)=(2,0)$ also corresponds to $\text{KP}_{p-f}^+$ whenever $f=\frac{p-\a-1}{2}$ is an integer in a range $|f|\leq p$. Namely, these are all equivalent to the known duality.

\begin{table}[tbp]
\renewcommand{\arraystretch}{1.1}
\centering
\begin{tabular}{|c|c|c|}
    \hline
    $(p,F,\a,f)$ & $(N,\WN)$ & $(\D_Q,\t)$ \\
    \hline
    (1,2,2,1) & (0,1) & $(\frac{3}{4},-\frac{3}{2})$ \\
    (1,2,0,-1) & (1,2) & $(\frac{3}{4},\frac{3}{2})$ \\
    (1,3,1,2) & (0,1) & $(\frac{2}{3},-1)$ \\
    (1,3,2,1) & (0,2) & $(\frac{5}{6},-\frac{3}{2})$ \\
    (1,3,2,2) & (0,1) & $(\frac{2}{3},-2)$ \\
    (1,3,0,-1) & (1,3) & $(\frac{5}{6},\frac{3}{2})$ \\
    (2,2,4,1) & (1,2) & $(\frac{1}{2},-\frac{5}{3})$ \\
    (2,2,0,-1) & (2,3) & $(\frac{1}{2},\frac{5}{3})$ \\
    (2,3,0,-1) & (3,4) & $(\frac{4}{9},\frac{5}{3})$ \\
    (3,2,0,-1) & (3,4) & $(\frac{3}{8},\frac{7}{4})$ \\
    \hline
\end{tabular}
\;\;\;
\begin{tabular}{|c|c|c|}
    \hline
    $(p,F,\a,f)$ & $(N,\WN)$ & $(\D_Q,\t)$ \\
    \hline
    $(1,2,2n\!-\!\!1,1)_{n=1,2}$ & (0,1) & $(\frac{3}{4},\texttt{-}n)$ \\
    $(1,2,2n\!-\!\!1,1)_{n=1,2}$ & (1,2) & $(\frac{3}{4},1\texttt{-}n)$ \\
    $(1,3,2n\!-\!\!1,1)_{n=1,2}$ & (0,2) & $(\frac{5}{6},\texttt{-}n)$ \\
    $(1,3,2n\!-\!\!1,2)_{n=1,2}$ & (0,1) & $(\frac{2}{3},\texttt{-}n\texttt{-}\frac{1}{2})$ \\
    $(1,3,2n\!-\!\!1,-1)_{n=1,2}$ & (1,3) & $(\frac{5}{6},1\texttt{-}n)$ \\
    $(2,2,2n,1)_{0\leq n \leq 3}$ & (1,2) & $(\frac{1}{2},\texttt{-}\frac{2n}{3})$ \\
    $(2,2,2n,-1)_{0\leq n \leq 3}$ & (2,3) & $(\frac{1}{2},\frac{2-2n}{3})$ \\
    $(2,3,2n,-1)_{0\leq n \leq 3}$ & (3,4) & $(\frac{4}{9},\frac{2-2n}{3})$ \\
    $(3,2,2n\!-\!\!1,-1)_{1\leq n \leq 5}$ & (3,4) & $(\frac{3}{8},1\texttt{-}\frac{n}{2})$ \\
    \hline
\end{tabular}
\caption{\label{tab: m12} Up to $p\leq 5$, $F\leq 3$, $N\leq 3$, $f\leq 10$, $\a\leq 2p$, and $N+\WN\leq 7$, we examine all the series expandable superconformal indices for candidate dualities with Type-1, $(m,n)=(1,0)$ for the LHS table, and $(m,n)=(2,0)$ for the RHS table respectively, however, we {\it don't} find any index matching except the cases that are equivalent to the known dualities, KP and $\text{KP}_{\g,\d}$.}
\end{table}
One may wonder what happens if we insist on the duality map of Type-1, not the Type-2 for $(m,n)=(1,0),(2,0)$. A quick answer is that we could {\it not} find any new duality, see table \ref{tab: m12} for the trials of the superconformal index matching that all failed. Let us describe the procedure of the scanning. By fixing an integer $f\in\mathbb{Z}$, i.e., fixing the target dual gauge rank, the FI parameter $\t$ is fixed to a specific value as given in \eqref{eq: case1 single term}. These are enough data to compute the indices of candidate dual theories for a given setup parametrized by $(p,F,N,\a)$. Up to $p\leq 5$, $F\leq 3$, $N\leq 3$, $\a\leq 2p$, and $N+\WN \leq 7$, we have checked all the possible cases with $\D_Q , R_0 >0$ for series expandability as in the table \ref{tab: m12}, that all failed to match the indices. We have intentionally excluded the cases that are equivalent to KP, $\text{KP}_{\g}^\pm$, and $\text{KP}_{\g,\d}$ to avoid the known duality interpretations. This result indicates that there is no new duality for $(m,n)=(1,0),(2,0)$ with the duality map of Type-1.

\begin{table}[tbp]
\renewcommand{\arraystretch}{1.1}
\centering
\begin{tabular}{|c|c|c|}
    \hline
    $(p,F,\a)$ & $(N,\WN)$ & $\D_Q$ \\
    \hline
    *(1,1,1) & (1,1) & 1/2 \\
    (1,2,1) & (1,2) & 3/4 \\
    (1,3,1) & (1,3) & 5/6 \\
    (1,3,1) & (2,2) & 1/2 \\
    (2,2,1) & (1,2) & 1/2 \\
    (2,2,2) & (2,3) & 1/2\\
    (2,3,0) & (2,3) & 4/9 \\
    (2,3,2) & (3,4) & 4/9 \\
    \hline
\end{tabular}
\;
\begin{tabular}{|c|c|c|}
    \hline
    $(p,F,\a)$ & $(N,\WN)$ & $\D_Q$ \\
    \hline
    *(3,1,1) & (1,1) & 1/4 \\
    *(3,1,3) & (2,2) & 1/4 \\
    *(3,2,0) & (2,2) & 1/4 \\
    (3,2,1) & (2,3) & 3/8 \\
    (3,2,3) & (3,4) & 3/8 \\
    (3,3,0) & (3,4) & 1/3 \\
    *(4,1,1) & (1,1) & 1/5 \\
    (4,2,2) & (3,4) & 3/10 \\
    (5,2,1) & (3,4) & 1/4 \\
    \hline
\end{tabular}
\caption{\label{tab: VV} Up to $p\leq 5$, $F\leq 3$, $N\leq 3$, $\a\leq 2p$, and $N+\WN\leq 7$, we examine all the series expandable superconformal indices of candidate dualities from $(m,n)=(1,1)$ monopole superpotential deformation, however, we could not find any index matching evidence except the cases that are equivalent to the known dualities, KP and $\text{KP}_{\g,\d}$. Here *-mark denotes the cases of trivial matching, not by duality. }
\end{table}

Now, let us consider the $(m,n)=(1,1)$ case, which has no way but to adopt the Type-1 duality map. Note that $\t$ in \eqref{eq: case1 single term} is a free parameter meaning the $U(1)_T$ topological symmetry is not broken as one can see in the superpotential $\D W_A = \Tr X^\a\, \hat V_0^+ \hat V_0^-$. Also, the dual gauge rank is fixed by $f = p - \a - 1$ which fixes the R-charge of the flavors $\D_Q$ from \eqref{eq: case1-2}. Thus, for a given theory parametrized by $(p,F,N,\a)$, we can read the index of the candidate dual theory. However, we could not find any instance of duality, see table \ref{tab: VV} for the list of the candidate dualities that we scanned but {\it failed} to work. The *-marked cases denote exceptional ones that the indices do match for an obvious reason. All such cases can happen for $N=\WN$. In fact, whenever $N = \WN$ the duality map of the Type-1 fixes the R-charge of the flavors as $\D_Q = \D$, so as that of the dual flavors, $2\D - \D_Q = \D$. For a given $(p,F,N)$, let the superconformal index as $\CI(a_i,b_i,w;q)$, where $a_i$ and $b_i$ are the fugacities for $SU(F)_t\times SU(F)_b$ flavor symmetries, while $w$ is the fugacity for the unbroken $U(1)_T$ topological symmetry, and $q$ is the expansion parameter that is the fugacity for the R-symmetry. Then the index matching of the *-marked cases can be represented as,
\begin{align}
    \CI(a_i,b_i,w;q) \overset{?}{=} \CI(a_i^{-1},b_i^{-1},w;q) \times \CI_{M}
    \label{eq: self dual ind}
\end{align}
where $\CI_{M}$ is the contribution from the mesonic singlets, $M_{j=0,\cdots,p-1}$. We put a question mark on the equality since this does not hold in general but only does for $F\leq 2$ as we will explain. To see this easily, let us express the index in terms of the single particle index by taking the plethystic log:
\begin{align}
    \CP\CI(a_i,b_i,w;q)-\CP\CI(a_i^{-1},b_i^{-1},w;q)
    \
    \overset{?}{=}
    \sum_{i,j=1}^F \sum_{k=1}^{p} ( a_i b_j - a_i^{-1}b_j^{-1} )q^{\frac{2k}{p+1}}
    \label{eq: self dual PL}
\end{align}
where $\CP\CI := (1-q^2)\text{PL}[\CI]$ reads the single particle index. The RHS corresponds to $(1-q^2)\text{PL}[\CI_M]$. Note that $a_i$ and $b_i$ satisfy the traceless relation from the $SU(F)$ groups, $\prod_{i=1}^F a_i = \prod_{i=1}^F b_i = 1$, so when $F=1$, \eqref{eq: self dual PL} trivially works:
\begin{align}
    \CP\CI(1,1,w;q)-\CP\CI(1,1,w;q)
    =
    0\,.
\end{align}
For $F=2$, by setting $a_1 = a_2^{-1} \equiv a$ and $b_1 = b_2^{-1} \equiv b$, one can see that the RHS vanishes,
\begin{align}
    \sum_{i,j=1}^2 \sum_{k=1}^{p} ( a_i b_j - a_i^{-1}b_j^{-1} )q^{\frac{2k}{p+1}} = 0
\end{align}
and from the localization formula, one can see that $\CI(a_i,b_i,w;q)$ is invariant under the permutations of $a_i$ and $b_i$, and in particular for $F=2$, the only non-trivial permutation $a_1\leftrightarrow a_2$ and $b_1\leftrightarrow b_2$ coincide to inversing fugacities, $a \leftrightarrow a^{-1}$ and $b \leftrightarrow b^{-1}$ respectively so that the LHS of \eqref{eq: self dual PL} also vanishes, confirming the equality. For $F>2$, both sides in \eqref{eq: self dual PL} becomes non--trivial and we could check they do not match in general. Since the mesonic singlet contribution, which is the only part that physically distinguishes the LHS and RHS theories, just vanishes for $F\leq 2$, what we are looking at regarding the index matchings from the *-marked cases is nothing but an index identity of a single theory. Hence, it is hard to say they represent duality and we exclude them from the duality searching\footnote{There were exactly the same situations as the *-marked for $(m,n)=(1,0),(2,0)$ with the map of Type-1, however, we intentionally excluded them in the duality searching at there.}.
Let us summarize the present subsection. By only assuming the possible chiral ring map, we tried to find new duality from a constraint by a single term of generic monopole deformation. Up to the quadratic powers of monopoles, we could not find any new duality and the index computation tells us that the only non-trivial index matchings always come from the identities of the known dualities, KP and $\text{KP}_\a^\pm$. In the next section, we continue the duality classification with two terms of generic monopole deformations.

\subsection{Two-term deformation}
Let us now consider a generic monopole deformation with two terms:
\begin{align}
    \D W_A = \Tr X^\a &\big(\hat V_0^+ \big)^m \big(\hat V_0^- \big)^n
    +
    \Tr X^\b \big(\hat V_0^+ \big)^k \big(\hat V_0^- \big)^l
    \nonumber\\
    &\;\longleftrightarrow\qquad
    \D W_B = \Tr x^\a \big(\hat v_0^+ \big)^m \big(\hat v_0^- \big)^n
    +
    \Tr x^\b \big(\hat v_0^+ \big)^k \big(\hat v_0^- \big)^l
    \label{eq: two generic monopole}
\end{align}
for non--negative integers, $\a,m,n,\b,k,l$. Since otherwise always equivalent to the known dualities, let us assume the monopole map of the Type-1 for searching new duality. Then the R-charge 2 conditions combined with constraints from the type-1 map \eqref{eq: case1-1} and \eqref{eq: case1-2} uniquely fix the dual gauge rank and the FI parameter as,
\begin{align}
    f = \frac{\text{det} 
    \left(
    \begin{array}{cc}
       m\!-\!n  \;& \; p\!+\!1\!-\!\a  \\
       k\!-\!l  \;& \; p\!+\!1\!-\!\b
    \end{array}
    \right)
    }{\text{det} 
    \left(
    \begin{array}{cc}
        m & n \\
        k & l
    \end{array}
    \right)
    }-2
    \;\;,\;\;
    \t = -\frac{\text{det} 
    \left(
    \begin{array}{cc}
       m\!+\!n  \;& \; p\!+\!1\!-\!\a  \\
       k\!+\!l  \;& \; p\!+\!1\!-\!\b
    \end{array}
    \right)
    }{\text{det} 
    \left(
    \begin{array}{cc}
        m & n \\
        k & l
    \end{array}
    \right)
    }\D
\end{align}
unless $ml-nk=0$, in which case either there is no solution for solving the quantum numbers, i.e., superconforaml symmetry is broken, or the two terms serve the same condition, effectively the same situation with a single term monopole deformation that is already considered in the previous subsection. Thus, we only consider $ml-nk\neq 0$ cases.
In this stage, any choice of parameters $(\a,m,n,\b,k,l)$ such that $f$ becomes an integer seems to give a candidate duality and one may wonder whether it indeed works. However, the answer is negative at least up to quadratic terms, i.e., $m+n\leq 2$ and $k+l\leq 2$. To avoid overlaps with known dualities, we first sort out the cases equivalent to $\text{KP}_{\widetilde{\a},\widetilde{\b}}$ which imposes the R-charge 2 conditions for some two dressing numbers $\widetilde{\a}$ and $\widetilde{\b}$ as,
\begin{align}
    R_0 + 2 \widetilde{\a}\D + \t  = 2
    \;\;,\;\;\;
    R_0 + 2 \widetilde{\b}\D - \t  = 2,.
\end{align}
Using $R_0 = (f+2)\D$, we can express $\widetilde{\a}$ and $\widetilde{\b}$ in terms of the deformation parameters,
\begin{align}
    \widetilde{\a} = p+1 
    +
    \frac{\text{det} 
    \left(
    \begin{array}{cc}
       n  \;& \; p\!+\!1\!-\!\a  \\
       l  \;& \; p\!+\!1\!-\!\b
    \end{array}
    \right)
    }{\text{det} 
    \left(
    \begin{array}{cc}
        m & n \\
        k & l
    \end{array}
    \right)
    }
    \;\;,\;\;\;\;
    \widetilde{\b} = p+1 
    -
    \frac{\text{det} 
    \left(
    \begin{array}{cc}
       m  \;& \; p\!+\!1\!-\!\a  \\
       k  \;& \; p\!+\!1\!-\!\b
    \end{array}
    \right)
    }{\text{det} 
    \left(
    \begin{array}{cc}
        m & n \\
        k & l
    \end{array}
    \right)
    }
    \label{eq: at bt}
\end{align}
with
\begin{align}
    f = 2p-\widetilde{\a}-\widetilde{\b}
    \;\;,\qquad
    \t = ( \widetilde{\b} - \widetilde{\a} ) \D \,.
\end{align}
Therefore, whenever the parameters $(\a,m,n,\b,k,l)$ make $\widetilde{\a}$ and $\widetilde{\b}$ to be integers in a range $0\leq \widetilde{\a},\widetilde{\b} \leq 2p$, the corresponding duality with superpotential deformation \eqref{eq: two generic monopole} is equivalent to $\text{KP}_{\widetilde{\a},\widetilde{\b}}$. Since we are looking for a new duality inequivalent to the known ones, our scope, then, would be:
\begin{align}
    f = 2p - \widetilde{\a} - \widetilde{\b} \in \mathbb{Z} \setminus \{0\}
    \;\;\;\;\text{and}\;\;\;
    \widetilde{\a} , \widetilde{\b} \notin \mathbb{Z}
    \label{eq: two monopole scope}
\end{align}
where the last non-integer condition ensures impossible to be interpreted as $\text{KP}_{\widetilde{\a},\widetilde{\b}}$. Note that regardless of the integrality of $\widetilde{\a}$ and $\widetilde{\b}$, the BPS observables of a candidate dual theories will always match if $f=0$ by the identities from the KP duality. Since such cases automatically hold no matter what values $\widetilde{\a}$ and $\widetilde{\b}$ are, we exclude $f=0$ cases from the searching scope for the purpose of looking for a new duality.
In this work, we examine generic monopole deformations \eqref{eq: two generic monopole} up to quadratic powers of monopoles, i.e., $m+n\leq 2$, $k+l\leq 2$. We leave further investigations of higher powers to forthcoming work. Without loss of generality, all the possible deformations are,\footnote{We omit $(m,n,k,l)=(1,1,1,1)$ case, that is $ml-nk=0$, since it will preserve supersymmetry if and only if $\a=\b$ which is equivalent to a single monopole deformed case that already discussed in the previous subsection.}
\begin{align}
    &(m,n,k,l)=(1,0,0,1)
    \;\;:\;\;
    \D W_A = \Tr X^\a \hat V_0^+ + \Tr X^\b \hat V_0^-
    \nonumber\\
    &(m,n,k,l)=(1,0,0,2)
    \;\;:\;\;
    \D W_A = \Tr X^\a \hat V_0^+ + \Tr X^\b \big(\hat V_0^-\big)^2
    \nonumber\\
    &(m,n,k,l)=(1,0,1,1)
    \;\;:\;\;
    \D W_A = \Tr X^\a \hat V_0^+ + \Tr X^\b \hat V_0^+ \hat V_0^-
    \nonumber\\
    &(m,n,k,l)=(2,0,0,2)
    \;\;:\;\;
    \D W_A = \Tr X^\a \big(\hat V_0^+\big)^2 + \Tr X^\b \big( \hat V_0^-\big)^2
    \nonumber\\
    &(m,n,k,l)=(2,0,1,1)
    \;\;:\;\;
    \D W_A = \Tr X^\a \big(\hat V_0^+\big)^2 + \Tr X^\b \hat V_0^+ \hat V_0^-
\end{align}
where we omit to write $\D W_B$. From \eqref{eq: at bt}, we read $\widetilde{\a}$ and $\widetilde{\b}$ as,
\begin{align}
    &(m,n,k,l)=(1,0,0,1)
    \;\;:\;\;
    \widetilde{\a} = \a \;,\; \widetilde{\b} = \b
    \nonumber\\
    &(m,n,k,l)=(1,0,0,2)
    \;\;:\;\;
    \widetilde{\a} = \a \;,\; \widetilde{\b} = \frac{p+1+\b}{2}
    \nonumber\\
    &(m,n,k,l)=(1,0,1,1)
    \;\;:\;\;
    \widetilde{\a} = \a \;,\; \widetilde{\b} = p+1-\a+\b
    \nonumber\\
    &(m,n,k,l)=(2,0,0,2)
    \;\;:\;\;
    \widetilde{\a} = \frac{p+1+\a}{2} \;,\; \widetilde{\b} = \frac{p+1+\b}{2}
    \nonumber\\
    &(m,n,k,l)=(2,0,1,1)
    \;\;:\;\;
    \widetilde{\a} = \frac{p+1+\a}{2} \;,\; \widetilde{\b} = \b+\frac{p+1-\a}{2}
\end{align}
and by insisting integrality of the dual gauge rank measure $f=2p-\widetilde{\a}-\widetilde{\b}$, the numbers $\widetilde{\a}$ and $\widetilde{\b}$ of the first three cases also become integers, thus, these cases always be equivalent to $\text{KP}_{\widetilde{\a},\widetilde{\b}}$ if $0\leq \widetilde{\a},\widetilde{\b} \leq 2p$, with no duality outside the range. On the other hand, the last two cases have possibilities for new dualities inequivalent to the known ones:
\begin{table}[tbp]
\renewcommand{\arraystretch}{1.1}
\centering
\begin{tabular}{|c|c|c|c|}
    \hline
    $p$ & $(\a,\b)$ & $F$ & $(N,\WN)$ \\
    \hline
    \multirow{5}{*}{1} & \multirow{5}{*}{(1,1)} & 2 & (1,2) \\
    &  & 3 & (1,3),(2,2) \\
    &  & 4 & (1,4),(2,3) \\
    &  & 5 & (1,5),(2,4),(3,3) \\
    &  & 6 & (1,6),(2,5),(3,4) \\
    \hline
    \multirow{5}{*}{2} & (0,0) & 2 & (1,2) \\
    & (0,4),(2,2) & 2 & (2,3) \\
    & (0,0) & 3 & (2,3) \\
    & (0,4),(2,2) & 3 & (3,4) \\
    & (0,0) & 4 & (2,5),(3,4) \\
    \hline
    \multirow{2}{*}{3} & (1,1) & 2 & (2,3) \\
    & (1,5),(3,3) & 2 & (3,4) \\
    \hline
    \multirow{2}{*}{4} & (0,0) & 2 & (2,3) \\
    & (0,4),(2,2) & 2 & (3,4) \\
    \hline
    5 & (1,1) & 2 & (3,4) \\ 
    \hline
    6 & (0,0) & 2 & (3,4) \\
    \hline
\end{tabular}
\;\;
\begin{tabular}{|c|c|c|c|}
    \hline
    $p$ & $(\a,\b)$ & $F$ & $(N,\WN)$ \\
    \hline
    \multirow{5}{*}{1} & \multirow{5}{*}{(1,1)} & 2 & (1,2) \\
    &  & 3 & (1,3),(2,2) \\
    &  & 4 & (1,4),(2,3) \\
    &  & 5 & (1,5),(2,4),(3,3) \\
    &  & 6 & (1,6),(2,5),(3,4) \\
    \hline
    \multirow{5}{*}{2} & $(2j,0)_{j\leq 2}$ & 2 & (1,2) \\
    & $(2j,2)_{j\leq 2}$ & 2 & (2,3) \\
    & $(2j,0)_{j\leq 2}$ & 3 & (2,3) \\
    & $(2j,2)_{j\leq 2}$ & 3 & (3,4) \\
    & $(2j,0)_{j\leq 2}$ & 4 & (2,5),(3,4) \\
    \hline
    \multirow{3}{*}{3} & $(2j+1,1)_{j\leq 3}$ & 2 & (2,3) \\
    & $(2j+1,3)_{j\leq 3}$ & 2 & (3,4) \\
    & $(2j+1,0)_{j\leq 3}$ & 3 & (3,4) \\
    \hline
    4 & $(2j,0)_{j\leq 5}$ & 2 & (2,3) \\
    \hline
    5 & $(2j+1,1)_{j\leq 6}$ & 2 & (3,4) \\
    \hline
\end{tabular}
\caption{\label{tab: S12} Up to $N+\WN \leq 7$, we have examined all the series of expandable superconformal indices in the $\mathcal{S}_1$ case as in the LHS table, and found {\it no} instances of index matching. Likewise, we check all the cases for $\mathcal{S}_2$ up to $N+\WN \leq 6$ and some for $N+\WN = 7$ as in the RHS table, {\it neither} found any index matching.}
\end{table}
\begin{itemize}
    \item $(m,n,k,l)=(2,0,0,2)$\\
    In this case, the integrality of $f$ insists $\a+\b \in 2\mathbb{Z}$ and whenever $p+\a$ is odd, $p+\b$ is also odd by integrality of $f$, making, $\widetilde{\a}$, $\widetilde{\b}$ simultaneously to be integers, so, $\text{KP}_{\widetilde{\a},\widetilde{\b}}$ interpretation is available if $0\leq \widetilde{\a},\widetilde{\b} \leq 2p$, outside of which range does not give duality. Therefore, to find new duality, we scan the rest cases, $p+\a \in 2\mathbb{Z}$:
    \begin{align}
        \mathcal{S}_1 :=
        \big\{(p,\a,\b)\,\big|\, \a+\b \neq 2p-2,\; p \!\!\!\!\!\mod 2 = \a \!\!\!\!\!\mod 2 = \b \!\!\!\!\!\mod 2 = 0\; \text{or}\; 1 \big\}
    \end{align}
    where the condition $\a+\b\neq 2p-2$ is to avoid the KP automatic cases.

    \item $(m,n,k,l)=(2,0,1,1)$\\
    The integrality of $f$ for this case does not constrain anything on $\a$ and $\b$. To avoid the $\text{KP}_{\widetilde{\a},\widetilde{\b}}$ equivalent duality, $p+\a$ should be even so that $\widetilde{\a}$ and $\widetilde{\b}$ are non-integer. Hence, what we need to scan for new duality is,
    \begin{align}
        \mathcal{S}_2 :=
        \big\{ (p,\a,\b)\,\big|\, \b\neq p-1,\; p\!\!\!\!\!\mod 2 = \a \!\!\!\!\! \mod 2 = 0 \;\text{or}\; 1 \big\}
    \end{align}
    where $\b \neq p-1$ is due to avoid the KP equivalent cases.
    
\end{itemize}
We have checked that all the series expandable superconformal indices from the candidate dual theories of $\mathcal{S}_1$ do not match up to $p\leq 6$, $F\leq 6$, $N + \WN \leq 7$, similarly for $\mathcal{S}_2$ neither, see table \ref{tab: S12} for the failure check of candidate dualities. This suggests the only possible monopole deformed KP dualities with generic monopole deformations \eqref{eq: two generic monopole} up to quadratic terms are always equivalent to the original KP, or $\text{KP}_{\widetilde{\a},\widetilde{\b}}$ with $0 \leq \widetilde{\a},\widetilde{\b} \leq 2p$. Namely, up to all the quadratic monopole superpotential deformations, the only possible dualities that we could find are always equivalent to the {\it linear} monopole superpotential deformed dualities. We wonder how far we can push this statement through the higher-power monopole deformations and it would be interesting if one can show this is always the case.
To summarize the section, we classify a landscape of 3d $\CN=2$ Seiberg--like dualities with an adjoint matter by only assuming a consistent duality map of chiral ring elements, together with constraints from generic monopole superpotential deformations. Up to quadratic powers of monopole superpotential deformation terms, we scan all possible candidate dualities by checking the index matching, however, we find that the working ones are always equivalent to KP, or $\text{KP}_\a^\pm$, or $\text{KP}_{\a,\b}$.

\section{Conclusion} \label{sec: conclusion}
In this work, we propose two 3d $\CN=2$ IR dualities of adjoint SQCD with linear monopole superpotential deformations:
\begin{equation}
\begin{tikzpicture}
  \tikzset{vertex/.style={circle,fill=white!25,minimum size=12pt,inner sep=2pt}}
  \tikzset{every loop/.style={}}
    \node[vertex] (eNc) at (-1,0) [shape=circle,draw=black,minimum size=2em] {$N$};
    \node[vertex] (eNf1) at (1,0.7) [shape=rectangle,draw=black,minimum height=2em, minimum width=2em] {$F$};
    \node[vertex] (eNf2) at (1,-0.7) [shape=rectangle,draw=black,minimum height=2em, minimum width=2em] {$F$};

    \draw[-to, min distance=1cm]  (eNc) edge [out=150, in=210] node {} (eNc);
    \draw[->-=.5] ([yshift= 2pt] eNc.east) to ([yshift= 0pt] eNf1.west);
    \draw[->-=.5] ([yshift= 0pt] eNf2.west) to ([yshift= -2pt] eNc.east);

    \node at (-1.6,0.6) {\tiny$X$};
    \node at (0,0.7) {\tiny$Q$};
    \node at (0,-0.7) {\tiny$\WQ$};
    \node at (0,-1.5) {$W_A= \Tr X^{p+1} + \D W_A$};

    \node at (3.5,0.4) {\scriptsize $\text{KP}_\a^\pm$ or $\text{KP}_{\a,\b}$};
    \node at (3.5,0) {$\longleftrightarrow$};

    \node[vertex] (mNc) at (7,0) [shape=circle,draw=black,minimum size=2em] {$\WN$};
    \node[vertex] (mNf1) at (9,0.7) [shape=rectangle,draw=black,minimum height=2em, minimum width=2em] {$F$};
    \node[vertex] (mNf2) at (9,-0.7) [shape=rectangle,draw=black,minimum height=2em, minimum width=2em] {$F$};

    \draw[-to, min distance=1cm]  (mNc) edge [out=150, in=210] node {} (mNc);
    \draw[->-=.5] ([yshift= 0pt] mNf1.west) to ([yshift= 2pt] mNc.east);
    \draw[->-=.5] ([yshift= -2pt] mNc.east) to ([yshift= 0pt] mNf2.west);
    \draw[->-=.5] ( mNf2.north) to ( mNf1.south);

    \node at (6.4,0.6) {\scriptsize$x$};
    \node at (8,0.7) {\tiny$\Wq$};
    \node at (8,-0.7) {\tiny$q$};
    \node at (9.5,0) {\tiny$M_i$};
    \node at (8,-1.5) {$W_B = \Tr x^{p+1} \!+\! \sum_{i=0}^{p-1} M_i \,\Wq x^{p\texttt{-}i\texttt{-}1} q + \D W_B$};

\end{tikzpicture}
\nonumber
\end{equation}
with
\begin{align}
    \text{KP}_\a^\pm
    \; &: \;\;
    \WN = pF-N-p+\a \;\;,\;\;\;\;\;\;
    \D W_A = \hat V_\a^\pm 
    \;\leftrightarrow \;
    \D W_B = \hat v_\a^\pm + \sum_{i=0}^{p-1} V_i^\mp\,\hat v_{p\texttt{-}i\texttt{-}1}^\mp
    \nonumber\\
    \text{KP}_{\a,\b}
     &: \;\;
    \WN = pF-N-2p+\a+\b \;\;,\;\;
    \D W_A = \hat V_\a^+ + \hat V_\b^-
    \;\leftrightarrow \;
    \D W_B = \hat v_\a^+ + \hat v_\b^+
\end{align}
for $0\leq \a,\b \leq 2p$. We confirm the perfect matching of the superconformal index and conduct the F-maximization to check the relevance of the monopole deformations for various cases. We also reassess the known monopole deformed KP dualities based on $\text{KP}_\a^\pm$, $\text{KP}_{\a,\b}$, and the original KP dualities.
Next, we classify a landscape of 3d $\CN=2$ Seiberg--like dualities with an adjoint matter by only assuming consistent chiral ring maps under duality, together with constraints by generic monopole superpotential deformations. Up to quadratic powers of monopole deformations, the only dualities arise if and only if they are equivalent to $\text{KP}_\a^\pm$, or $\text{KP}_{\a,\b}$, or automatic from KP. 
It is interesting to observe that among the generic possible monopole deformations, only those involving at most linear monopole superpotential terms seem to consistently yield dualities. A possible hint for this phenomenon arises from the deconfined picture of adjoint SQCD. The KP duality can be derived from this deconfined picture under the assumption of Aharony duality. Similarly, as partially illustrated in equation \eqref{eq: p=2 KPa quiver}, the $\text{KP}_\alpha^\pm$ dualities can be obtained by further incorporating the BBP duality, which shifts the dual gauge rank by one relative to the Aharony case. This rank-shifting property is crucial: in order to generate a new duality that is inequivalent to KP, such a shift of the dual gauge rank must occur during the derivation. To our knowledge, the BBP duality is the only known 3d $\mathcal{N}=2$ Seiberg-like duality involving unitary gauge groups that realizes such a shift. However, a key requirement for applying the BBP duality is the presence of {\it linear} monopole superpotential terms. Consequently, if one attempts to derive a potential duality with non-linear monopole deformations from the deconfined adjoint SQCD picture, the necessary application of the BBP duality may be obstructed. In such cases, the derivation is likely to fail due to the incompatibility with the BBP step. Therefore, if we take the deconfined approach as the governing structure for KP-type dualities, this constraint provides a compelling reason why arbitrary non-linear monopole deformations typically do not lead to valid new dualities, even when naive quantum numbers analysis suggests one.

Conversely, since the confinement of $D_p^{3d}[SU(N)]$ has been derived solely based on the Aharony and BBP dualities, our observation that all consistent dualities precisely coincide with those derivable ones from the deconfined picture suggests that the Aharony and BBP dualities constitute the fundamental basis of all other 3d $\CN=2$ Seiberg--like dualities of unitary gauge groups, at least with a single adjoint matter field. Similar messages were also discussed in the context of $SL(2,\mathbb{Z})$ dual of 3d $\CN=4$ quiver theories as well \cite{Benvenuti:2023qtv,Hwang:2020wpd,Hwang:2021ulb,Bottini:2022vpy,Comi:2022aqo,Giacomelli:2023zkk,Giacomelli:2024laq} where the basic movements consist of the Aharony and BBP dualities.
\\
\\
Let us finish by addressing some possible future directions:
\paragraph{Proof of $\text{KP}_\a^\pm$ via deconfined picture} As we showed in section \ref{sec: linear} for $p=2$ case, the $\text{KP}_\a^\pm$ duality can be derived using the deconfined picture by only assuming Aharony and BBP dualities. The strategy is to use the two basic dualities successively in the deconfined $D_p^{3d}[SU(N)]$ quiver tail such that it becomes $D_p^{3d}[SU(\WN)]$, upon confining it gives rise to the expected dual theory. Indeed, we could derive $\text{KP}_0^\pm$ in this way which will be discussed in more detail in future work \cite{future}, and full generalization to $0\leq \a \leq 2p$ would be straightforward at least conceptually. Given that all the superconformal index matchings are numerical, this direction will provide a more substantial analytic check for $\text{KP}_\a^\pm$ duality.

\paragraph{Relevance check for extended dressings} The relevance of the monopole superpotential deformation is crucial for our dualities $\text{KP}_\a^\pm$ and $\text{KP}_{\a,\b}$ if we think them as obtained by the deformation from the adjoint SQCD of $\Tr X^{p+1}$ superpotential. Unfortunately, we could not check the relevance of the monopole operators $\hat V_\a^\pm$ with extended dressings $\a \geq p$ except for a few cases. However, there still is a possibility of some particular sequence of RG flows that allows $\hat V_{\a>p}^\pm$ as superpotential term, also possibly starting with another UV theory. The analogous situation for the BBP duality was cleverly resolved by coupling copies of auxiliary Ising-SCFTs to 3d SQCD \cite{Benini:2017dud}. Since $\text{KP}_\a^\pm$ and $\text{KP}_{\a,\b}$ are natural extensions of the BBP duality, we believe such resolution may exist and it would be interesting if one could work out.

\paragraph{More than quadratic power deformations} According to the duality classification in section \ref{sec: bootstrap}, $\text{KP}_\a^\pm$, $\text{KP}_{\a,\b}$, and KP itself are the only possible dualities that we can find from generic monopole superpotential deformations up to quadratic powers of monopoles. From the reasoning based on the deconfined picture, we believe this is always the case even if we increase the powers of monopole operators more than quadratic. This check will provide strong evidence that the Aharony and BBP dualities are all we need to derive the other 3d $\CN=2$ Seiberg--like dualities of unitary gauge groups at least involving only adjoint matters.

\section*{Acknowledgements}

We are grateful to Chiung Hwang for helpful discussions. SK especially thanks him for collaborations with closely related topics where the key ingredients in this work originated from. We also thank Jaewon Song, Sungjay Lee, and Ki-Hong Lee for helpful discussions and Dongmin Gang, Heeyeon Kim, and Cyril Closset for discussions with related topics and valuable insights. The work of QJ is supported by the National Research Foundation of Korea (NRF) grant NRF2023R1A2C1004965 and RS-2024-00405629. The work of QJ is also supported by Jang Young-Sil Fellow Program at the Korea Advanced Institute of Science and Technology. The work of SK is supported by KIAS Individual Grant PG092102 at Korea Institute for Advanced Study.

\newpage
\appendix

\section{Supersymmetric observables} \label{app: index f-max}
In this appendix, we review supersymmetric observables in 3d $\CN=2$ gauge theories that are used for various analyses in this paper. We also provide the results here.

\subsection{Superconformal index and duality check} \label{app: index result}
The superconformal index has played a crucial role in various analyses of supersymmetric quantum field theories in various contexts. One of the most important features is that it is an RG invariant observable so that one can compute it in the UV description via the localization technique to see non--trivial aspects in the IR. In this paper, we use it as a non--trivial check for our proposed dualities \eqref{eq: duality proposal}. We follow the convention of the 3d superconformal index in \cite{Bhattacharya:2008zy,Bhattacharya:2008bja}:
\begin{align}
    \CI = \Tr (-1)^F e^{\b\{Q,S\}} q^{E+j} \prod_{a} t_a^{F_a}
\end{align}
where $F$ is the fermion number, $\b$ is the inverse temperature, $Q$($S=Q^\dagger$) is a particularly chosen supercharge such that only the BPS states saturating the bound,
\begin{align}
    \{ Q, S \} = E - j - R \geq 0
\end{align}
contribute to the index, making $\CI$ to  be $\b$ independent. In the UV gauge theory description, $\CI$ counts gauge invariant BPS operators graded by $E$, $j$, and $R$ which are the energy, angular momentum, and the R-charge respectively, which are the Cartan generators of the bosonic part of the 3d $\CN=2$ superconformal group $SO(3,2)\times SO(2)_R$, with $F_a$ being theory dependent flavor symmetry charges. For 3d $\CN=2$ SQCD of $U(N)$ gauge group with $F$ flavors $(Q,\WQ)$ and an adjoint $X$ of superpotential $\Tr X^{p+1}$, the flavor symmetry group is $U(1)_T \times U(1)_A \times SU(F)_t \times SU(F)_b$ whose fugacities in the index would be $w,\eta,a_i$, and $b_i$ respectively. For explicit index formula, see \cite{Kapustin:2011vz} for the Coulomb branch localization and \cite{Hwang:2015wna} for the Higgs branch localization.
Here, we provide the results of the superconformal index for duality check of $\text{KP}_\a^+$ and $\text{KP}_{\a,\b}$ with various cases. For series expansion, we consider the cases $0 < \D_Q < \frac{2}{p+1} $ and $R_0,r_0 > 0$ where,
\begin{align}
    R_0 = F(1-\D_Q) - \frac{2}{p+1}(N-1)
    \;,\;\;
    r_0 = F\big(1-(\frac{2}{p+1}-\D_Q)\big) - \frac{2}{p+1}(\WN-1)
\end{align}
are the R-charges of the bare monopole operators on both sides of the duality. 
However, we sometimes cross the border to cover cases beyond it. We write down the single particle index by taking the plethystic log to show more transparent meanings, such as confinement to Wess--Zumino theories. We exclude trivial cases guaranteed by index identities from the KP duality, i.e., $\text{KP}_p^+$, $\text{KP}_{\a,p-\a}$, as well as $\text{KP}_{\a,\a}$ which already have been treated in \cite{Hwang:2022jjs}.
\subsubsection*{$\text{KP}_\a^+$ for $\a < p$}
\begin{tabularx}{\textwidth}{ 
  |c|c|c |>{\centering\arraybackslash}X|}
\hline
    $p$ & $\alpha$ & $(F,N,\widetilde{N},\D_Q)$ & Single-particle index\\
\hline
    \multirow{31}{*}[-2cm]{2} & \multirow{13}{*}[-1.5cm]{0} & $(1,0,0,\frac{1}{3})$ & 0\\ 
    \cline{3-4}&&$(2,0,2,\frac{1}{3})$&0\\
    \cline{3-4}&&$(2,1,1,\frac{1}{3})$&$\left(1+\frac{1}{\eta ^4}+4 \eta ^2\right) q^{2/3}+\left(-1-\frac{4}{\eta ^2}-\eta ^4\right) q^{4/3}$\\
    \cline{3-4}&&$(2,2,0,\frac{1}{3})$&$\frac{1}{\eta ^4 q^{2/3}}+\frac{1}{\eta ^4}+\left(-\frac{4}{\eta ^2}+4 \eta ^2\right) q^{2/3}+\left(-\frac{4}{\eta ^2}+4 \eta ^2\right) q^{4/3}-\eta ^4 q^2-\eta ^4 q^{8/3}$\\
    \cline{3-4}&&$(3,0,4,\frac{1}{2})$&0\\
    \cline{3-4}&&$(3,1,3,\frac{7}{18})$&$q^{2/3}+9 \eta ^2 q^{7/9}-q^{4/3}-9 \eta ^4 q^{14/9}+\frac{q^{5/3}}{\eta ^6}-17 q^2+\mathcal{O}(q^{7/3})$\\
    \cline{3-4}&&$(3,2,2,\frac{1}{3})$&$\left(1+\frac{1}{\eta ^6}+9 \eta ^2\right) q^{2/3}+\left(\frac{1}{\eta ^6}+9 \eta ^2\right) q^{4/3}+\left(-18-\frac{1}{\eta ^6}-\frac{9}{\eta ^4}-\frac{9}{\eta ^2}-9 \eta ^2-9 \eta ^4-\eta ^6\right) q^2 + \mathcal{O}(q^{10/3})$\\
    \cline{3-4}&&$(3,3,1,\frac{5}{18})$&$\frac{1}{\eta ^6 \sqrt[3]{q}}+\frac{\sqrt[3]{q}}{\eta ^6}+9 \eta ^2 q^{5/9}+q^{2/3}+9 \eta ^2 q^{11/9}-q^{4/3}-\frac{9 q^{13/9}}{\eta ^2}-\frac{9 q^{14/9}}{\eta ^4}-17 q^2+\mathcal{O}(q^{7/3})$\\
    \cline{3-4}&&$(3,4,0,\frac{1}{6})$&$\frac{1}{\eta ^6 q}+\frac{1}{\eta ^6 \sqrt[3]{q}}+9 \eta ^2 \sqrt[3]{q}+\left(-\frac{9}{\eta ^2}+9 \eta ^2\right) q-\frac{9 q^{5/3}}{\eta ^2}-\eta ^6 q^{7/3}$\\
    \cline{3-4}&&$(4,2,4,\frac{5}{12})$&$q^{2/3}+16 \eta ^2 q^{5/6}+\frac{q^{4/3}}{\eta ^8}+16 \eta ^2 q^{3/2}+\left(-32+\frac{1}{\eta ^8}\right) q^2+\mathcal{O}(q^{13/6})$\\
    \cline{3-4}&&$(4,3,3,\frac{1}{3})$&$\left(1+\frac{1}{\eta ^8}+16 \eta ^2\right) q^{2/3}+\left(\frac{1}{\eta ^8}+16 \eta ^2\right) q^{4/3}-31 q^2 + \mathcal{O}(q^{8/3})$\\
    \cline{3-4}&&$(4,4,2,\frac{1}{4})$&$\frac{1}{\eta ^8}+16 \eta ^2 \sqrt{q}+\left(1+\frac{1}{\eta ^8}\right) q^{2/3}+16 \eta ^2 q^{7/6}-32 q^2 + \mathcal{O}(q^{13/6})$\\
    \cline{3-4}&&$(5,4,4,\frac{1}{3})$&$\left(1+\frac{1}{\eta ^{10}}+25 \eta ^2\right) q^{2/3}+\left(\frac{1}{\eta ^{10}}+25 \eta ^2\right) q^{4/3}-49 q^2+\mathcal{O}(q^{8/3})$\\
    \cline{2-4}&\multirow{12}{*}[-2cm]{1}& $(1,0,1,\frac{1}{3})$&0\\
    \cline{3-4}&&$(1,1,0,\frac{1}{3})$&$\frac{1}{\eta ^2}+\eta ^2 q^{2/3}-\frac{q^{4/3}}{\eta ^2}-\eta ^2 q^2$\\
    \cline{3-4}&&$(2,0,3,\frac{1}{2})$&0\\
    \cline{3-4}&&$(2,1,2,\frac{1}{3})$&$\left(1+4 \eta ^2\right) q^{2/3}+\left(\frac{1}{\eta ^4}-\eta ^4\right) q^{4/3}+\left(-8-\frac{4}{\eta ^2}-4 \eta ^2\right) q^2+\mathcal{O}(q^{8/3})$\\
    \cline{3-4}&&$(2,2,1,\frac{1}{3})$&$\frac{1}{\eta ^4}+\left(1+\frac{1}{\eta ^4}+4 \eta ^2\right) q^{2/3}+\left(-\frac{1}{\eta ^4}-\frac{4}{\eta ^2}+4 \eta ^2\right) q^{4/3}+\left(-8-\frac{4}{\eta ^2}-4 \eta ^2-\eta ^4\right) q^2+\mathcal{O}(q^{8/3})$\\
    \cline{3-4}&&$(2,3,0,\frac{1}{6})$&$\frac{1}{\eta ^4 q^{2/3}}+\frac{1}{\eta ^4}+4 \eta ^2 \sqrt[3]{q}+\left(-\frac{4}{\eta ^2}+4 \eta ^2\right) q-\frac{4 q^{5/3}}{\eta ^2}-\eta ^4 q^2-\eta ^4 q^{8/3}$\\
    \cline{3-4}&&$(3,1,4,\frac{1}{2})$&$q^{2/3}+9 \eta ^2 q+\frac{q^{5/3}}{\eta ^6}+\left(-18-9 \eta ^4\right) q^2+\mathcal{O}(q^{7/3})$\\
    \cline{3-4}&&$(3,2,3,\frac{7}{18})$&$q^{2/3}+9 \eta ^2 q^{7/9}+\frac{q}{\eta ^6}+q^{4/3}+9 \eta ^2 q^{13/9}+\frac{q^{5/3}}{\eta ^6}-18 q^2+\mathcal{O}(q^{19/9})$\\
    \cline{3-4}&&$(3,3,2,\frac{5}{18})$&$\frac{\sqrt[3]{q}}{\eta ^6}+9 \eta ^2 q^{5/9}+q^{2/3}+\frac{q}{\eta ^6}+9 \eta ^2 q^{11/9}+q^{4/3}-18 q^2+\mathcal{O}(q^{19/9})$\\
    \cline{3-4}&&$(3,4,1,\frac{1}{6})$&$\frac{1}{\eta ^6 \sqrt[3]{q}}+\left(\frac{1}{\eta ^6}+9 \eta ^2\right) \sqrt[3]{q}+q^{2/3}+9 \eta ^2 q-\frac{9 q^{5/3}}{\eta ^2}+\left(-18-\frac{9}{\eta ^4}\right) q^2+\mathcal{O}(q^{7/3})$\\
    \cline{3-4}&&$(4,3,4,\frac{5}{12})$&$\left(1+\frac{1}{\eta ^8}\right) q^{2/3}+16 \eta ^2 q^{5/6}+\left(1+\frac{1}{\eta ^8}\right) q^{4/3}+16 \eta ^2 q^{3/2}-31 q^2+\mathcal{O}(q^{8/3})$\\
    \cline{3-4}&&$(4,4,3,\frac{1}{4})$&$16 \eta ^2 \sqrt{q}+\left(1+\frac{1}{\eta ^8}\right) q^{2/3}+16 \eta ^2 q^{7/6}+\left(1+\frac{1}{\eta ^8}\right) q^{4/3}-31 q^2+\mathcal{O}(q^{8/3})$\\
\hline
\end{tabularx}

\newpage

\begin{tabularx}{\textwidth}{ 
  |c|c|c |>{\centering\arraybackslash}X|}
\hline
    $p$ & $\alpha$ & $(F,N,\widetilde{N},\D_Q)$ & Single-particle index\\
\hline
\multirow{28}{*}[-3cm]{3} & \multirow{8}{*}[-2.5cm]{0}&$(1,0,0,\frac{1}{4})$&0\\
\cline{3-4}&&$(2,0,3,\frac{1}{4})$&0\\
\cline{3-4}&&$(2,1,2,\frac{1}{4})$&$\left(1+4 \eta ^2\right) \sqrt{q}+\left(\frac{1}{\eta ^4}-\eta ^4\right) q+\left(-1-\frac{4}{\eta ^2}\right) q^{3/2}$\\
\cline{3-4}&&$(2,2,1,\frac{1}{4})$&$\frac{1}{\eta ^4}+\left(1+\frac{1}{\eta ^4}+4 \eta ^2\right) \sqrt{q}+\left(-\frac{4}{\eta ^2}+4 \eta ^2\right) q+\left(-1-\frac{4}{\eta ^2}-\eta ^4\right) q^{3/2}-\eta ^4 q^2$\\
\cline{3-4}&&$(2,3,0,\frac{1}{4})$&$\frac{1}{\eta ^4 q}+\frac{1}{\eta ^4 \sqrt{q}}+\frac{1}{\eta ^4}+\left(-\frac{4}{\eta ^2}+4 \eta ^2\right) \sqrt{q}+\left(-\frac{4}{\eta ^2}+4 \eta ^2\right) q+\left(-\frac{4}{\eta ^2}+4 \eta ^2\right) q^{3/2}-\eta ^4 q^2-\eta ^4 q^{5/2}-\eta ^4 q^3$\\
\cline{3-4}&&$(3,2,4,\frac{1}{4})$&$\left(1+9 \eta ^2\right) \sqrt{q}+\left(1+9 \eta ^2\right) q+\left(-1+\frac{1}{\eta ^6}-9 \eta ^4-\eta ^6\right) q^{3/2}+\left(-18+\frac{1}{\eta ^6}-9 \eta ^2-9 \eta ^4\right) q^2+\mathcal{O}(q^{5/2})$\\
\cline{3-4}&&$(3,3,3,\frac{1}{4})$&$\left(1+\frac{1}{\eta ^6}+9 \eta ^2\right) \sqrt{q}+\left(1+\frac{1}{\eta ^6}+9 \eta ^2\right) q+\left(\frac{1}{\eta ^6}+9 \eta ^2\right) q^{3/2}+\left(-19-\frac{1}{\eta ^6}-\frac{9}{\eta ^4}-\frac{9}{\eta ^2}-9 \eta ^2-9 \eta ^4-\eta ^6\right) q^2+\mathcal{O}(q^{5/2})$\\
\cline{3-4}&&$(3,4,2,\frac{1}{4})$&$\frac{1}{\eta ^6 \sqrt{q}}+\frac{1}{\eta ^6}+\left(1+\frac{1}{\eta ^6}+9 \eta ^2\right) \sqrt{q}+\left(1+9 \eta ^2\right) q+\left(-1-\frac{1}{\eta ^6}-\frac{9}{\eta ^4}-\frac{9}{\eta ^2}+9 \eta ^2\right) q^{3/2}+\left(-18-\frac{9}{\eta ^4}-\frac{9}{\eta ^2}\right) q^2+\mathcal{O}(q^{5/2})$\\
\cline{2-4}&\multirow{9}{*}[-2.5cm]{1}&$(1,0,1,\frac{1}{4})$&0\\
\cline{3-4}&&$(1,1,0,\frac{1}{4})$&$\frac{1}{\eta ^2}+\eta ^2 \sqrt{q}-\frac{q^{3/2}}{\eta ^2}-\eta ^2 q^2$\\
\cline{3-4}&&$(2,0,4,\frac{3}{8})$&0\\
\cline{3-4}&&$(2,1,3,\frac{1}{4})$&$\left(1+4 \eta ^2\right) \sqrt{q}-\eta ^4 q+\frac{q^{3/2}}{\eta ^4}+\left(-8-\frac{4}{\eta ^2}-4 \eta ^2\right) q^2+\mathcal{O}(q^{5/2})$\\
\cline{3-4}&&$(2,2,2,\frac{1}{4})$&$\left(1+\frac{1}{\eta ^4}+4 \eta ^2\right) \sqrt{q}+\left(1+\frac{1}{\eta ^4}+4 \eta ^2\right) q+\left(-1-\frac{4}{\eta ^2}-\eta ^4\right) q^{3/2}+\left(-8-\frac{1}{\eta ^4}-\frac{4}{\eta ^2}-4 \eta ^2-\eta ^4\right) q^2+\mathcal{O}(q^3)$\\
\cline{3-4}&&$(2,3,1,\frac{1}{4})$&$\frac{1}{\eta ^4 \sqrt{q}}+\frac{1}{\eta ^4}+\left(1+\frac{1}{\eta ^4}+4 \eta ^2\right) \sqrt{q}+\left(-\frac{1}{\eta ^4}-\frac{4}{\eta ^2}+4 \eta ^2\right) q+\left(-\frac{4}{\eta ^2}+4 \eta ^2\right) q^{3/2}+\left(-8-\frac{4}{\eta ^2}-4 \eta ^2-\eta ^4\right) q^2+\mathcal{O}(q^{5/2})$\\
\cline{3-4}&&$(2,4,0,\frac{1}{8})$&$\frac{1}{\eta ^4 q}+\frac{1}{\eta ^4 \sqrt{q}}+\frac{1}{\eta ^4}+4 \eta ^2 \sqrt[4]{q}+\left(-\frac{4}{\eta ^2}+4 \eta ^2\right) q^{3/4}+\left(-\frac{4}{\eta ^2}+4 \eta ^2\right) q^{5/4}-\frac{4 q^{7/4}}{\eta ^2}-\eta ^4 q^2-\eta ^4 q^{5/2}-\eta ^4 q^3$\\
\cline{3-4}&&$(3,3,4,\frac{1}{4})$&$\left(1+9 \eta ^2\right) \sqrt{q}+\left(1+\frac{1}{\eta ^6}+9 \eta ^2\right) q+\left(1+\frac{1}{\eta ^6}+9 \eta ^2\right) q^{3/2}+\left(-18+\frac{1}{\eta ^6}-9 \eta ^2-9 \eta ^4-\eta ^6\right) q^2+\mathcal{O}(q^{5/2})$\\
\cline{3-4}&&$(3,4,3,\frac{1}{4})$&$\left(1+9 \eta ^2\right) \sqrt{q}+\left(1+\frac{1}{\eta ^6}+9 \eta ^2\right) q+\left(1+\frac{1}{\eta ^6}+9 \eta ^2\right) q^{3/2}+\left(-18+\frac{1}{\eta ^6}-9 \eta ^2-9 \eta ^4-\eta ^6\right) q^2+\mathcal{O}(q^{5/2})$\\
\cline{2-4}&\multirow{4}{*}{2}&$(1,0,2,\frac{1}{4})$&0\\
\cline{3-4}&&$(1,1,1,\frac{1}{4})$&$\left(1+\frac{1}{\eta ^2}+\eta ^2\right) \sqrt{q}+\left(-1-\frac{1}{\eta ^2}-\eta ^2\right) q^{3/2}$\\
\cline{3-4}&&$(1,2,0,\frac{1}{4})$&$\frac{1}{\eta ^2 \sqrt{q}}+\frac{1}{\eta ^2}+\eta ^2 \sqrt{q}+\left(-\frac{1}{\eta ^2}+\eta ^2\right) q-\frac{q^{3/2}}{\eta ^2}-\eta ^2 q^2-\eta ^2 q^{5/2}$\\
\hline
\end{tabularx}
\newpage
\begin{tabularx}{\textwidth}{ 
  |c|c|c |>{\centering\arraybackslash}X|}
\hline
    $p$ & $\alpha$ & $(F,N,\widetilde{N},\D_Q)$ & Single-particle index\\
\hline
    \multirow{5}{*}[-2.5cm]{3} & \multirow{5}{*}[-2.5cm]{2} & $(2,1,4,\frac{3}{8})$&$\sqrt{q}+4 \eta ^2 q^{3/4}+q+\left(-1+\frac{1}{\eta ^4}-\eta ^4\right) q^{3/2}-4 \eta ^2 q^{7/4}-8 q^2+\mathcal{O}(q^{9/4})$\\
\cline{3-4}&&$(2,2,3,\frac{1}{4})$&$\left(1+4 \eta ^2\right) \sqrt{q}+\left(2+\frac{1}{\eta ^4}+4 \eta ^2\right) q+\left(\frac{1}{\eta ^4}-\eta ^4\right) q^{3/2}+\left(-10-\frac{4}{\eta ^2}-8 \eta ^2-2 \eta ^4\right) q^2+\mathcal{O}(q^{5/2})$\\
\cline{3-4}&&$(2,3,2,\frac{1}{4})$&$\frac{1}{\eta ^4}+\left(1+\frac{1}{\eta ^4}+4 \eta ^2\right) \sqrt{q}+\left(2+\frac{1}{\eta ^4}+4 \eta ^2\right) q+\left(-\frac{1}{\eta ^4}-\frac{4}{\eta ^2}+4 \eta ^2\right) q^{3/2}+\left(-10-\frac{2}{\eta ^4}-\frac{8}{\eta ^2}-4 \eta ^2-\eta ^4\right) q^2+\mathcal{O}(q^{5/2})$\\
\cline{3-4}&&$(2,4,1,\frac{1}{8})$&$\frac{1}{\eta ^4 \sqrt{q}}+\frac{1}{\eta ^4}+4 \eta ^2 \sqrt[4]{q}+\left(1+\frac{1}{\eta ^4}\right) \sqrt{q}+4 \eta ^2 q^{3/4}+q+\left(-\frac{4}{\eta ^2}+4 \eta ^2\right) q^{5/4}+\left(-1-\frac{1}{\eta ^4}\right) q^{3/2}-\frac{8 q^{7/4}}{\eta ^2}+\left(-8-\eta ^4\right) q^2+\mathcal{O}(q^{9/4})$\\
\cline{3-4}&&$(3,4,4,\frac{1}{4})$&$\left(1+\frac{1}{\eta ^6}+9 \eta ^2\right) \sqrt{q}+\left(2+\frac{1}{\eta ^6}+9 \eta ^2\right) q+\left(1+\frac{1}{\eta ^6}+9 \eta ^2\right) q^{3/2}-17 q^2+\mathcal{O}(q^{5/2})$\\
\hline
    \multirow{27}{*}{4} & \multirow{7}{*}[-2.5cm]{0} & $(1,0,0,\frac{1}{5})$&0\\
    \cline{3-4}&&$(2,0,4,\frac{1}{5})$&0\\
    \cline{3-4}&&$(2,1,3,\frac{1}{5})$&$\left(1+4 \eta ^2\right) q^{2/5}-\eta ^4 q^{4/5}+\frac{q^{6/5}}{\eta ^4}+\left(-1-\frac{4}{\eta ^2}\right) q^{8/5}$\\
    \cline{3-4}&&$(2,2,2,\frac{1}{5})$&$\left(1+\frac{1}{\eta ^4}+4 \eta ^2\right) q^{2/5}+\left(1+\frac{1}{\eta ^4}+4 \eta ^2\right) q^{4/5}+\left(-1-\frac{4}{\eta ^2}-\eta ^4\right) q^{6/5}+\left(-1-\frac{4}{\eta ^2}-\eta ^4\right) q^{8/5}$\\
    \cline{3-4}&&$(2,3,1,\frac{1}{5})$&$\frac{1}{\eta ^4 q^{2/5}}+\frac{1}{\eta ^4}+\left(1+\frac{1}{\eta ^4}+4 \eta ^2\right) q^{2/5}+\left(-\frac{4}{\eta ^2}+4 \eta ^2\right) q^{4/5}+\left(-\frac{4}{\eta ^2}+4 \eta ^2\right) q^{6/5}+\left(-1-\frac{4}{\eta ^2}-\eta ^4\right) q^{8/5}-\eta ^4 q^2-\eta ^4 q^{12/5}$\\
    \cline{3-4}&&$(2,4,0,\frac{1}{5})$&$\frac{1}{\eta ^4 q^{6/5}}+\frac{1}{\eta ^4 q^{4/5}}+\frac{1}{\eta ^4 q^{2/5}}+\frac{1}{\eta ^4}+\left(-\frac{4}{\eta ^2}+4 \eta ^2\right) q^{2/5}+\left(-\frac{4}{\eta ^2}+4 \eta ^2\right) q^{4/5}+\left(-\frac{4}{\eta ^2}+4 \eta ^2\right) q^{6/5}+\left(-\frac{4}{\eta ^2}+4 \eta ^2\right) q^{8/5}-\eta ^4 q^2-\eta ^4 q^{12/5}-\eta ^4 q^{14/5}-\eta ^4 q^{16/5}$\\
    \cline{3-4}&&$(3,4,4,\frac{1}{5})$&$\left(1+\frac{1}{\eta ^6}+9 \eta ^2\right) q^{2/5}+\left(1+\frac{1}{\eta ^6}+9 \eta ^2\right) q^{4/5}+\left(1+\frac{1}{\eta ^6}+9 \eta ^2\right) q^{6/5}+\left(\frac{1}{\eta ^6}+9 \eta ^2\right) q^{8/5}+\left(-19-\frac{1}{\eta ^6}-\frac{9}{\eta ^4}-\frac{9}{\eta ^2}-9 \eta ^2-9 \eta ^4-\eta ^6\right) q^2 + \mathcal{O}(q^{12/5})$\\
    \cline{2-4}&\multirow{6}{*}[-1cm]{1}&$(1,0,1,\frac{1}{5})$&0\\
    \cline{3-4}&&$(1,1,0,\frac{1}{5})$&$\frac{1}{\eta ^2}+\eta ^2 q^{2/5}-\frac{q^{8/5}}{\eta ^2}-\eta ^2 q^2$\\
    \cline{3-4}&&$(2,1,4,\frac{1}{5})$&$\left(1+4 \eta ^2\right) q^{2/5}-\eta ^4 q^{4/5}+\frac{q^{8/5}}{\eta ^4}+\left(-8-\frac{4}{\eta ^2}-4 \eta ^2\right) q^2+\mathcal{O}(q^{12/5})$\\
    \cline{3-4}&&$(2,2,3,\frac{1}{5})$&$\left(1+4 \eta ^2\right) q^{2/5}+\left(1+\frac{1}{\eta ^4}+4 \eta ^2\right) q^{4/5}+\left(\frac{1}{\eta ^4}-\eta ^4\right) q^{6/5}+\left(-1-\frac{4}{\eta ^2}-\eta ^4\right) q^{8/5}+\left(-8-\frac{4}{\eta ^2}-4 \eta ^2\right) q^2+\mathcal{O}(q^{12/5})$\\
    \cline{3-4}&&$(2,3,2,\frac{1}{5})$&$\frac{1}{\eta ^4}+\left(1+\frac{1}{\eta ^4}+4 \eta ^2\right) q^{2/5}+\left(1+\frac{1}{\eta ^4}+4 \eta ^2\right) q^{4/5}+\left(-\frac{4}{\eta ^2}+4 \eta ^2\right) q^{6/5}+\left(-1-\frac{1}{\eta ^4}-\frac{4}{\eta ^2}-\eta ^4\right) q^{8/5}+\left(-8-\frac{4}{\eta ^2}-4 \eta ^2-\eta ^4\right) q^2+\mathcal{O}(q^{12/5})$\\
\hline
\end{tabularx}

\newpage

\begin{tabularx}{\textwidth}{ 
  |c|c|c |>{\centering\arraybackslash}X|}
\hline
    $p$ & $\alpha$ & $(F,N,\widetilde{N},\D_Q)$ & Single-particle index\\
\hline
    \multirow{13}{*}[-4.5cm]{4} & \multirow{1}{*}{1} & $(2,4,1,\frac{1}{5})$&$\frac{1}{\eta ^4 q^{4/5}}+\frac{1}{\eta ^4 q^{2/5}}+\frac{1}{\eta ^4}+\left(1+\frac{1}{\eta ^4}+4 \eta ^2\right) q^{2/5}+\left(-\frac{1}{\eta ^4}-\frac{4}{\eta ^2}+4 \eta ^2\right) q^{4/5}+\left(-\frac{4}{\eta ^2}+4 \eta ^2\right) q^{6/5}+\left(-\frac{4}{\eta ^2}+4 \eta ^2\right) q^{8/5}+\left(-8-\frac{4}{\eta ^2}-4 \eta ^2-\eta ^4\right) q^2+\mathcal{O}(q^{12/5})$\\
    \cline{2-4}&\multirow{4}{*}[-3cm]{2}&$(1,0,2,\frac{1}{5})$&0\\
    \cline{3-4}&&$(1,1,1,\frac{1}{5})$&$\left(1+\frac{1}{\eta ^2}+\eta ^2\right) q^{2/5}-q^{4/5}+q^{6/5}+\left(-1-\frac{1}{\eta ^2}-\eta ^2\right) q^{8/5}$\\
    \cline{3-4}&&$(1,2,0,\frac{1}{5})$&$\frac{1}{\eta ^2 q^{2/5}}+\frac{1}{\eta ^2}+\eta ^2 q^{2/5}+\eta ^2 q^{4/5}-\frac{q^{6/5}}{\eta ^2}-\frac{q^{8/5}}{\eta ^2}-\eta ^2 q^2-\eta ^2 q^{12/5}$\\
    \cline{3-4}&&$(2,2,4,\frac{1}{5})$&$\left(1+4 \eta ^2\right) q^{2/5}+\left(1+4 \eta ^2\right) q^{4/5}+\left(1+\frac{1}{\eta ^4}-\eta ^4\right) q^{6/5}+\left(\frac{1}{\eta ^4}-\eta ^4\right) q^{8/5}+\left(-10-\frac{4}{\eta ^2}-8 \eta ^2-\eta ^4\right) q^2+\mathcal{O}(q^{12/5})$\\
    \cline{3-4}&&$(2,3,3,\frac{1}{5})$&$\left(1+\frac{1}{\eta ^4}+4 \eta ^2\right) q^{2/5}+\left(1+\frac{1}{\eta ^4}+4 \eta ^2\right) q^{4/5}+\left(2+\frac{1}{\eta ^4}+4 \eta ^2\right) q^{6/5}+\left(-1-\frac{4}{\eta ^2}-\eta ^4\right) q^{8/5}+\left(-9-\frac{1}{\eta ^4}-\frac{4}{\eta ^2}-4 \eta ^2-\eta ^4\right) q^2+\mathcal{O}(q^{12/5})$\\
    \cline{3-4}&&$(2,4,2,\frac{1}{5})$&$\frac{1}{\eta ^4 q^{2/5}}+\frac{1}{\eta ^4}+\left(1+\frac{1}{\eta ^4}+4 \eta ^2\right) q^{2/5}+\left(1+\frac{1}{\eta ^4}+4 \eta ^2\right) q^{4/5}+\left(1-\frac{1}{\eta ^4}-\frac{4}{\eta ^2}+4 \eta ^2\right) q^{6/5}+\left(-\frac{1}{\eta ^4}-\frac{4}{\eta ^2}+4 \eta ^2\right) q^{8/5}+\left(-10-\frac{1}{\eta ^4}-\frac{8}{\eta ^2}-4 \eta ^2-\eta ^4\right) q^2+\mathcal{O}(q^{12/5})$\\
    \cline{2-4}&\multirow{6}{*}[-2.5cm]{3}&$(1,0,3,\frac{3}{10})$&0\\
    \cline{3-4}&&$(1,1,2,\frac{1}{5})$&$\left(1+\eta ^2\right) q^{2/5}+\left(1+\frac{1}{\eta ^2}\right) q^{4/5}+\left(-1-\eta ^2\right) q^{6/5}+\left(-1-\frac{1}{\eta ^2}\right) q^{8/5}$\\
    \cline{3-4}&&$(1,2,1,\frac{1}{5})$&$\frac{1}{\eta ^2}+\left(1+\frac{1}{\eta ^2}+\eta ^2\right) q^{2/5}+\left(1+\eta ^2\right) q^{4/5}+\left(-1-\frac{1}{\eta ^2}\right) q^{6/5}+\left(-1-\frac{1}{\eta ^2}-\eta ^2\right) q^{8/5}-\eta ^2 q^2$\\
    \cline{3-4}&&$(1,3,0,\frac{1}{10})$&$\frac{1}{\eta ^2 q^{3/5}}+\frac{1}{\eta ^2 \sqrt[5]{q}}+\left(\frac{1}{\eta ^2}+\eta ^2\right) \sqrt[5]{q}+\eta ^2 q^{3/5}+\left(-\frac{1}{\eta ^2}+\eta ^2\right) q-\frac{q^{7/5}}{\eta ^2}+\left(-\frac{1}{\eta ^2}-\eta ^2\right) q^{9/5}-\eta ^2 q^{11/5}-\eta ^2 q^{13/5}$\\
    \cline{3-4}&&$(2,3,4,\frac{1}{5})$&$\left(1+4 \eta ^2\right) q^{2/5}+\left(2+\frac{1}{\eta ^4}+4 \eta ^2\right) q^{4/5}+\left(2+\frac{1}{\eta ^4}+4 \eta ^2\right) q^{6/5}+\left(\frac{1}{\eta ^4}-\eta ^4\right) q^{8/5}+\left(-10-\frac{4}{\eta ^2}-8 \eta ^2-2 \eta ^4\right) q^2+\mathcal{O}(q^{12/5})$\\
    \cline{3-4}&&$(2,4,3,\frac{1}{5})$&$\frac{1}{\eta ^4}+\left(1+\frac{1}{\eta ^4}+4 \eta ^2\right) q^{2/5}+\left(2+\frac{1}{\eta ^4}+4 \eta ^2\right) q^{4/5}+\left(2+\frac{1}{\eta ^4}+4 \eta ^2\right) q^{6/5}+\left(-\frac{1}{\eta ^4}-\frac{4}{\eta ^2}+4 \eta ^2\right) q^{8/5}+\left(-10-\frac{2}{\eta ^4}-\frac{8}{\eta ^2}-4 \eta ^2-\eta ^4\right) q^2+\mathcal{O}(q^{12/5})$\\
\hline
    \multirow{9}{*}{5} & \multirow{4}{*}[-1cm]{0} & $(1,0,0,\frac{1}{6})$&0\\
    \cline{3-4}&&$(2,1,4,\frac{1}{6})$&$\left(1+4 \eta ^2\right) \sqrt[3]{q}-\eta ^4 q^{2/3}+\frac{q^{4/3}}{\eta ^4}+\left(-1-\frac{4}{\eta ^2}\right) q^{5/3}$\\
    \cline{3-4}&&$(2,2,3,\frac{1}{6})$&$\left(1+4 \eta ^2\right) \sqrt[3]{q}+\left(1+\frac{1}{\eta ^4}+4 \eta ^2\right) q^{2/3}+\left(\frac{1}{\eta ^4}-\eta ^4\right) q+\left(-1-\frac{4}{\eta ^2}-\eta ^4\right) q^{4/3}+\left(-1-\frac{4}{\eta ^2}\right) q^{5/3}$\\
    \cline{3-4}&&$(2,3,2,\frac{1}{6})$&$\frac{1}{\eta ^4}+\left(1+\frac{1}{\eta ^4}+4 \eta ^2\right) \sqrt[3]{q}+\left(1+\frac{1}{\eta ^4}+4 \eta ^2\right) q^{2/3}+\left(-\frac{4}{\eta ^2}+4 \eta ^2\right) q+\left(-1-\frac{4}{\eta ^2}-\eta ^4\right) q^{4/3}+\left(-1-\frac{4}{\eta ^2}-\eta ^4\right) q^{5/3}-\eta ^4 q^2$\\
\hline
\end{tabularx}

\newpage

\begin{tabularx}{\textwidth}{ 
  |c|c|c |>{\centering\arraybackslash}X|}
\hline
    $p$ & $\alpha$ & $(F,N,\widetilde{N},\D_Q)$ & Single-particle index\\
\hline
    \multirow{15}{*}[-5cm]{5} & \multirow{1}{*}{0} & $(2,4,1,\frac{1}{6})$&$\frac{1}{\eta ^4 q^{2/3}}+\frac{1}{\eta ^4 \sqrt[3]{q}}+\frac{1}{\eta ^4}+\left(1+\frac{1}{\eta ^4}+4 \eta ^2\right) \sqrt[3]{q}+\left(-\frac{4}{\eta ^2}+4 \eta ^2\right) q^{2/3}+\left(-\frac{4}{\eta ^2}+4 \eta ^2\right) q+\left(-\frac{4}{\eta ^2}+4 \eta ^2\right) q^{4/3}+\left(-1-\frac{4}{\eta ^2}-\eta ^4\right) q^{5/3}-\eta ^4 q^2-\eta ^4 q^{7/3}-\eta ^4 q^{8/3}$\\
    \cline{2-4}&\multirow{5}{*}[-2cm]{1}&$(1,0,1,\frac{1}{6})$&0\\
    \cline{3-4}&&$(1,1,0,\frac{1}{6})$&$\frac{1}{\eta ^2}+\eta ^2 \sqrt[3]{q}-\frac{q^{5/3}}{\eta ^2}-\eta ^2 q^2$\\
    \cline{3-4}&&$(2,2,4,\frac{1}{6})$&$\left(1+4 \eta ^2\right) \sqrt[3]{q}+\left(1+4 \eta ^2\right) q^{2/3}+\left(\frac{1}{\eta ^4}-\eta ^4\right) q+\left(\frac{1}{\eta ^4}-\eta ^4\right) q^{4/3}+\left(-1-\frac{4}{\eta ^2}\right) q^{5/3}+\left(-8-\frac{4}{\eta ^2}-4 \eta ^2\right) q^2+\mathcal{O}(q^{8/3})$\\
    \cline{3-4}&&$(2,3,3,\frac{1}{6})$&$\left(1+\frac{1}{\eta ^4}+4 \eta ^2\right) \sqrt[3]{q}+\left(1+\frac{1}{\eta ^4}+4 \eta ^2\right) q^{2/3}+\left(1+\frac{1}{\eta ^4}+4 \eta ^2\right) q+\left(-1-\frac{4}{\eta ^2}-\eta ^4\right) q^{4/3}+\left(-1-\frac{4}{\eta ^2}-\eta ^4\right) q^{5/3}+\left(-8-\frac{1}{\eta ^4}-\frac{4}{\eta ^2}-4 \eta ^2-\eta ^4\right) q^2+\mathcal{O}(q^3)$\\
    \cline{3-4}&&$(2,4,2,\frac{1}{6})$&$\frac{1}{\eta ^4 \sqrt[3]{q}}+\frac{1}{\eta ^4}+\left(1+\frac{1}{\eta ^4}+4 \eta ^2\right) \sqrt[3]{q}+\left(1+\frac{1}{\eta ^4}+4 \eta ^2\right) q^{2/3}+\left(-\frac{4}{\eta ^2}+4 \eta ^2\right) q+\left(-\frac{1}{\eta ^4}-\frac{4}{\eta ^2}+4 \eta ^2\right) q^{4/3}+\left(-1-\frac{4}{\eta ^2}-\eta ^4\right) q^{5/3}+\left(-8-\frac{4}{\eta ^2}-4 \eta ^2-\eta ^4\right) q^2+\mathcal{O}(q^{7/3})$\\
    \cline{2-4}&\multirow{5}{*}[-2cm]{2}&$(1,0,2,\frac{1}{6})$&0\\
    \cline{3-4}&&$(1,1,1,\frac{1}{6})$&$\left(1+\frac{1}{\eta ^2}+\eta ^2\right) \sqrt[3]{q}-q^{2/3}+q^{4/3}+\left(-1-\frac{1}{\eta ^2}-\eta ^2\right) q^{5/3}$\\
    \cline{3-4}&&$(1,2,0,\frac{1}{6})$&$\frac{1}{\eta ^2 \sqrt[3]{q}}+\frac{1}{\eta ^2}+\eta ^2 \sqrt[3]{q}+\eta ^2 q^{2/3}-\frac{q^{4/3}}{\eta ^2}-\frac{q^{5/3}}{\eta ^2}-\eta ^2 q^2-\eta ^2 q^{7/3}$\\
    \cline{3-4}&&$(2,3,4,\frac{1}{6})$&$\left(1+4 \eta ^2\right) \sqrt[3]{q}+\left(1+\frac{1}{\eta ^4}+4 \eta ^2\right) q^{2/3}+\left(1+\frac{1}{\eta ^4}+4 \eta ^2\right) q+\left(1+\frac{1}{\eta ^4}-\eta ^4\right) q^{4/3}+\left(-1-\frac{4}{\eta ^2}-\eta ^4\right) q^{5/3}+\left(-9-\frac{4}{\eta ^2}-4 \eta ^2-\eta ^4\right) q^2+\mathcal{O}(q^{7/3})$\\
    \cline{3-4}&&$(2,4,3,\frac{1}{6})$&$\frac{1}{\eta ^4}+\left(1+\frac{1}{\eta ^4}+4 \eta ^2\right) \sqrt[3]{q}+\left(1+\frac{1}{\eta ^4}+4 \eta ^2\right) q^{2/3}+\left(1+\frac{1}{\eta ^4}+4 \eta ^2\right) q+\left(1-\frac{4}{\eta ^2}+4 \eta ^2\right) q^{4/3}+\left(-1-\frac{1}{\eta ^4}-\frac{4}{\eta ^2}-\eta ^4\right) q^{5/3}+\left(-9-\frac{1}{\eta ^4}-\frac{4}{\eta ^2}-4 \eta ^2-\eta ^4\right) q^2+\mathcal{O}(q^{7/3})$\\
    \cline{2-4}&\multirow{5}{*}[-1.5cm]{3}&$(1,0,3,\frac{1}{6})$&0\\
    \cline{3-4}&&$(1,1,2,\frac{1}{6})$&$\left(1+\eta ^2\right) \sqrt[3]{q}+\frac{q^{2/3}}{\eta ^2}-\eta ^2 q^{4/3}+\left(-1-\frac{1}{\eta ^2}\right) q^{5/3}$\\
    \cline{3-4}&&$(1,2,1,\frac{1}{6})$&$\frac{1}{\eta ^2}+\left(1+\frac{1}{\eta ^2}+\eta ^2\right) \sqrt[3]{q}+\eta ^2 q^{2/3}-\frac{q^{4/3}}{\eta ^2}+\left(-1-\frac{1}{\eta ^2}-\eta ^2\right) q^{5/3}-\eta ^2 q^2$\\
    \cline{3-4}&&$(1,3,0,\frac{1}{6})$&$\frac{1}{\eta ^2 q^{2/3}}+\frac{1}{\eta ^2 \sqrt[3]{q}}+\frac{1}{\eta ^2}+\eta ^2 \sqrt[3]{q}+\eta ^2 q^{2/3}+\left(-\frac{1}{\eta ^2}+\eta ^2\right) q-\frac{q^{4/3}}{\eta ^2}-\frac{q^{5/3}}{\eta ^2}-\eta ^2 q^2-\eta ^2 q^{7/3}-\eta ^2 q^{8/3}$\\
    \cline{3-4}&&$(2,4,4,\frac{1}{6})$&$\left(1+\frac{1}{\eta ^4}+4 \eta ^2\right) \sqrt[3]{q}+\left(1+\frac{1}{\eta ^4}+4 \eta ^2\right) q^{2/3}+\left(2+\frac{1}{\eta ^4}+4 \eta ^2\right) q+\left(2+\frac{1}{\eta ^4}+4 \eta ^2\right) q^{4/3}+\left(-1-\frac{4}{\eta ^2}-\eta ^4\right) q^{5/3}+\left(-9-\frac{1}{\eta ^4}-\frac{4}{\eta ^2}-4 \eta ^2-\eta ^4\right) q^2+\mathcal{O}(q^{7/3})$\\
\hline
\end{tabularx}

\newpage

\begin{tabularx}{\textwidth}{ 
  |c|c|c |>{\centering\arraybackslash}X|}
\hline
    $p$ & $\alpha$ & $(F,N,\widetilde{N},\D_Q)$ & Single-particle index\\
    \hline
    \multirow{3}{*}[-1cm]{5} & \multirow{3}{*}[-1cm]{4} &$(1,1,3,\frac{1}{6})$&$\left(1+\eta ^2\right) \sqrt[3]{q}+q^{2/3}+\left(\frac{1}{\eta ^2}-\eta ^2\right) q-q^{4/3}+\left(-1-\frac{1}{\eta ^2}\right) q^{5/3}$\\
    \cline{3-4}&&$(1,2,2,\frac{1}{6})$&$\left(1+\frac{1}{\eta ^2}+\eta ^2\right) \sqrt[3]{q}+\left(2+\frac{1}{\eta ^2}+\eta ^2\right) q^{2/3}+\left(-2-\frac{1}{\eta ^2}-\eta ^2\right) q^{4/3}+\left(-1-\frac{1}{\eta ^2}-\eta ^2\right) q^{5/3}$\\
    \cline{3-4}&&$(1,3,1,\frac{1}{6})$&$\frac{1}{\eta ^2 \sqrt[3]{q}}+\frac{1}{\eta ^2}+\left(1+\frac{1}{\eta ^2}+\eta ^2\right) \sqrt[3]{q}+\left(1+\eta ^2\right) q^{2/3}+\left(-\frac{1}{\eta ^2}+\eta ^2\right) q+\left(-1-\frac{1}{\eta ^2}\right) q^{4/3}+\left(-1-\frac{1}{\eta ^2}-\eta ^2\right) q^{5/3}-\eta ^2 q^2-\eta ^2 q^{7/3}$\\
\hline
\end{tabularx}

\newpage
\subsubsection*{$\text{KP}_\a^+$ for $\a > p$}
\begin{tabularx}{\textwidth}{ 
  |c|c|c |>{\centering\arraybackslash}X|}
\hline
    $p$ & $\alpha$ & $(F,N,\widetilde{N},\D_Q)$ & Single-particle index\\
\hline
    \multirow{16}{*}{2} &\multirow{6}{*}[-1.5cm]{3}&$(1,1,2,\frac{1}{2})$&$1+q^{2/3}-q^{4/3}-q^2$\\
    \cline{3-4}&&$(1,2,1,\frac{1}{6})$&$1+\left(\frac{1}{\eta ^2}+\eta ^2\right) \sqrt[3]{q}+q^{2/3}-q^{4/3}+\left(-\frac{1}{\eta ^2}-\eta ^2\right) q^{5/3}-q^2$\\
    \cline{3-4}&&$(2,2,3,\frac{1}{2})$&$1+\left(2+\frac{1}{\eta ^4}\right) q^{2/3}+4 \eta ^2 q-q^{4/3}-\frac{4 q^{5/3}}{\eta ^2}+\left(-9-\frac{1}{\eta ^4}-\eta ^4\right) q^2+\mathcal{O}(q^{7/3})$\\
    \cline{3-4}&&$(2,3,2,\frac{1}{6})$&$1+4 \eta ^2 \sqrt[3]{q}+\left(2+\frac{1}{\eta ^4}\right) q^{2/3}+4 \eta ^2 q+\left(-1+\frac{1}{\eta ^4}-\eta ^4\right) q^{4/3}+\left(-\frac{4}{\eta ^2}-4 \eta ^2\right) q^{5/3}+\left(-9-\frac{1}{\eta ^4}-\eta ^4\right) q^2+\mathcal{O}(q^{7/3})$\\
\cline{3-4} && $(3,3,4,\frac{4}{9})$ & $1+\left(2+\frac{1}{\eta ^6}\right) q^{2/3}+9 \eta ^2 q^{8/9}+\frac{q^{4/3}}{\eta ^6}+9 \eta ^2 q^{14/9}+\left(-19-\frac{1}{\eta ^6}\right) q^2 + \mathcal{O}(q^{\frac{20}{9}})$\\
\cline{3-4}&&$(3,4,3,\frac{2}{9})$&$1+9 \eta ^2 q^{4/9}+\left(2+\frac{1}{\eta ^6}\right) q^{2/3}+9 \eta ^2 q^{10/9}+\frac{q^{4/3}}{\eta ^6}+\left(-19-\eta ^6\right) q^2 + \mathcal{O}(q^{20/9})$\\
\cline{2-4}&\multirow{2}{*}[-0.25cm]{4}&$(1,2,2,\frac{1}{6})$&$\frac{1}{q^{2/3}}+1-q^2-q^{8/3}$\\
\cline{3-4}&&$(2,3,3,\frac{1}{6})$&$\frac{1}{q^{2/3}}+1+4 \eta ^2 \sqrt[3]{q}+\left(1-\eta ^4\right) q^{2/3}+\left(-1+\frac{1}{\eta ^4}\right) q^{4/3}-\frac{4 q^{5/3}}{\eta
   ^2}-q^2-q^{8/3}$\\
\hline
\multirow{4}{*}[-3cm]{3}&\multirow{3}{*}[-1.25cm]{4} & $(1,2,2,\frac{1}{4})$&$1+\left(2+\frac{1}{\eta ^2}+\eta ^2\right) \sqrt{q}+\left(-2-\frac{1}{\eta ^2}-\eta ^2\right) q^{3/2}-q^2$\\
\cline{3-4}&&$(2,3,4,\frac{3}{8})$&$1+\left(2+\frac{1}{\eta ^4}\right) \sqrt{q}+4 \eta ^2 q^{3/4}+\left(2+\frac{1}{\eta ^4}\right) q+4 \eta ^2 q^{5/4}-q^{3/2}-\frac{4 q^{7/4}}{\eta ^2}+\left(-11-\frac{2}{\eta ^4}-\eta ^4\right) q^2+\mathcal{O}(q^{9/4})$\\
\cline{3-4}&&$(2,4,3,\frac{1}{8})$&$1+4 \eta ^2 \sqrt[4]{q}+\left(2+\frac{1}{\eta ^4}\right) \sqrt{q}+4 \eta ^2 q^{3/4}+\left(2+\frac{1}{\eta ^4}\right) q+4 \eta ^2 q^{5/4}+\left(-1+\frac{1}{\eta ^4}-\eta ^4\right) q^{3/2}+\left(-\frac{4}{\eta ^2}-4 \eta ^2\right) q^{7/4}+\left(-11-\frac{1}{\eta ^4}-2 \eta ^4\right) q^2+\mathcal{O}(q^{9/4})$\\
\cline{2-4}&\multirow{1}{*}{5}&$(2,4,4)$&$\frac{1}{\sqrt{q}}+1+4 \eta^2 q^{1/4}+2\sqrt{q}+4\eta^2 q^{3/4}+\left(1+\frac{1}{\eta^4}-\eta^4 \right)q+\left(-2+\frac{1}{\eta^4}-\eta^4\right)q^{3/2}+\left(-\frac{3}{\eta^2}-4\eta^2\right)q^{7/4}-9q^2+\mathcal{O}(q^{9/4})$\\
\cline{2-4}&\multirow{1}{*}{6}&$(3,6,6,\frac{1}{12})$& $\frac{1}{q}+\frac{1}{\sqrt{q}}+1+9 \eta ^2 \sqrt[6]{q}+\sqrt{q}+9 \eta ^2 q^{2/3}+\left(1-\eta ^6\right) q+9 \eta ^2 q^{7/6}-9 \eta ^4 q^{4/3}+\left(\frac{1}{\eta ^6}-\eta ^6\right) q^{3/2}-9 \eta ^2 q^{5/3}-9 \eta ^4 q^{11/6}+\left(-20+\frac{1}{\eta ^6}+\eta ^6\right)
   q^2+\mathcal{O}(q^{13/6})$\\
\hline
    \multirow{3}{*}[-0.5cm]{4} &\multirow{2}{*}[-0.5cm]{5}& $(1,2,3,\frac{3}{10})$&$1+2 q^{2/5}+\left(\frac{1}{\eta ^2}+\eta ^2\right) q^{3/5}+q^{4/5}-q^{6/5}+\left(-\frac{1}{\eta ^2}-\eta ^2\right) q^{7/5}-2 q^{8/5}-q^2$\\
    \cline{3-4}&&$(1,3,2,\frac{1}{10})$&$1+\left(\frac{1}{\eta ^2}+\eta ^2\right) \sqrt[5]{q}+2 q^{2/5}+\left(\frac{1}{\eta ^2}+\eta ^2\right) q^{3/5}+q^{4/5}-q^{6/5}+\left(-\frac{1}{\eta ^2}-\eta ^2\right) q^{7/5}-2 q^{8/5}+\left(-\frac{1}{\eta ^2}-\eta ^2\right) q^{9/5}-q^2$\\
    \cline{2-4}&\multirow{1}{*}{6}&$(1,3,3,\frac{1}{10})$&$\frac{1}{q^{2/5}}+1+\eta ^2 \sqrt[5]{q}+2 q^{2/5}+\frac{q^{3/5}}{\eta ^2}-\eta ^2 q^{7/5}-2 q^{8/5}-\frac{q^{9/5}}{\eta ^2}-q^2-q^{12/5}$\\
\hline
    \multirow{1}{*}{5} &\multirow{1}{*}{6}&$(1,3,3,\frac{1}{6})$&$1+\left(2+\frac{1}{\eta ^2}+\eta ^2\right) \sqrt[3]{q}+\left(2+\frac{1}{\eta ^2}+\eta ^2\right) q^{2/3}+\left(-2-\frac{1}{\eta ^2}-\eta ^2\right) q^{4/3}+\left(-2-\frac{1}{\eta ^2}-\eta ^2\right) q^{5/3}-q^2$\\
\hline

\end{tabularx}

\subsubsection*{$\text{KP}_{\a,\b}$ for $\a < p$ and $\b < p$}
\begin{tabularx}{\textwidth}{ 
  |c|c|c |>{\centering\arraybackslash}X|}
\hline
    $p$ & $(\alpha,\beta)$ & $(F,N,\widetilde{N})$ & Single-particle index\\
\hline
\multirow{5}{*}{2} & \multirow{5}{*}{$(0,1)$} & $(2,0,1)$ & $0$\\
\cline{3-4}&&$(2,1,0)$ & $4 q^{1/3} - 4q^{5/3}$ \\
\cline{3-4}&&$(3,1,2)$ & $q^{2/3} + 9 q^{8/9} - 9 q^{16/9} - 17 q^2 + \mathcal{O}(q^{20/9})$\\
\cline{3-4}&&$(3,2,1)$ & $9 q^{4/9} + q^{2/3} + 9q^{10/9} - 9 q^{14/9} - 9q^{16/9} - 17q^2 + \mathcal{O}(q^{20/9})$\\
\cline{3-4}&& $(4,2,3)$ & $q^{2/3} + 16 q^{5/6} + q^{4/3} + 16 q^{3/2} - 31 q^2 + \mathcal{O}(q^{13/6})$\\
\cline{3-4}&& $(4,3,2)$ & $16 q^{1/2} + q^{2/3} + 16 q^{7/6} + q^{4/3} - 31 q^2 + \mathcal{O}(q^{13/6})$\\
\hline
\multirow{12}{*}{3} & \multirow{5}{*}{$(0,1)$} &$(2,0,1)$ & $0$\\
\cline{3-4}&&$(2,1,0)$ & $4 q^{1/4} - 4 q^{7/4}$\\
\cline{3-4}&&$(3,1,3)$ & $q^{1/2} + 9 q^{5/6} - 9q^{5/3} - 17 q^2 +\mathcal{O}(q^{7/3})$\\
\cline{3-4}&&$(3,2,2)$ & $10 q^{1/2} + 10 q - 10 q^{3/2} - 35 q^2 + \mathcal{O}(q^{3})$\\
\cline{3-4}&&$(3,3,1)$ & $9q^{1/6} + q^{1/2} + 9q^{2/3} + 9q^{7/6} - 9q^{4/3} - 9q^{5/3} - 9q^{11/6} - 17 q^2 +\mathcal{O}(q^{7/3})$\\
\cline{2-4}&\multirow{3}{*}{$(0,2)$} &$(2,1,1)$ & $5 q^{1/2} - 5 q^{3/2}$\\
\cline{3-4}&&$(3,2,3)$ & $q^{1/2} + 9q^{2/3} + 2q + 9q^{7/6} - 9q^{11/6} - 19 q^2 +\mathcal{O}(q^{13/6})$\\
\cline{3-4}&&$(3,3,2)$ & $9 q^{1/3} + q^{1/2} + 9q^{5/6} + 2q + 9 q^{4/3} - 9q^{5/3} - 9q^{11/6} - 19 q^2+\mathcal{O}(q^{13/6})$\\
\cline{2-4}&\multirow{4}{*}{$(1,2)$} &$(1,0,0)$ & $0$\\
\cline{3-4}&&$(2,1,2)$ & $q^{1/2} + 4 q^{3/4} + q - q^{3/2} - 4 q^{7/4} - 8q^2+\mathcal{O}(q^{9/4})$\\
\cline{3-4}&&$(2,2,1)$ & $4q^{1/4} + q^{1/2} + 4 q^{3/4} + q - q^{3/2} - 8q^{7/4} - 8q^2+\mathcal{O}(q^{9/4})$\\
\cline{3-4}&&$(3,3,3)$ & $10q^{1/2} + 11q + 11q^{3/2} - 36 q^2 +\mathcal{O}(q^{5/2})$\\
\hline
\multirow{14}{*}[-.5cm]{4}&\multirow{4}{*}[-0.5cm]{$(0,1)$}&$(2,0,1)$ & $0$\\
\cline{3-4}&&$(2,1,0)$ & $4 q^{1/5} - 4 q^{9/5}$\\
\cline{3-4}&&$(3,2,3)$ & $q^{2/5} + 9q^{8/15} + q^{4/5} + 9q^{14/15}-9q^{22/15}-q^{8/5}-9q^{28/15}-17q^2+\mathcal{O}(q^{32/15})$\\
\cline{3-4}&&$(3,3,2)$ & $9q^{4/15} + q^{2/5} + 9q^{2/3} + q^{4/5} + 9q^{16/15}-9q^{4/3}-q^{8/5}-9q^{26/15}-9q^{28/15}-17q^2+\mathcal{O}(q^{32/15})$\\
\cline{2-4}&\multirow{2}{*}{$(0,2)$}&$(2,1,1)$ & $5q^{2/5}-q^{4/5}+q^{6/5}-5q^{8/5}$\\
\cline{3-4}&&$(3,3,3)$ & $10q^{2/5}+10q^{4/5}+11q^{6/5}-10q^{8/5}-36q^2+\mathcal{O}(q^{12/5})$\\
\cline{2-4}&\multirow{2}{*}{$(0,3)$}&$(2,1,2)$ & $q^{2/5}+4q^{3/5}+q^{4/5}-q^{6/5}-4q^{7/5}-q^{8/5}$\\
\cline{3-4}&&$(2,2,1)$ & $4q^{1/5}+q^{2/5}+4q^{3/5}+q^{4/5}-q^{6/5}-4q^{7/5}-q^{8/5}-4q^{9/5}$\\
\cline{2-4}&\multirow{2}{*}{$(1,2)$}&$(2,1,2)$ & $q^{2/5}+4q^{3/5}-4q^{9/5}-8q^2+\mathcal{O}(q^{11/5})$\\
\cline{3-4}&&$(2,2,1)$ & $4q^{1/5}+q^{2/5}+4q^{3/5}-8q^{9/5}-8q^2+\mathcal{O}(q^{11/5})$\\
\cline{2-4}&\multirow{2}{*}{$(1,3)$}&$(1,0,0)$ & 0\\
\cline{3-4}&&$(2,2,2)$ & $5q^{2/5}+6q^{4/5}-6q^{8/5}-16q^2+\mathcal{O}(q^{12/5})$\\
\cline{2-4}&\multirow{2}{*}[-0.3cm]{$(2,3)$}&$(2,2,3)$ & $q^{2/5}+4q^{3/5}+2q^{4/5}+4q+2q^{6/5}-q^{8/5}-4q^{9/5}-11q^2+\mathcal{O}(q^{11/5})$\\
\cline{3-4}&&$(2,3,2)$ & $4q^{1/5}+q^{2/5}+4q^{3/5}+2q^{4/5}+4q+2q^{6/5}-q^{8/5}-8q^{9/5}-11q^2+\mathcal{O}(q^{11/5})$\\
\hline

\end{tabularx}
\newpage
\begin{tabularx}{\textwidth}{ 
  |c|c|c |>{\centering\arraybackslash}X|}
\hline
    $p$ & $(\alpha,\beta)$ & $(F,N,\widetilde{N})$ & Single-particle index\\
\hline
\multirow{18}{*}{5} & \multirow{3}{*}{$(0,1)$}&$(2,0,1)$ & $0$\\
\cline{3-4}&&$(2,1,0)$ & $4q^{1/6} - 4q^{11/6}$\\
\cline{3-4}&&$(3,3,3)$ & $10q^{1/3}+10q^{2/3}+10q-10q^{4/3}-10q^{5/3}-35q^2+\mathcal{O}(q^{3})$\\
\cline{2-4}&\multirow{1}{*}{$(0,2)$}&$(2,1,1)$ & $5q^{1/3}-q^{2/3}+q^{4/3}-5q^{5/3}$\\
\cline{2-4}&\multirow{2}{*}{$(0,3)$}&$(2,1,2)$ & $q^{1/3}+4q^{1/2}-4q^{3/2}-q^{5/3}$\\
\cline{3-4}&&$(2,2,1)$ & $4q^{1/6}+q^{1/3}+4q^{1/2}-4q^{3/2}-q^{5/3}-4q^{11/6}$\\
\cline{2-4}&\multirow{1}{*}{$(0,4)$}&$(2,2,2)$ & $5q^{1/3}+6q^{2/3}-6q^{4/3}-5q^{5/3}+\mathcal{O}( q^{7/3})$\\
\cline{2-4}&\multirow{2}{*}{$(1,2)$}&$(2,1,2)$ & $q^{1/3}+4q^{1/2}-q+q^{4/3}-4q^{11/6}-8q^2+\mathcal{O}(q^{13/6})$\\
\cline{3-4}&&$(2,2,1)$ & $4q^{1/6}+q^{1/3}+4q^{1/2}-q+q^{4/3}-8q^{11/6}-8q^2+\mathcal{O}(q^{13/6})$\\
\cline{2-4}&\multirow{1}{*}{$(1,3)$}&$(2,2,2)$ & $5q^{1/3}+5q^{2/3}-5q^{5/3}-16q^2+\mathcal{O}(q^{8/3})$\\
\cline{2-4}&\multirow{3}{*}[-0.5cm]{$(1,4)$}&$(1,0,0)$ & $0$\\
\cline{3-4}&&$(2,2,3)$ & $q^{1/3}+4q^{1/2}+2q^{2/3}+4q^{5/6}+q-q^{4/3}-4q^{3/2}-2q^{5/3}-4q^{11/6}-8q^2+\mathcal{O}(q^{13/6})$\\
\cline{3-4}&&$(2,3,2)$ & $4q^{1/6}+q^{1/3}+4q^{1/2}+2q^{2/3}+4q^{5/6}+q-q^{4/3}-4q^{3/2}-2q^{5/3}-8q^{11/6}-8q^2+\mathcal{O}(q^{13/6})$\\
\cline{2-4}&\multirow{3}{*}[-0.3cm]{$(2,3)$}&$(1,0,0)$ & $0$\\
\cline{3-4}&&$(2,2,3)$ & $q^{1/3}+4q^{1/2}+q^{2/3}+4q^{5/6}+q+q^{4/3}-q^{5/3}-4q^{11/6}-10q^2+\mathcal{O}(q^{13/6})$\\
\cline{3-4}&&$(2,3,2)$ & $4q^{1/6}+q^{1/3}+4q^{1/2}+q^{2/3}+4q^{5/6}+q+q^{4/3}-q^{5/3}-8q^{11/6}-10q^2+\mathcal{O}(q^{13/6})$\\
\cline{2-4}&\multirow{1}{*}{$(2,4)$}&$(2,3,3)$ & $5q^{1/3}+6q^{2/3}+6q+q^{4/3}-6q^{5/3}-18q^2+\mathcal{O}(q^{7/3})$\\
\cline{2-4}&\multirow{1}{*}{$(3,4)$}&$(1,1,1)$ & $2q^{1/3}+q^{2/3}-q^{4/3}-2q^{5/3}+\mathcal{O}(q^{19/6})$\\
\hline
\end{tabularx}
\newpage
\subsubsection*{$\text{KP}_{\a,\b}$ for $\a \geq p$ or $\b \geq p$}
\begin{tabularx}{\textwidth}{ 
  |c|c|c |>{\centering\arraybackslash}X|}
\hline
    $p$ & $(\alpha,\beta)$ & $(F,N,\widetilde{N})$ & Single-particle index\\
\hline
    \multirow{18}{*}{1} & \multirow{10}{*}{(0,1)} & $(2,1,0)$ & $4 \sqrt{q}-4 q^{3/2}$\\ 
    \cline{3-4}&&$(3,2,0)$&$9 \sqrt[3]{q}-9 q^{5/3}$\\
    \cline{3-4}&&$(4,0,3)$&0\\
    \cline{3-4}&&$(4,1,2)$&$q+16 q^{5/4}-31 q^2-16 q^{9/4}-36 q^{5/2}+16 q^{11/4}+30 q^3+\mathcal{O}(q^{13/4})$\\
    \cline{3-4}&&$(5,0,4)$&0\\
    \cline{3-4}&&$(5,1,3)$&$q+25 q^{7/5}-49 q^2-25 q^{12/5}+\mathcal{O}(q^{13/5})$\\
    \cline{3-4}&&$(6,0,5)$&0\\
    \cline{3-4}&&$(6,1,4)$&$q+36 q^{3/2}-71 q^2-155 q^3+\mathcal{O}(q^{7/2})$\\
    \cline{3-4}&&$(6,2,3)$&$q+36 q^{7/6}-71 q^2+36 q^{17/6}+\mathcal{O}(q^{19/6})$\\
    \cline{3-4}&&$(7,1,5)$&$q+49 q^{11/7}-97 q^2+49 q^{17/7}-49 q^{18/7}+96 q^3+\mathcal{O}(q^{22/7})$\\
    \cline{2-4}&\multirow{8}{*}{$(1,2)$} &$(2,1,2)$&$1-q^2$\\
    \cline{3-4}&&$(3,1,3)$&$1-q^2$\\
    \cline{3-4}&&$(4,2,3)$&$1+q+16 q^{5/4}-32 q^2-16 q^{9/4}-36 q^{5/2}+16 q^{11/4}+30 q^3+\mathcal{O}(q^{13/4})$\\
    \cline{3-4}&&$(4,1,4)$&$1-q^2$\\
    \cline{3-4}&&$(5,1,5)$&$1-q^2$\\
    \cline{3-4}&&$(5,2,4)$&$1+q+25 q^{7/5}-50 q^2-25 q^{12/5}+25 q^{13/5}-100 q^{14/5}+48 q^3+\mathcal{O}(q^{17/5})$\\
    \cline{3-4}&&$(6,2,5)$&$1+q+36 q^{3/2}-72 q^2-155 q^3+\mathcal{O}(q^{7/2})$\\
    \cline{3-4}&&$(6,3,4)$&$1+q+36 q^{7/6}-72 q^2+36 q^{17/6}+\mathcal{O}(q^{19/6})$\\
\hline
    \multirow{10}{*}[-1.5cm]{2} & \multirow{1}{*}{$(0,2)$} & $(3,1,3)$ & $2 q^{2/3}+9 q^{10/9}-q^{4/3}-9 q^{16/9}-17 q^2-9 q^{20/9}+16 q^{8/3}+54 q^{26/9}+\mathcal{O}(q^{28/9})$\\ 
    \cline{2-4}&\multirow{2}{*}[-0.25cm]{$(0,3)$} &$(2,1,2)$&$1+q^{2/3}-q^{4/3}-q^2$\\
    \cline{3-4}&&$(3,2,3)$&$1+2 q^{2/3}+9 q^{8/9}-q^{4/3}-9 q^{16/9}-18 q^2-9 q^{20/9}+16 q^{8/3}+54 q^{26/9}+\mathcal{O}(q^{28/9})$\\
    \cline{2-4}&\multirow{3}{*}[-0.5cm]{$(1,2)$}&$(2,1,2)$&$2 q^{2/3}+4 q-4 q^{5/3}-10 q^2-4 q^{7/3}+14 q^{8/3}+24 q^3+\mathcal{O}(q^{10/3})$\\
    \cline{3-4}&&$(3,2,3)$&$2 q^{2/3}+9 q^{8/9}+2 q^{4/3}+9 q^{14/9}-19 q^2-18 q^{20/9}-18 q^{22/9}-21 q^{8/3}-18 q^{26/9}+\mathcal{O}(q^{10/3})$\\
    \cline{3-4}&&$(4,3,4)$&$2 q^{2/3}+16 q^{5/6}+2 q^{4/3}+16 q^{3/2}-31 q^2-34 q^{8/3}-32 q^{17/6}-72 q^3+\mathcal{O}(q^{19/6})$\\
    \cline{2-4}&\multirow{2}{*}[-0.25cm]{$(1,4)$}&$(2,2,3)$&$\frac{1}{q^{2/3}}+1-q^2-q^{8/3}-4 q^3+\mathcal{O}(q^{10/3})$\\
    \cline{3-4}&&$(3,3,4)$&$\frac{1}{q^{2/3}}+1+q^{2/3}+9 q^{8/9}-9 q^{16/9}-18 q^2-9 q^{20/9}+15 q^{8/3}+54 q^{26/9}+\mathcal{O}(q^{28/9})$\\
    \cline{2-4}&\multirow{2}{*}[-0.25cm]{$(2,3)$}&$(2,2,3)$&$1+3 q^{2/3}+4 q-q^{4/3}-4 q^{5/3}-11 q^2-4 q^{7/3}+14 q^{8/3}+24 q^3+\mathcal{O}(q^{10/3})$\\
    \cline{3-4}&&$(3,3,4)$&$1+3 q^{2/3}+9 q^{8/9}+q^{4/3}+9 q^{14/9}-20 q^2-18 q^{20/9}-18 q^{22/9}-21 q^{8/3}-18 q^{26/9}+\mathcal{O}(q^{10/3})$\\
\hline
    \multirow{4}{*}[-0.5cm]{3}& \multirow{1}{*}{$(0,3)$} & $(2,1,2)$ & $2 \sqrt{q}+4 q^{3/4}-4 q^{5/4}-2 q^{3/2}$\\ 
    \cline{2-4}&\multirow{1}{*}{$(0,5)$}&$(2,2,3)$&$\frac{1}{\sqrt{q}}+1+\sqrt{q}-q^{3/2}-q^2-q^{5/2}+4 q^{11/4}+\mathcal{O}(q^3)$\\
    \cline{2-4}&\multirow{1}{*}{$(1,4)$}&$(2,2,3)$&$1+2 \sqrt{q}+4 q^{3/4}+q-2 q^{3/2}-4 q^{7/4}-9 q^2-4 q^{9/4}+6 q^{5/2}+12 q^{11/4}+14 q^3+\mathcal{O}(q^{13/4})$\\
    \cline{2-4}&\multirow{1}{*}{$(2,3)$}&$(2,2,3)$&$2 \sqrt{q}+4 q^{3/4}+3 q+4 q^{5/4}-4 q^{7/4}-13 q^2-16 q^{9/4}-12 q^{5/2}+27 q^3+\mathcal{O}(q^{13/4})$\\
\hline
\end{tabularx}
\newpage
\begin{tabularx}{\textwidth}{ 
  |c|c|c |>{\centering\arraybackslash}X|}
\hline
    $p$ & $(\alpha,\beta)$ & $(F,N,\widetilde{N})$ & Single-particle index\\
\hline
    \multirow{2}{*}[-0.25cm]{3}& \multirow{1}{*}{$(2,5)$} & $(2,3,4)$ & $\frac{1}{\sqrt{q}}+1+2 \sqrt{q}+4 q^{3/4}+q-2 q^{3/2}-4 q^{7/4}-9 q^2-4 q^{9/4}+5 q^{5/2}+12 q^{11/4}+13 q^3+\mathcal{O}(q^{13/4})$\\  
    \cline{2-4}&\multirow{1}{*}{$(3,4)$}&$(2,3,4)$&$1+3 \sqrt{q}+4 q^{3/4}+3 q+4 q^{5/4}-q^{3/2}-4 q^{7/4}-14 q^2-16 q^{9/4}-12 q^{5/2}+27 q^3+\mathcal{O}(q^{13/4})$\\
\hline \multirow{6}{*}[-0.25cm]{4}& \multirow{1}{*}{$(0,5)$} & $(2,2,3)$ & $1+2 q^{2/5}+4 q^{3/5}+q^{4/5}-q^{6/5}-4 q^{7/5}-2 q^{8/5}-q^2$\\ 
\cline{2-4}&\multirow{1}{*}{$(1,4)$}&$(2,2,3)$&$2 q^{2/5}+4 q^{3/5}+2 q^{4/5}+4 q-4 q^{7/5}-2 q^{8/5}-4 q^{9/5}-10 q^2-4 q^{11/5}+14 q^{14/5}+24 q^3+\mathcal{O}(q^{16/5})$\\
\cline{2-4}&\multirow{1}{*}{$(0,7)$}&$(2,3,4)$&$\frac{1}{q^{4/5}}+\frac{1}{q^{2/5}}+1+q^{2/5}-q^{4/5}-q^{6/5}-q^{8/5}-10 q^2+\mathcal{O}(q^{11/5})$\\
\cline{2-4}&\multirow{1}{*}{$(1,6)$}&$(2,3,4)$&$\frac{1}{q^{2/5}}+1+2 q^{2/5}+4 q^{3/5}+q^{4/5}-q^{6/5}-2 q^{8/5}-4 q^{9/5}-10 q^2+\mathcal{O}(q^{11/5})$\\
\cline{2-4}&\multirow{1}{*}{$(2,5)$}&$(2,3,4)$&$1+2 q^{2/5}+4 q^{3/5}+2 q^{4/5}+4 q+2 q^{6/5}-2 q^{8/5}-4 q^{9/5}-12 q^2+\mathcal{O}(q^{11/5})$\\
\cline{2-4}&\multirow{1}{*}{$(3,4)$}&$(2,3,4)$&$2 q^{2/5}+4 q^{3/5}+3 q^{4/5}+4 q+3 q^{6/5}+4 q^{7/5}-4 q^{9/5}-13 q^2+\mathcal{O}(q^{11/5})$\\
\hline \multirow{4}{*}[-0.5cm]{5}& \multirow{1}{*}{$(0,5)$} & $(2,2,3)$ & $2 \sqrt[3]{q}+4 \sqrt{q}+2 q^{2/3}+4 q^{5/6}-4 q^{7/6}-2 q^{4/3}-4 q^{3/2}-2 q^{5/3}$\\ 
\cline{2-4}&\multirow{1}{*}{$(0,7)$}&$(2,3,4)$&$\frac{1}{\sqrt[3]{q}}+1+2 \sqrt[3]{q}+4 \sqrt{q}+q^{2/3}-2 q^{4/3}-4 q^{3/2}-3 q^{5/3}-9 q^2+\mathcal{O}(q^{13/6})$\\
\cline{2-4}&\multirow{1}{*}{$(1,6)$}&$(2,3,4)$&$1+2 \sqrt[3]{q}+4 \sqrt{q}+2 q^{2/3}+4 q^{5/6}+q-q^{4/3}-4 q^{3/2}-3 q^{5/3}-4 q^{11/6}-10 q^2+\mathcal{O}(q^{13/6})$\\
\cline{2-4}&\multirow{1}{*}{$(2,5)$}&$(2,3,4)$&$2 \sqrt[3]{q}+4 \sqrt{q}+2 q^{2/3}+4 q^{5/6}+2 q+4 q^{7/6}+q^{4/3}-4 q^{3/2}-2 q^{5/3}-4 q^{11/6}-11 q^2+\mathcal{O}(q^{13/6})$\\
\hline \multirow{2}{*}[-0.25cm]{6}& \multirow{1}{*}{$(0,7)$} & $(2,3,4)$ & $1+2 q^{2/7}+4 q^{3/7}+2 q^{4/7}+4 q^{5/7}+q^{6/7}-q^{8/7}-4 q^{9/7}-2 q^{10/7}-4 q^{11/7}-3 q^{12/7}-2 q^2+\mathcal{O}(q^{15/7})$\\ 
\cline{2-4}&\multirow{1}{*}{$(1,6)$}&$(2,3,4)$&$2 q^{2/7}+4 q^{3/7}+2 q^{4/7}+4 q^{5/7}+2 q^{6/7}+4 q-4 q^{9/7}-2 q^{10/7}-4 q^{11/7}-2 q^{12/7}-4 q^{13/7}-10 q^2+\mathcal{O}(q^{15/7})$\\
\hline \multirow{1}{*}{7}& \multirow{1}{*}{$(0,7)$} & $(2,3,4)$ & $2 \sqrt[4]{q}+4 q^{3/8}+2 \sqrt{q}+4 q^{5/8}+2 q^{3/4}+4 q^{7/8}-4 q^{9/8}-2 q^{5/4}-4 q^{11/8}-2 q^{3/2}-4 q^{13/8}-2 q^{7/4}-q^2+\mathcal{O}(q^{9/4})$\\ 
\hline
\end{tabularx}
\\
\\
\\

\subsection{\texorpdfstring{$S_b^3$}{Sb3} partition function and diagrammatic expression} \label{app: ptf quiver diagram}
Let us review the localization formula of the squashed three-sphere partition function \cite{Hama:2011ea} and its identities from the basic $3d$ $\CN=2$ IR dualities. We also provide a recipe for reading the formula from a given quiver diagram.
The squashed three-sphere partition function of a $3d$ $\CN=2$ gauge theory with gauge group $G$ and a set of chiral multiplets, collectively denoted by $\Phi$, having $R$-charge $r_\Phi$ and charged under the gauge and global symmetries with weights $\r_\Phi$ and $\phi_\Phi$ of representation $R_\Phi$ and $R_\phi$ respectively, with real mass parameter $M$, is given by an integral over the Coulomb branch parameter $\s$ as follows:
\begin{align}
    Z_{S_b^3} = \frac{1}{|W|} 
    \int \prod_{j=1}^{\text{rank}(G)}\!\!\! d \s_j\;
    e^{2\pi i \z \Tr \s}
    \frac{
    \prod_{\Phi} \prod_{\r_\Phi \in R_\Phi} \prod_{\phi_\Phi \in R_\phi} s_b (\frac{i Q}{2}(1-r_\Phi) - \rho_\Phi(\s) - \phi_\Phi(M) )
    }
    {
    \prod_{\a\in \text{root}} s_b(\a(\s) + \frac{i Q}{2})
    }
    \label{eq: S_b^3 ptf integral formula}
\end{align}
where $W$ is the Weyl group of $G$, $Q\equiv b + b^{-1}$ parametrizes the squashing parameter $b$, $\a$ runs over the roots of the gauge group $G$, and $\z$ is the Fayet-Illiopoulos parameter for the $U(1)$ factors in $G$. The double-sine function $s_b(x)$ is defined by
\begin{align}
    s_b(x) := \prod_{l,n\geq 0}
    \frac{l b + n b^{-1} + Q/2 - i x}{l b + n b^{-1} + Q/2 + i x} \,,
\end{align}
which satisfies a property $s_b(x) s_b(-x) = 1$ reflecting a cancellation of the contributions of two coupled massive chiral fields. There is an alternative way for the partition function evaluation by using geometric manipulations, so-called the A-model approach \cite{Closset:2018ghr,Closset:2019hyt,Closset:2023vos,Closset:2023jiq,Closset:2023bdr,Closset:2023izb,Closset:2024sle,Closset:2025lqt}, which does not involve integrals but summing the fibering and the handle gluing operators over Bethe vacua at rational values of the squared squashing parameter $b^2 \in \mathbb{Q}$.
\\

Let us present the partition function identities of the Aharony and BBP dualities:
\paragraph{Aharony duality}
\begin{align}
    &\frac{1}{N!} \int \prod_{j=1}^{N} d\s_j
    e^{\pi i \z \sum_{j=1}^{N} \s_j}
    \frac{\prod_{j=1}^{N} \prod_{a=1}^{F}s_b( \s_j - m_a + i Q/2 )\prod_{b=1}^{F} s_b(-\s_j - \Wm_b + i Q/2) }
    {\prod_{i\neq j}^{N} s_b(\s_i - \s_j + i Q/2)}
    \nonumber\\
    &=s_b\Big( \frac{iQ}{2}(N-F)+\frac{\sum_{a=1}^{F} m_a +\sum_{b=1}^{F} \Wm_b}{2} \pm \frac{\z}{2} \Big)
    \prod_{a,b=1}^{F} s_b (-m_a - \Wm_b + iQ/2)
    \nonumber\\
    &\quad\times
    \frac{1}{(F - N)!} \int \prod_{j=1}^{F - N}d\s_j
    e^{\pi i (-\z) \sum_{j=1}^{F_N} \s_j}
    \frac{\prod_{j=1}^{F-N} \prod_{a=1}^{F}s_b( -\s_j + m_a)\prod_{b=1}^{F} s_b(\s_j + \Wm_b) }
    {\prod_{i\neq j}^{F-N} s_b(\s_i - \s_j + i Q/2)}
    \label{eq: Aharony int}
\end{align}
where we have used a shorthand notation $s_b(x \pm y )\equiv s_b(x+y)s_b(x-y)$.

\paragraph{One-monopole BBP duality}
\begin{align}
    &\frac{1}{N} \int \prod_{j=1}^{N} d\s_j
    e^{\pi i \z \sum_{j=1}^{N} \s_j}
    \frac{\prod_{j=1}^{N} \prod_{a=1}^{F}s_b( \s_j - m_a + i Q/2 )\prod_{b=1}^{F} s_b(-\s_j - \Wm_b + i Q/2) }
    {\prod_{i\neq j}^{N} s_b(\s_i - \s_j + i Q/2)}
    \nonumber\\
    &=s_b\Big( \frac{iQ}{2}(N\!-\!F)+\frac{\sum_{a=1}^{F} m_a \!+\!\sum_{b=1}^{F} \Wm_b}{2} \!+\! \frac{\z}{2} \Big)
    \prod_{a,b=1}^{F}\!\! s_b (iQ/2 -m_a - \Wm_b )
    e^{-\frac{\pi i}{2}\sum_{a=1}^{F} (m_a^2 - \Wm_a^2) }
    \nonumber\\
    &\quad\times
    \frac{1}{\WN !} \int \prod_{j=1}^{\WN}d\s_j
    e^{\pi i (-\z+iQ) \sum_{j=1}^{\WN} \s_j}
    \frac{\prod_{j=1}^{\WN} \prod_{a=1}^{F}s_b( -\s_j + m_a)\prod_{b=1}^{F} s_b(\s_j + \Wm_b) }
    {\prod_{i\neq j}^{\WN} s_b(\s_i - \s_j + i Q/2)}
    \label{eq: BBP int}
\end{align}
where $\WN = F - N - 1$ and the condition for monopole superpotential is given as
\begin{align}
    \sum_{a=1}^{F} m_a + \sum_{b=1}^{F}\Wm_b = \z + iQ(F - N - 1).
    \label{eq: BBP1 condition}
\end{align}

\paragraph{Partition function from a quiver diagram}
It is convenient to use a quiver diagram for the integral expressions of the partition function. Let us provide a recipe for it:
\begin{equation}
    \begin{tikzpicture}[baseline={([yshift=-.5ex]current bounding box.center)}]
  \tikzset{vertex/.style={circle,fill=white!25,minimum size=12pt,inner sep=2pt}}
  \tikzset{every loop/.style={}}
    \node[vertex] (N) at (0,0) [shape=circle,draw=black,minimum size=2em] {$N$};
    \node at (0,-0.6) {\scriptsize $\eta$};
    \end{tikzpicture}
    \;\;\Longleftrightarrow \;\;
    \frac{1}{N!} \int \prod_{i=1}^{N} d\s_i \; e^{\pi i\; \frac{iQ}{2}\eta \sum_{i=1}^{N} \s_i} \frac{1}{\prod_{i\neq j}^{N} s_b ( \s_i - \s_j + \frac{iQ}{2} )} \,,
    \label{eq: circular node}
\end{equation}
\begin{equation}
\begin{tikzpicture}[baseline={([yshift=-.5ex]current bounding box.center)}]
  \tikzset{vertex/.style={circle,fill=white!25,minimum size=12pt,inner sep=2pt}}
  \tikzset{every loop/.style={}}
    \node[vertex] (N) at (0,0) {$N$};
    \node[vertex] (M) at (1.5,0) {$M$};
    \node at (0.7,0.3) {$\D$};
    \draw[->-=.55] ([yshift= 0pt] N.east) to ([yshift= 0pt] M.west);
    \end{tikzpicture}
    \;\;\Longleftrightarrow\;\;
    \prod_{i=1}^{N} \prod_{j=1}^{M} 
    s_b \Big( y_j - x_i + \frac{iQ}{2}(1-\D) \Big) \,,
    \qquad\qquad
    \label{eq: arrow}
\end{equation}
\begin{equation}
    +\big\{\D\big\}
    \;\;\Longleftrightarrow\;\;
    s_b \Big(\frac{iQ}{2}(1-\D) \Big) \,.
    \qquad\qquad\qquad
    \label{eq: singlet}
\end{equation}
The circular node \eqref{eq: circular node} represents $U(N)$ gauge group and we associate it with an integration over the Coulomb branch parameter $\s_j$ with a vector multiplet contribution and the FI parameter $\eta$ written under the node.\footnote{For convenience, we use a different normalization for the FI parameter compared to the one in \eqref{eq: S_b^3 ptf integral formula}, which are related by $\z = \frac{iQ}{2}\eta$.} An arrow in \eqref{eq: arrow} between two nodes, either circular (gauge) or square (flavor), represents a chiral multiplet contribution in the (anti-)fundamental representation of the $U(N)$ ($U(M)$) node whose fugacities are $x_i$ $(y_j)$ with real mass parameter $\D$.\footnote{Similar to the FI parameter, the real mass parameter here is normalized as $\D = \frac{2}{iQ} m+r$ where $m$ is an original real mass parameter appearing in \eqref{eq: S_b^3 ptf integral formula} and $r$ is the R-charge of the chiral multiplet.} We also put a list of real masses with curly-bracket as in \eqref{eq: singlet} to represent extra singlet contributions. For example, the Aharony and BBP dualities in a general linear quiver can be drawn instead of the integrals \eqref{eq: Aharony int} and \eqref{eq: BBP int} as:
\begin{equation}
\begin{tikzpicture}
  \tikzset{vertex/.style={circle,fill=white!25,minimum size=12pt,inner sep=2pt}}
  \tikzset{every loop/.style={}}
    \node[vertex] (n1) at (0-1.5,0) [shape=circle,draw=black,minimum size=2em] {$N_1$};
    \node[vertex] (n2) at (1.5-1.5,0) [shape=circle,draw=black,minimum height=2em, minimum width=2em] {$N_2$};
    \node[vertex] (n3) at (3-1.5,0) [shape=circle,draw=black,minimum height=2em, minimum width=2em] {$N_3$};

    \draw[->-=.5] ([yshift= 2pt] n1.east) to ([yshift= 2pt] n2.west);
    \draw[->-=.5] ([yshift= -2pt] n2.west) to ([yshift= -2pt] n1.east);
    \draw[->-=.5] ([yshift= 2pt] n2.east) to ([yshift= 2pt] n3.west);
    \draw[->-=.5] ([yshift= -2pt] n3.west) to ([yshift= -2pt] n2.east);

    \node at (-0.7-1.5,0) {$\cdots$};
    \node at (3.7-1.5,0) {$\cdots$};
    \node at (0.75-1.5,0.4) {\scriptsize $\D_1$};
    \node at (2.25-1.5,0.4) {\scriptsize $\D_2$};
    \node at (0-1.5,-0.7) {\scriptsize $\th$};
    \node at (1.5-1.5,-0.7) {\scriptsize $\xi$};
    \node at (3-1.5,-0.7) {\scriptsize $\l$};
    
    \node[vertex] (N1) at (0+5,0+3-0.5) [shape=circle,draw=black,minimum size=2em] {$N_1$};
    \node[vertex] (N2) at (1.5+5,0+3-0.5) [shape=circle,draw=black,minimum height=2em, minimum width=2em] {\scriptsize $N^{(0)}\!$};
    \node[vertex] (N3) at (3+5,0+3-0.5) [shape=circle,draw=black,minimum height=2em, minimum width=2em] {$N_3$};

    \draw[-to, min distance=1cm]  (N1) edge [out=120, in=60] node {} (N1);
    \draw[-to, min distance=1cm]  (N3) edge [out=120, in=60] node {} (N3);
    \draw[->-=.5] ([yshift= 2pt] N1.east) to ([yshift= 2pt] N2.west);
    \draw[->-=.5] ([yshift= -2pt] N2.west) to ([yshift= -2pt] N1.east);
    \draw[->-=.5] ([yshift= 2pt] N2.east) to ([yshift= 2pt] N3.west);
    \draw[->-=.5] ([yshift= -2pt] N3.west) to ([yshift= -2pt] N2.east);
    \draw[->-=.5] ([xshift= 6pt,yshift= -2pt] N1.north) to [bend right=-60] ([xshift= -6pt,yshift= -2pt] N3.north);
    \draw[->-=.5] ([xshift= -9pt,yshift= -4pt] N3.north) to [bend right=60] ([xshift= 9pt,yshift= -4pt] N1.north);

    \node at (-0.7+5,0+3-0.5) {$\cdots$};
    \node at (3.7+5,0+3-0.5) {$\cdots$};
    \node at (0+5,1.2+3-0.5) {\scriptsize $2\D_1$};
    \node at (3+5,1.2+3-0.5) {\scriptsize $2\D_2$};
    \node at (0.75+5,0.25+3-0.5) {\scriptsize $1\texttt{-}\D_1$};
    \node at (2.25+5,0.25+3-0.5) {\scriptsize $1\texttt{-}\D_2$};
    \node at (1.5+5,1.2+3-0.5) {\scriptsize $\D_1 \texttt{+} \D_2$};
    \node at (0+5,-0.7+3-0.5) {\scriptsize $\th+\xi$};
    \node at (1.5+5,-0.7+3-0.5) {\scriptsize $-\xi$};
    \node at (3+5,-0.7+3-0.5) {\scriptsize $\l+\xi$};

    \node at (6.3,1.5-0.5) {\scriptsize $+\{N_1(1\texttt{-}\D_1) \texttt{+} N_3 (1\texttt{-}\D_2) \texttt{-}N_2 \texttt{+} 1 \pm \xi\}$};

    \node[vertex] (b1) at (5,0-3+0.5) [shape=circle,draw=black,minimum size=2em] {$N_1$};
    \node[vertex] (b2) at (6.5,0-3+0.5) [shape=circle,draw=black,minimum height=2em, minimum width=2em] {\scriptsize $N^{(1)}\!$};
    \node[vertex] (b3) at (8,0-3+0.5) [shape=circle,draw=black,minimum height=2em, minimum width=2em] {$N_3$};

    \draw[-to, min distance=1cm]  (b1) edge [out=120, in=60] node {} (b1);
    \draw[-to, min distance=1cm]  (b3) edge [out=120, in=60] node {} (b3);
    \draw[->-=.5] ([yshift= 2pt] b1.east) to ([yshift= 2pt] b2.west);
    \draw[->-=.5] ([yshift= -2pt] b2.west) to ([yshift= -2pt] b1.east);
    \draw[->-=.5] ([yshift= 2pt] b2.east) to ([yshift= 2pt] b3.west);
    \draw[->-=.5] ([yshift= -2pt] b3.west) to ([yshift= -2pt] b2.east);
    \draw[->-=.5] ([xshift= 6pt,yshift= -2pt] b1.north) to [bend right=-60] ([xshift= -6pt,yshift= -2pt] b3.north);
    \draw[->-=.5] ([xshift= -9pt,yshift= -4pt] b3.north) to [bend right=60] ([xshift= 9pt,yshift= -4pt] b1.north);

    \node at (4.3,0-3+0.5) {$\cdots$};
    \node at (8.7,0-3+0.5) {$\cdots$};
    \node at (5,1.2-3+0.5) {\scriptsize $2\D_1$};
    \node at (8,1.2-3+0.5) {\scriptsize $2\D_2$};
    \node at (5.75,0.25-3+0.5) {\scriptsize $1\texttt{-}\D_1$};
    \node at (7.25,0.25-3+0.5) {\scriptsize $1\texttt{-}\D_2$};
    \node at (1.5+5,1.2-3+0.5) {\scriptsize $\D_1 \texttt{+} \D_2$};
    \node at (5,-0.7-3+0.5) {\scriptsize $\th\texttt{+}\xi \mp \D_1$};
    \node at (6.5,-0.7-3+0.5) {\scriptsize $-\xi \pm 1$};
    \node at (8,-0.7-3+0.5) {\scriptsize $\l\texttt{+}\xi \mp \D_2$};

    \node at (9,0.8-3+0.5) {\scriptsize $+\{2\mp 2\xi\}$};

    
    \node at (3,1.5) 
    {\begin{tikzpicture}
      \node [rotate=45] {\textcolor{black}{$\Longleftrightarrow$}};    
    \end{tikzpicture}
    };
    \node at (3,2) {\scriptsize Aharony};
    
    \node at (3,-1.5) 
    {\begin{tikzpicture}
      \node [rotate=135] {\textcolor{black}{$\Longleftrightarrow$}};    
    \end{tikzpicture}
    };
    \node at (3,-1) {\scriptsize $\text{BBP}^\pm$};

    \node at (-0.5,3.5) {$N^{(\a)} \equiv N_1+N_3 - N_2 - \a$};

\end{tikzpicture}
\label{eq: basic duality in quiver}
\end{equation}
where, for $\text{BBP}^\pm$, the real mass parameters should satisfy the constraint from the monopole superpotential term, $N_1(1-\D_1) + N_3(1-\D_2) - N_2 + 1 \pm \xi = 2$.

\subsection{\texorpdfstring{$S^3$}{S3} partition function and F-maximization} \label{app: f-max result}
The proposed duality in the main paper assumes relevance of the monopole superpotential deformation that can be checked using the F-maximization \cite{Jafferis:2010un}, which states that the three-sphere free energy of a 3d $\CN=2$ theory,
\begin{align}
    F_{S^3} = - \log |Z_{S^3}|\,,
\end{align}
is maximized at the superconformal fixed point. Namely, the superconformal R-charges, or the conformal dimensions of the chiral operators are determined by maximizing the free energy $F_{S^3}$. Since only the operator of the conformal dimension less than two can trigger an RG flow to a distinct new IR fixed point, we call the operator as relevant. We follow the convention of \cite{Pufu:2016zxm} for the computation of the $S^3$-partition function $Z_{S^3}$:
\begin{align}
    Z_{S^3} = \frac{1}{|W|} 
    \int \prod_{i=1}^{\text{rank}(G)}\!\!\! d \s_i
    \,
    e^{-2\pi \t \Tr \s}
    \prod_{\a\in\text{root}} \big( 2 \sinh(\pi \a(\s))\big)
    \prod_{\Phi} \prod_{\r_\Phi \in R_\Phi}
    e^{l\big(1-r_\Phi + i \r_\Phi(\s)\big)}
\end{align}
with
\begin{align}
    l(z) := -z\, \log(1-e^{2\pi i\, z})
    + \frac{i}{2} \Big(
    \pi z^2 + \frac{1}{\pi} \text{Li}_2 \big( e^{2\pi i\, z} \big)
    \Big) - \frac{\pi i }{12}
\end{align}
where $\t$ is the FI parameter that corresponds to the mixing parameter for $U(1)_T$ topological symmetry, $\Phi$ runs for chiral multiplets of R-charge $r_\Phi$ with representation $R_\Phi$ whose weights are $\r_\Phi$. A subtlety in the F-maximization procedure is that one needs to take care about the decoupling operator hitting the unitary bound of the conformal dimension 1/2. Once there arises a decoupling operator, it should be extracted from the theory by introducing a flip field that couples to it in the superpotential and repeat the procedure until there is no decoupling operator in the IR fixed point. 
In the following, we perform F--maximization for 3d $\CN=2$ SQCD of $U(N)$ gauge group with $F$ flavors $(Q,\WQ)$ and an adjoint, $X$, of superpotential $\Tr X^{p+1}$ for various choices of $p$, $N$, and $F$, with $N\leq 3$. If we denote the superconformal R-charge of flavors as $\D_Q$, then the superconformal R-charge of the dressed monopole operator becomes,
\begin{align}
    R_\a^\pm :=
    R[\hat V_\a^\pm] = F(1-\D_Q) - \frac{2}{p+1}(N-1-\a) \pm \t\,.
\end{align}
where $\t$ is the mixing parameter of the $U(1)_T$ topological symmetry. We begin by assuming a fixed point with superpotential $W= \Tr X^{p+1}$, which we denote $\CT$. At $\CT$, we check the relevance of the monopole operators $\hat V_\a^+$. If $\hat V_\a^+$ is a relevant operator, deform $\CT$ by $\D W_1 = \hat V_\a^+$ to get a new fixed point by RG flow, say $\CT_\a$, in which case we achieve the IR fixed point for $\text{KP}_\a^+$ duality. If this is the case, we check the relevance of $\hat V_\b^-$ further here. If $\hat V_\b^-$ is also relevant, we can get a fixed point for the $\text{KP}_{\a,\b}$ duality by deforming with it, say $\CT_{\a,\b}$. Here we enumerate detailed processes for various cases:
\\
\paragraph{$\ast \;\; U(1)$ gauge group}
\begin{itemize}
    \item $(p,F)=(2,2)\;:$
    At $\CT$, $R_0^+ = 1.1829$ so that we can flow to $\CT_0$ and $\CT_1$. At $\CT_0$ and $\CT_1$, we find $R_0^- = 0.78983$, $1.1113$ respectively so that we achieve fixed points $\CT_{0,0},\CT_{1,0},\CT_{0,1}$, and $\CT_{1,1}$.

    \item $(p,F)=(2,3)\;:$
    At $\CT$, $R_0^+ = 1.6891$ so that we can flow to $\CT_0$ where we find $R_0^- = 1.5439$, achieving a fixed point $\CT_{0,0}$.

    \item $(p,F)=(3,2)\;:$
    At $\CT$, $R_0^+ = 1.1829$ so that we can flow to $\CT_0$ and $\CT_1$ where we find $R_0^- = 0.7898,1.03139$ respectively, achieving fixed points $\{\CT_{0,\b}\}_{\b\leq 2}$ and $\{\CT_{1,\b}\}_{\b\leq 1}$.

    \item $(p,F)=(3,3)\;:$
    At $\CT$, $R_0^+ = 1.6891$ so that we can flow to $\CT_0$ where we find $R_0^- = 1.5439$, achieving a fixed point $\CT_{0,0}$.

    \item $(p,F)=(4,2)\;:$
    At $\CT$, $R_0^+ = 1.1829$ so that we can flow to $\{\CT_\a\}_{\a=0,1,2}$ where we find $R_0^- = 0.7898,0.9833,1.1748$ respectively, achieving fixed points $\{\CT_{0,\b}\}_{\b\leq 3}$,$\{\CT_{1,\b}\}_{\b\leq 2}$ and $\{\CT_{2,\b}\}_{\b\leq 2}$.

    \item $(p,F)=(5,2)\;:$
    At $\CT$, $R_0^+ = 1.1829$ so that we can flow to $\{\CT_\a\}_{\a=0,1,2}$ where we find $R_0^- = 0.7898,0.9511,1.1113$ respectively, achieving fixed points $\{\CT_{0,\b}\}_{\b\leq 3}$,$\{\CT_{1,\b}\}_{\b\leq 3}$ and $\{\CT_{2,\b}\}_{\b\leq 2}$.
\end{itemize}
\paragraph{$\ast \;\; U(2)$ gauge group}
\begin{itemize}
    \item $(p,F)=(2,2)\;:$
    At $\CT$, $R_0^+ = 0.7322$ so that we can flow to $\{\CT_\a\}_{\a=0,1}$. For $\a=0$, we need to flip $\WQ Q$ and $\WQ X Q$ to get $R_0^- = 3.1201$ so that no further relevant monopole deformation is possible. For $\a=1$, we should flip $\WQ Q$ to get $R_0^- = 0.5509$, achieving fixed points $\{\CT_{1,\b}\}_{\b\leq 2}$.

    \item $(p,F)=(2,3)\;:$
    At $\CT$, $R_0^+ = 1.2313$ so that we can flow to $\{\CT_\a\}_{\a=0,1}$ where we find $R_0^- = 0.8637,1.1818$ respectively, achieving fixed points $\{\CT_{0,\b}\}_{\b\leq 1}$ and $\{\CT_{1,\b}\}_{\b\leq 1}$.

    \item $(p,F)=(2,4)\;:$
    At $\CT$, $R_0^+ = 1.7312$ so that we can flow to $\CT_0$ where we find $R_0^- = 1.6026$, achieving a fixed points $\CT_{0,0}$.

    \item $(p,F)=(3,2)\;:$
    At $\CT$, $R_0^+ = 0.8515$ so that we can flow to $\{\CT_\a\}_{\a=0,1,2}$. For $\a=0$, we could not perform the F--maximization for computational limit. For $\a=1$, $\WQ Q$ should be flipped and $R_0^- = 0.5788$ to get fixed points $\{\CT_{0,\b}\}_{\b\leq 2}$. For $\a=2$, $R_0^- = 0.7796$ to get fixed points $\{\CT_{2,\b}\}_{\b\leq 1}$

    \item $(p,F)=(3,3)\;:$
    At $\CT$, $R_0^+ = 1.3592$ so that we can flow to $\{\CT_\a\}_{\a=0,1}$ where we find $R_0^- = 1.0535,1.2919$ respectively, achieving fixed points $\{\CT_{0,\b}\}_{\a\leq 1}$ and $\{\CT_{1,\b}\}_{\a\leq 1}$.
    
    \item $(p,F)=(4,2)\;:$
    At $\CT$, $R_0^+ = 0.9224$ so that we can flow to $\{\CT_\a\}_{\a=0,1,2}$. For $\a=0$, we need to flip $\WQ Q$ to get $R_0^- = 0.5916$ so that we can flow to fixed points $\{\CT_{0,\b}\}_{\b\leq 3}$. For $\a=1$, $R_0^- = 0.5977$ so that we can flow to fixed points $\{\CT_{1,\b}\}_{\b\leq 3}$. For $\a=2$, $R_0^- = 0.7888$ so that we can flow to fixed points $\{\CT_{2,\b}\}_{\b\leq 3}$.

    \item $(p,F)=(4,3)\;:$
    At $\CT$, $R_0^+ = 1.4336$ so that we can flow to $\{\CT_\a\}_{\a=0,1}$ where we find $R_0^- = 1.1647,1.3549$ respectively, achieving fixed points $\{\CT_{0,\b}\}_{\a\leq 2}$ and $\{\CT_{1,\b}\}_{\a\leq 1}$.

    \item $(p,F)=(5,2)\;:$
    At $\CT$, $R_0^+ = 0.9688$ so that we can flow to $\{\CT_\a\}_{\a=0,1,2,3}$. For $\a=0$, we need to flip $\WQ Q$ to get $R_0^- = 0.5945$ so that we can flow to fixed points $\{\CT_{0,\b}\}_{\b\leq 4}$. For $\a=1$, $R_0^- = 0.6344$ so that we can flow to fixed points $\{\CT_{1,\b}\}_{\b\leq 4}$. For $\a=2$, $R_0^- = 0.7940$ so that we can flow to fixed points $\{\CT_{2,\b}\}_{\b\leq 3}$. For $\a=3$, $R_0^- = 0.9539$ so that we can flow to fixed points $\{\CT_{3,\b}\}_{\b\leq 3}$.
    
\end{itemize}
\paragraph{$\ast \;\; U(3)$ gauge group}
\begin{itemize}
    \item $(p,F)=(2,3)\;:$
    At $\CT$, $R_0^+ = 0.7885$ so that we can flow to $\{\CT_\a\}_{\a=0,1}$. For $\a=0$, we need to flip $\WQ Q$ to get $R_0^- = 0.6114$ so that we can flow to fixed points $\{\CT_{0,\b}\}_{\b\leq 2}$. For $\a=1$, we need to flip $\WQ Q$ to get $R_0^- = 0.5720$ so that we can flow to fixed points $\{\CT_{1,\b}\}_{\b\leq 2}$.
    
    \item $(p,F)=(2,4)\;:$
    At $\CT$, $R_0^+ = 1.2794$ so that we can flow to $\{\CT_\a\}_{\a=0,1}$ where we find $R_0^- = 0.9372,1.2528$ respectively, achieving fixed points $\{\CT_{0,\b}\}_{\a\leq 1}$ and $\{\CT_{1,\b}\}_{\a\leq 1}$.

    \item $(p,F)=(3,2)\;:$
    At $\CT$, we need to flip $\WQ Q$ to get $R_0^+ = 0.5256$ so that we can flow to $\{\CT_\a\}_{\a=0,1,2}$. For $\a=0,1$ we could not compute F--maximization for computational limit. For $\a=2$, we need to flip $\WQ Q$ to get $R_0^- = 0.5060$, achieving fixed points $\{\CT_{2,\b}\}_{\b\leq 2}$.

    \item $(p,F)=(3,3)\;:$
    At $\CT$, $R_0^+ = 1.0311$ so that we can flow to $\{\CT_\a\}_{\a=0,1}$. For $\a=0$, we need to flip $\WQ Q$ to get $R_0^- = 0.6449$ so that we can flow to fixed points $\{\CT_{0,\b}\}_{\b\leq 2}$. For $\a=1$, $R_0^- = 0.8086$ so that we can flow to fixed points $\{\CT_{1,\b}\}_{\b\leq 2}$.

    \item $(p,F)=(4,2)\;:$
    At $\CT$, $R_0^+ = 0.6624$ so that we can flow to $\{\CT_\a\}_{\a=0,1,2,3}$. For $\a=0,1$, we could not compute F--maximization for computational limit. For $\a=2$, we should flip $\WQ Q$ to get $R_0^- = 0.5378$ so that we can flow to fixed points $\{\CT_{2,\b}\}_{\b\leq 3}$. For $\a=3$, $R_0^- = 0.5967$ so that we can flow to fixed points $\{\CT_{3,\b}\}_{\b\leq 3}$.

    \item $(p,F)=(4,3)\;:$
    At $\CT$, $R_0^+ = 1.1762$ so that we can flow to $\{\CT_\a\}_{\a=0,1,2}$ where we find $R_0^- = 0.7860,0.9746$, and $1.1646$ respectively, achieving fixed points $\{\CT_{0,\b}\}_{\a\leq 3}$, $\{\CT_{1,\b}\}_{\a\leq 2}$, and $\{\CT_{2,\b}\}_{\a\leq 2}$ respectively.

    \item $(p,F)=(5,2)\;:$
    At $\CT$, $R_0^+ = 0.7536$ so that we can flow to $\{\CT_\a\}_{\a=0,1,2,3}$. For $\a=0$, we could not compute F--maximization for computational limit. For $\a=1$, we need to flip $\WQ Q$ and $\hat V_0^-$ to get $R_1^- = 0.8226$ so that we can flow to fixed points $\{\CT_{1,\b}\}_{1\leq \b \leq 4}$. For $\a=2$, we need to flip $\WQ Q$ to get $R_0^- = 0.5634$ so that we can flow to fixed points $\{\CT_{2,\b}\}_{\b \leq 4}$. For $\a=3$, $R_0^- = 0.6368$ so that we can flow to fixed points $\{\CT_{3,\b}\}_{\b \leq 4}$.

    \item $(p,F)=(5,3)\;:$
    At $\CT$, $R_0^+ = 1.2704$ so that we can flow to $\{\CT_\a\}_{\a=0,1,2}$ where we find $R_0^- = 0.9255,1.0836$, and $1.2416$ respectively, achieving fixed points $\{\CT_{0,\b}\}_{\b\leq 3}$, $\{\CT_{1,\b}\}_{\b\leq 2}$, and $\{\CT_{2,\b}\}_{\b\leq 2}$ respectively.
    
\end{itemize}

\newpage
\bibliographystyle{ytphys}
\bibliography{ref}

\providecommand{\href}[2]{#2}\begingroup\raggedright\begin{thebibliography}{10}

\bibitem{Seiberg:1994pq}
N.~Seiberg, ``{Electric - magnetic duality in supersymmetric nonAbelian gauge
  theories},'' \href{http://dx.doi.org/10.1016/0550-3213(94)00023-8}{{\em Nucl.
  Phys. B} {\bfseries 435} (1995) 129--146},
  \href{http://arxiv.org/abs/hep-th/9411149}{{\ttfamily arXiv:hep-th/9411149}}.

\bibitem{Kutasov:1995np}
D.~Kutasov and A.~Schwimmer, ``{On duality in supersymmetric Yang-Mills
  theory},'' \href{http://dx.doi.org/10.1016/0370-2693(95)00676-C}{{\em Phys.
  Lett. B} {\bfseries 354} (1995) 315--321},
  \href{http://arxiv.org/abs/hep-th/9505004}{{\ttfamily arXiv:hep-th/9505004}}.

\bibitem{Brodie:1996vx}
J.~H. Brodie, ``{Duality in supersymmetric SU(N(c)) gauge theory with two
  adjoint chiral superfields},''
  \href{http://dx.doi.org/10.1016/0550-3213(96)00416-6}{{\em Nucl. Phys. B}
  {\bfseries 478} (1996) 123--140},
  \href{http://arxiv.org/abs/hep-th/9605232}{{\ttfamily arXiv:hep-th/9605232}}.

\bibitem{Giveon:2008zn}
A.~Giveon and D.~Kutasov, ``{Seiberg Duality in Chern-Simons Theory},''
  \href{http://dx.doi.org/10.1016/j.nuclphysb.2008.09.045}{{\em Nucl. Phys. B}
  {\bfseries 812} (2009) 1--11},
  \href{http://arxiv.org/abs/0808.0360}{{\ttfamily arXiv:0808.0360 [hep-th]}}.

\bibitem{Benini:2011mf}
F.~Benini, C.~Closset, and S.~Cremonesi, ``{Comments on 3d Seiberg-like
  dualities},'' \href{http://dx.doi.org/10.1007/JHEP10(2011)075}{{\em JHEP}
  {\bfseries 10} (2011) 075}, \href{http://arxiv.org/abs/1108.5373}{{\ttfamily
  arXiv:1108.5373 [hep-th]}}.

\bibitem{Kapustin:2011gh}
A.~Kapustin, ``{Seiberg-like duality in three dimensions for orthogonal gauge
  groups},'' \href{http://arxiv.org/abs/1104.0466}{{\ttfamily arXiv:1104.0466
  [hep-th]}}.

\bibitem{Hwang:2011qt}
C.~Hwang, H.~Kim, K.-J. Park, and J.~Park, ``{Index computation for 3d
  Chern-Simons matter theory: test of Seiberg-like duality},''
  \href{http://dx.doi.org/10.1007/JHEP09(2011)037}{{\em JHEP} {\bfseries 09}
  (2011) 037}, \href{http://arxiv.org/abs/1107.4942}{{\ttfamily arXiv:1107.4942
  [hep-th]}}.

\bibitem{Hwang:2011ht}
C.~Hwang, K.-J. Park, and J.~Park, ``{Evidence for Aharony duality for
  orthogonal gauge groups},''
  \href{http://dx.doi.org/10.1007/JHEP11(2011)011}{{\em JHEP} {\bfseries 11}
  (2011) 011}, \href{http://arxiv.org/abs/1109.2828}{{\ttfamily arXiv:1109.2828
  [hep-th]}}.

\bibitem{Hanany:2015via}
A.~Hanany, C.~Hwang, H.~Kim, J.~Park, and R.-K. Seong, ``{Hilbert Series for
  Theories with Aharony Duals},''
  \href{http://dx.doi.org/10.1007/JHEP11(2015)132}{{\em JHEP} {\bfseries 11}
  (2015) 132}, \href{http://arxiv.org/abs/1505.02160}{{\ttfamily
  arXiv:1505.02160 [hep-th]}}. [Addendum: JHEP 04, 064 (2016)].

\bibitem{Hwang:2018uyj}
C.~Hwang, H.~Kim, and J.~Park, ``{On 3d Seiberg-Like Dualities with Two
  Adjoints},'' \href{http://dx.doi.org/10.1002/prop.201800064}{{\em Fortsch.
  Phys.} {\bfseries 66} no.~11-12, (2018) 1800064},
  \href{http://arxiv.org/abs/1807.06198}{{\ttfamily arXiv:1807.06198
  [hep-th]}}.

\bibitem{Hwang:2020ddr}
C.~Hwang, S.~Pasquetti, and M.~Sacchi, ``{Flips, dualities and symmetry
  enhancements},'' \href{http://dx.doi.org/10.1007/JHEP05(2021)094}{{\em JHEP}
  {\bfseries 05} (2021) 094}, \href{http://arxiv.org/abs/2010.10446}{{\ttfamily
  arXiv:2010.10446 [hep-th]}}.

\bibitem{Bottini:2021vms}
L.~E. Bottini, C.~Hwang, S.~Pasquetti, and M.~Sacchi, ``{4d S-duality wall and
  SL(2, \ensuremath{\mathbb{Z}}) relations},''
  \href{http://dx.doi.org/10.1007/JHEP03(2022)035}{{\em JHEP} {\bfseries 03}
  (2022) 035}, \href{http://arxiv.org/abs/2110.08001}{{\ttfamily
  arXiv:2110.08001 [hep-th]}}.

\bibitem{Benvenuti:2024mpn}
S.~Benvenuti, R.~Comi, and S.~Pasquetti, ``{Star-triangle dualities and
  supersymmetric improved bifundamentals},''
  \href{http://arxiv.org/abs/2410.19049}{{\ttfamily arXiv:2410.19049
  [hep-th]}}.

\bibitem{Benvenuti:2024seb}
S.~Benvenuti, R.~Comi, S.~Pasquetti, G.~Pedde~Ungureanu, S.~Rota, and A.~Shri,
  ``{Planar Abelian Mirror Duals of $\mathcal{N}=2$ SQCD$_3$},''
  \href{http://arxiv.org/abs/2411.05620}{{\ttfamily arXiv:2411.05620
  [hep-th]}}.

\bibitem{Benvenuti:2025huk}
S.~Benvenuti, R.~Comi, S.~Pasquetti, G.~Pedde~Ungureanu, S.~Rota, and A.~Shri,
  ``{A Chiral-Planar dualization algorithm for $3d$$\mathcal{N}=2$
  Chern-Simons-matter theories},''
  \href{http://arxiv.org/abs/2505.02913}{{\ttfamily arXiv:2505.02913
  [hep-th]}}.

\bibitem{Benvenuti:2025qnq}
S.~Benvenuti, R.~Comi, S.~Pasquetti, G.~Pedde~Ungureanu, S.~Rota, and A.~Shri,
  ``{Planar Abelian Duals of Chern-Simons QCD},''
  \href{http://arxiv.org/abs/2506.05465}{{\ttfamily arXiv:2506.05465
  [hep-th]}}.

\bibitem{Hayashi:2025guk}
H.~Hayashi, T.~Nosaka, and T.~Okazaki, ``{Abelian dualities and line defect
  indices for 3d gauge theories},''
  \href{http://arxiv.org/abs/2506.01278}{{\ttfamily arXiv:2506.01278
  [hep-th]}}.

\bibitem{Okazaki:2021pnc}
T.~Okazaki and D.~J. Smith, ``{Seiberg-like dualities for orthogonal and
  symplectic 3d $ \mathcal{N} $ = 2 gauge theories with boundaries},''
  \href{http://dx.doi.org/10.1007/JHEP07(2021)231}{{\em JHEP} {\bfseries 07}
  (2021) 231}, \href{http://arxiv.org/abs/2105.07450}{{\ttfamily
  arXiv:2105.07450 [hep-th]}}.

\bibitem{Okazaki:2021gkk}
T.~Okazaki and D.~J. Smith, ``{Web of Seiberg-like dualities for 3D N=2
  quivers},'' \href{http://dx.doi.org/10.1103/PhysRevD.105.086023}{{\em Phys.
  Rev. D} {\bfseries 105} no.~8, (2022) 086023},
  \href{http://arxiv.org/abs/2112.07347}{{\ttfamily arXiv:2112.07347
  [hep-th]}}.

\bibitem{Cho:2024civ}
M.~Cho, K.~Maruyoshi, E.~Nardoni, and J.~Song, ``{Large landscape of 4d
  superconformal field theories from small gauge theories},''
  \href{http://dx.doi.org/10.1007/JHEP11(2024)010}{{\em JHEP} {\bfseries 11}
  (2024) 010}, \href{http://arxiv.org/abs/2408.02953}{{\ttfamily
  arXiv:2408.02953 [hep-th]}}.

\bibitem{Kang:2024inx}
M.~J. Kang, C.~Lawrie, K.-H. Lee, and J.~Song, ``{Landscape of 4d N=1 SCFTs
  with a=c},'' \href{http://dx.doi.org/10.1103/PhysRevD.111.086002}{{\em Phys.
  Rev. D} {\bfseries 111} no.~8, (2025) 086002},
  \href{http://arxiv.org/abs/2412.17895}{{\ttfamily arXiv:2412.17895
  [hep-th]}}.

\bibitem{Amariti:2025zgj}
A.~Amariti, F.~Mantegazza, S.~Rota, and A.~Zanetti, ``{A - BCD dualities},''
  \href{http://arxiv.org/abs/2510.19966}{{\ttfamily arXiv:2510.19966
  [hep-th]}}.

\bibitem{Aharony:2013dha}
O.~Aharony, S.~S. Razamat, N.~Seiberg, and B.~Willett, ``{3d dualities from 4d
  dualities},'' \href{http://dx.doi.org/10.1007/JHEP07(2013)149}{{\em JHEP}
  {\bfseries 07} (2013) 149}, \href{http://arxiv.org/abs/1305.3924}{{\ttfamily
  arXiv:1305.3924 [hep-th]}}.

\bibitem{Aharony:1997gp}
O.~Aharony, ``{IR duality in d = 3 N=2 supersymmetric USp(2N(c)) and U(N(c))
  gauge theories},''
  \href{http://dx.doi.org/10.1016/S0370-2693(97)00530-3}{{\em Phys. Lett. B}
  {\bfseries 404} (1997) 71--76},
  \href{http://arxiv.org/abs/hep-th/9703215}{{\ttfamily arXiv:hep-th/9703215}}.

\bibitem{Borokhov:2002ib}
V.~Borokhov, A.~Kapustin, and X.-k. Wu, ``{Topological disorder operators in
  three-dimensional conformal field theory},''
  \href{http://dx.doi.org/10.1088/1126-6708/2002/11/049}{{\em JHEP} {\bfseries
  11} (2002) 049}, \href{http://arxiv.org/abs/hep-th/0206054}{{\ttfamily
  arXiv:hep-th/0206054}}.

\bibitem{Borokhov:2002cg}
V.~Borokhov, A.~Kapustin, and X.-k. Wu, ``{Monopole operators and mirror
  symmetry in three-dimensions},''
  \href{http://dx.doi.org/10.1088/1126-6708/2002/12/044}{{\em JHEP} {\bfseries
  12} (2002) 044}, \href{http://arxiv.org/abs/hep-th/0207074}{{\ttfamily
  arXiv:hep-th/0207074}}.

\bibitem{Kapustin:2005py}
A.~Kapustin, ``{Wilson-'t Hooft operators in four-dimensional gauge theories
  and S-duality},'' \href{http://dx.doi.org/10.1103/PhysRevD.74.025005}{{\em
  Phys. Rev. D} {\bfseries 74} (2006) 025005},
  \href{http://arxiv.org/abs/hep-th/0501015}{{\ttfamily arXiv:hep-th/0501015}}.

\bibitem{Gang:2018huc}
D.~Gang and M.~Yamazaki, ``{Three-dimensional gauge theories with supersymmetry
  enhancement},'' \href{http://dx.doi.org/10.1103/PhysRevD.98.121701}{{\em
  Phys. Rev. D} {\bfseries 98} no.~12, (2018) 121701},
  \href{http://arxiv.org/abs/1806.07714}{{\ttfamily arXiv:1806.07714
  [hep-th]}}.

\bibitem{Gang:2021hrd}
D.~Gang, S.~Kim, K.~Lee, M.~Shim, and M.~Yamazaki, ``{Non-unitary TQFTs from 3D
  $ \mathcal{N} $ = 4 rank 0 SCFTs},''
  \href{http://dx.doi.org/10.1007/JHEP08(2021)158}{{\em JHEP} {\bfseries 08}
  (2021) 158}, \href{http://arxiv.org/abs/2103.09283}{{\ttfamily
  arXiv:2103.09283 [hep-th]}}.

\bibitem{Gang:2023rei}
D.~Gang, H.~Kim, and S.~Stubbs, ``{Three-Dimensional Topological Field Theories
  and Nonunitary Minimal Models},''
  \href{http://dx.doi.org/10.1103/PhysRevLett.132.131601}{{\em Phys. Rev.
  Lett.} {\bfseries 132} no.~13, (2024) 131601},
  \href{http://arxiv.org/abs/2310.09080}{{\ttfamily arXiv:2310.09080
  [hep-th]}}.

\bibitem{Gang:2024loa}
D.~Gang, H.~Kim, B.~Park, and S.~Stubbs, ``{Three Dimensional Topological Field
  Theories and Nahm Sum Formulas},''
  \href{http://arxiv.org/abs/2411.06081}{{\ttfamily arXiv:2411.06081
  [hep-th]}}.

\bibitem{Creutzig:2024ljv}
T.~Creutzig, N.~Garner, and H.~Kim, ``{Mirror Symmetry and Level-rank Duality
  for 3d $\mathcal{N} = 4$ Rank 0 SCFTs},''
  \href{http://arxiv.org/abs/2406.00138}{{\ttfamily arXiv:2406.00138
  [hep-th]}}.

\bibitem{Ferrari:2023fez}
A.~E.~V. Ferrari, N.~Garner, and H.~Kim, ``{Boundary vertex algebras for 3d
  $\mathcal{N}=4$ rank-0 SCFTs},''
  \href{http://dx.doi.org/10.21468/SciPostPhys.17.2.057}{{\em SciPost Phys.}
  {\bfseries 17} no.~2, (2024) 057},
  \href{http://arxiv.org/abs/2311.05087}{{\ttfamily arXiv:2311.05087
  [hep-th]}}.

\bibitem{Gang:2022kpe}
D.~Gang and D.~Kim, ``{Generalized non-unitary Haagerup-Izumi modular data from
  3D S-fold SCFTs},'' \href{http://dx.doi.org/10.1007/JHEP03(2023)185}{{\em
  JHEP} {\bfseries 03} (2023) 185},
  \href{http://arxiv.org/abs/2211.13561}{{\ttfamily arXiv:2211.13561
  [hep-th]}}.

\bibitem{Gang:2023ggt}
D.~Gang, D.~Kim, and S.~Lee, ``{A non-unitary bulk-boundary correspondence:
  Non-unitary Haagerup RCFTs from S-fold SCFTs},''
  \href{http://dx.doi.org/10.21468/SciPostPhys.17.2.064}{{\em SciPost Phys.}
  {\bfseries 17} no.~2, (2024) 064},
  \href{http://arxiv.org/abs/2310.14877}{{\ttfamily arXiv:2310.14877
  [hep-th]}}.

\bibitem{Baek:2024tuo}
S.~Baek and D.~Gang, ``{3D bulk field theories for 2D non-unitary $ \mathcal{N}
  $ = 1 supersymmetric minimal models},''
  \href{http://dx.doi.org/10.1007/JHEP01(2025)027}{{\em JHEP} {\bfseries 01}
  (2025) 027}, \href{http://arxiv.org/abs/2405.05746}{{\ttfamily
  arXiv:2405.05746 [hep-th]}}.

\bibitem{Gang:2024tlp}
D.~Gang, H.~Kang, and S.~Kim, ``{Non-hyperbolic 3-manifolds and 3D field
  theories for 2D Virasoro minimal models},''
  \href{http://arxiv.org/abs/2405.16377}{{\ttfamily arXiv:2405.16377
  [hep-th]}}.

\bibitem{Baek:2025uev}
S.~Baek and H.~Kang, ``{Non-hyperbolic 3-manifolds and bulk field theories for
  supersymmetric/$W_N$ minimal models},''
  \href{http://arxiv.org/abs/2511.04524}{{\ttfamily arXiv:2511.04524
  [hep-th]}}.

\bibitem{Jeong:2025xid}
K.~Jeong and S.~Lee, ``{QFT Realization of Non-Unitary
  $\mathfrak{sl}(2,\mathbb{C})$ WRT Invariants and Their Galois
  Conjugations},'' \href{http://arxiv.org/abs/2511.16380}{{\ttfamily
  arXiv:2511.16380 [hep-th]}}.

\bibitem{Dedushenko:2023cvd}
M.~Dedushenko, ``{On the 4d/3d/2d view of the SCFT/VOA correspondence},''
  \href{http://arxiv.org/abs/2312.17747}{{\ttfamily arXiv:2312.17747
  [hep-th]}}.

\bibitem{Gaiotto:2024ioj}
D.~Gaiotto and H.~Kim, ``{3D TFTs from 4d $ \mathcal{N} $ = 2 BPS particles},''
  \href{http://dx.doi.org/10.1007/JHEP03(2025)173}{{\em JHEP} {\bfseries 03}
  (2025) 173}, \href{http://arxiv.org/abs/2409.20393}{{\ttfamily
  arXiv:2409.20393 [hep-th]}}.

\bibitem{Go:2025ixu}
B.~Go, Q.~Jia, H.~Kim, and S.~Kim, ``{From BPS Spectra of Argyres-Douglas
  Theories to Families of 3d TFTs},''
  \href{http://arxiv.org/abs/2502.15133}{{\ttfamily arXiv:2502.15133
  [hep-th]}}.

\bibitem{Kim:2024dxu}
H.~Kim and J.~Song, ``{A Family of Vertex Operator Algebras from
  Argyres-Douglas Theory},'' \href{http://arxiv.org/abs/2412.20015}{{\ttfamily
  arXiv:2412.20015 [hep-th]}}.

\bibitem{ArabiArdehali:2024ysy}
A.~Arabi~Ardehali, M.~Dedushenko, D.~Gang, and M.~Litvinov, ``{Bridging 4D QFTs
  and 2D VOAs via 3D high-temperature EFTs},''
  \href{http://arxiv.org/abs/2409.18130}{{\ttfamily arXiv:2409.18130
  [hep-th]}}.

\bibitem{ArabiArdehali:2024vli}
A.~Arabi~Ardehali, D.~Gang, N.~J. Rajappa, and M.~Sacchi, ``{3d SUSY
  enhancement and non-semisimple TQFTs from four dimensions},''
  \href{http://arxiv.org/abs/2411.00766}{{\ttfamily arXiv:2411.00766
  [hep-th]}}.

\bibitem{Kim:2025rog}
M.~Kim and S.~Kim, ``{3D TFTs and boundary VOAs from BPS spectra of $(G,G')$
  Argyres-Douglas theories},''
  \href{http://arxiv.org/abs/2511.23194}{{\ttfamily arXiv:2511.23194
  [hep-th]}}.

\bibitem{Csaki:1996sm}
C.~Csaki, M.~Schmaltz, and W.~Skiba, ``{A Systematic approach to confinement in
  N=1 supersymmetric gauge theories},''
  \href{http://dx.doi.org/10.1103/PhysRevLett.78.799}{{\em Phys. Rev. Lett.}
  {\bfseries 78} (1997) 799--802},
  \href{http://arxiv.org/abs/hep-th/9610139}{{\ttfamily arXiv:hep-th/9610139}}.

\bibitem{Amariti:2025lem}
A.~Amariti, F.~Mantegazza, and S.~Rota, ``{Rank-two tensors and deconfinement
  in 3d $\mathcal{N}=2$$SU(N)$ gauge theories},''
  \href{http://arxiv.org/abs/2504.21654}{{\ttfamily arXiv:2504.21654
  [hep-th]}}.

\bibitem{Amariti:2024gco}
A.~Amariti and F.~Mantegazza, ``{Confinement for 3d $\mathcal{N}=2$$SU(N)$ with
  a Symmetric tensor},'' \href{http://arxiv.org/abs/2405.11972}{{\ttfamily
  arXiv:2405.11972 [hep-th]}}.

\bibitem{Bajeot:2022kwt}
S.~Bajeot and S.~Benvenuti, ``{S-confinements from deconfinements},''
  \href{http://dx.doi.org/10.1007/JHEP11(2022)071}{{\em JHEP} {\bfseries 11}
  (2022) 071}, \href{http://arxiv.org/abs/2201.11049}{{\ttfamily
  arXiv:2201.11049 [hep-th]}}.

\bibitem{Benvenuti:2021nwt}
S.~Benvenuti and G.~Lo~Monaco, ``{A toolkit for ortho-symplectic dualities},''
  \href{http://dx.doi.org/10.1007/JHEP09(2023)002}{{\em JHEP} {\bfseries 09}
  (2023) 002}, \href{http://arxiv.org/abs/2112.12154}{{\ttfamily
  arXiv:2112.12154 [hep-th]}}.

\bibitem{Hwang:2021xyw}
C.~Hwang, S.~S. Razamat, E.~Sabag, and M.~Sacchi, ``{Rank $Q$ E-string on
  spheres with flux},''
  \href{http://dx.doi.org/10.21468/SciPostPhys.11.2.044}{{\em SciPost Phys.}
  {\bfseries 11} no.~2, (2021) 044},
  \href{http://arxiv.org/abs/2103.09149}{{\ttfamily arXiv:2103.09149
  [hep-th]}}.

\bibitem{Bajeot:2023gyl}
S.~Bajeot, S.~Benvenuti, and M.~Sacchi, ``{S-confining gauge theories and
  supersymmetry enhancements},''
  \href{http://dx.doi.org/10.1007/JHEP08(2023)042}{{\em JHEP} {\bfseries 08}
  (2023) 042}, \href{http://arxiv.org/abs/2305.10274}{{\ttfamily
  arXiv:2305.10274 [hep-th]}}.

\bibitem{Hwang:2024hhy}
C.~Hwang and S.~Kim, ``{S-confinement of 3d Argyres-Douglas theories and the
  Seiberg-like duality with an adjoint matter},''
  \href{http://arxiv.org/abs/2407.11129}{{\ttfamily arXiv:2407.11129
  [hep-th]}}.

\bibitem{Benvenuti:2024glr}
S.~Benvenuti, R.~Comi, S.~Pasquetti, and M.~Sacchi, ``{Deconfinements,
  Kutasov-Schwimmer dualities and D$_{p}$[SU(N)] theories},''
  \href{http://dx.doi.org/10.1007/JHEP04(2025)056}{{\em JHEP} {\bfseries 04}
  (2025) 056}, \href{http://arxiv.org/abs/2407.11134}{{\ttfamily
  arXiv:2407.11134 [hep-th]}}.

\bibitem{Okazaki:2023kpq}
T.~Okazaki and D.~J. Smith, ``{3d exceptional gauge theories and boundary
  confinement},'' \href{http://dx.doi.org/10.1007/JHEP11(2023)199}{{\em JHEP}
  {\bfseries 11} (2023) 199}, \href{http://arxiv.org/abs/2308.14428}{{\ttfamily
  arXiv:2308.14428 [hep-th]}}.

\bibitem{Benini:2017dud}
F.~Benini, S.~Benvenuti, and S.~Pasquetti, ``{SUSY monopole potentials in 2+1
  dimensions},'' \href{http://dx.doi.org/10.1007/JHEP08(2017)086}{{\em JHEP}
  {\bfseries 08} (2017) 086}, \href{http://arxiv.org/abs/1703.08460}{{\ttfamily
  arXiv:1703.08460 [hep-th]}}.

\bibitem{Kim:2013cma}
H.~Kim and J.~Park, ``{Aharony Dualities for 3d Theories with Adjoint
  Matter},'' \href{http://dx.doi.org/10.1007/JHEP06(2013)106}{{\em JHEP}
  {\bfseries 06} (2013) 106}, \href{http://arxiv.org/abs/1302.3645}{{\ttfamily
  arXiv:1302.3645 [hep-th]}}.

\bibitem{Amariti:2018wht}
A.~Amariti and L.~Cassia, ``{USp(2N$_{c}$) SQCD$_{3}$ with antisymmetric:
  dualities and symmetry enhancements},''
  \href{http://dx.doi.org/10.1007/JHEP02(2019)013}{{\em JHEP} {\bfseries 02}
  (2019) 013}, \href{http://arxiv.org/abs/1809.03796}{{\ttfamily
  arXiv:1809.03796 [hep-th]}}.

\bibitem{Amariti:2019rhc}
A.~Amariti, L.~Cassia, I.~Garozzo, and N.~Mekareeya, ``{Branes, partition
  functions and quadratic monopole superpotentials},''
  \href{http://dx.doi.org/10.1103/PhysRevD.100.046001}{{\em Phys. Rev. D}
  {\bfseries 100} no.~4, (2019) 046001},
  \href{http://arxiv.org/abs/1901.07559}{{\ttfamily arXiv:1901.07559
  [hep-th]}}.

\bibitem{Hwang:2022jjs}
C.~Hwang, S.~Kim, and J.~Park, ``{Monopole deformations of 3d Seiberg-like
  dualities with adjoint matters},''
  \href{http://dx.doi.org/10.1007/JHEP11(2022)111}{{\em JHEP} {\bfseries 11}
  (2022) 111}, \href{http://arxiv.org/abs/2202.09000}{{\ttfamily
  arXiv:2202.09000 [hep-th]}}.

\bibitem{Benvenuti:2023qtv}
S.~Benvenuti, R.~Comi, and S.~Pasquetti, ``{Mirror dualities with four
  supercharges},'' \href{http://dx.doi.org/10.1007/JHEP10(2024)234}{{\em JHEP}
  {\bfseries 10} (2024) 234}, \href{http://arxiv.org/abs/2312.07667}{{\ttfamily
  arXiv:2312.07667 [hep-th]}}.

\bibitem{Hwang:2020wpd}
C.~Hwang, S.~Pasquetti, and M.~Sacchi, ``{4d mirror-like dualities},''
  \href{http://dx.doi.org/10.1007/JHEP09(2020)047}{{\em JHEP} {\bfseries 09}
  (2020) 047}, \href{http://arxiv.org/abs/2002.12897}{{\ttfamily
  arXiv:2002.12897 [hep-th]}}.

\bibitem{Hwang:2021ulb}
C.~Hwang, S.~Pasquetti, and M.~Sacchi, ``{Rethinking mirror symmetry as a local
  duality on fields},''
  \href{http://dx.doi.org/10.1103/PhysRevD.106.105014}{{\em Phys. Rev. D}
  {\bfseries 106} no.~10, (2022) 105014},
  \href{http://arxiv.org/abs/2110.11362}{{\ttfamily arXiv:2110.11362
  [hep-th]}}.

\bibitem{Bottini:2022vpy}
L.~E. Bottini, C.~Hwang, S.~Pasquetti, and M.~Sacchi, ``{Dualities from
  dualities: the sequential deconfinement technique},''
  \href{http://dx.doi.org/10.1007/JHEP05(2022)069}{{\em JHEP} {\bfseries 05}
  (2022) 069}, \href{http://arxiv.org/abs/2201.11090}{{\ttfamily
  arXiv:2201.11090 [hep-th]}}.

\bibitem{Comi:2022aqo}
R.~Comi, C.~Hwang, F.~Marino, S.~Pasquetti, and M.~Sacchi, ``{The SL(2,
  \ensuremath{\mathbb{Z}}) dualization algorithm at work},''
  \href{http://dx.doi.org/10.1007/JHEP06(2023)119}{{\em JHEP} {\bfseries 06}
  (2023) 119}, \href{http://arxiv.org/abs/2212.10571}{{\ttfamily
  arXiv:2212.10571 [hep-th]}}.

\bibitem{Giacomelli:2023zkk}
S.~Giacomelli, C.~Hwang, F.~Marino, S.~Pasquetti, and M.~Sacchi, ``{Probing bad
  theories with the dualization algorithm. Part I},''
  \href{http://dx.doi.org/10.1007/JHEP04(2024)008}{{\em JHEP} {\bfseries 04}
  (2024) 008}, \href{http://arxiv.org/abs/2309.05326}{{\ttfamily
  arXiv:2309.05326 [hep-th]}}.

\bibitem{Giacomelli:2024laq}
S.~Giacomelli, C.~Hwang, F.~Marino, S.~Pasquetti, and M.~Sacchi, ``{Probing bad
  theories with the dualization algorithm. Part II.},''
  \href{http://dx.doi.org/10.1007/JHEP07(2024)165}{{\em JHEP} {\bfseries 07}
  (2024) 165}, \href{http://arxiv.org/abs/2401.14456}{{\ttfamily
  arXiv:2401.14456 [hep-th]}}.

\bibitem{Cecotti:2012jx}
S.~Cecotti and M.~Del~Zotto, ``{Infinitely many N=2 SCFT with ADE flavor
  symmetry},'' \href{http://dx.doi.org/10.1007/JHEP01(2013)191}{{\em JHEP}
  {\bfseries 01} (2013) 191}, \href{http://arxiv.org/abs/1210.2886}{{\ttfamily
  arXiv:1210.2886 [hep-th]}}.

\bibitem{Cecotti:2013lda}
S.~Cecotti, M.~Del~Zotto, and S.~Giacomelli, ``{More on the N=2 superconformal
  systems of type $D_p(G)$},''
  \href{http://dx.doi.org/10.1007/JHEP04(2013)153}{{\em JHEP} {\bfseries 04}
  (2013) 153}, \href{http://arxiv.org/abs/1303.3149}{{\ttfamily arXiv:1303.3149
  [hep-th]}}.

\bibitem{Maruyoshi:2023mnv}
K.~Maruyoshi, E.~Nardoni, and J.~Song, ``{Dualities of Adjoint SQCD and
  Supersymmetry Enhancement},''
  \href{http://arxiv.org/abs/2306.08867}{{\ttfamily arXiv:2306.08867
  [hep-th]}}.

\bibitem{Closset:2020afy}
C.~Closset, S.~Giacomelli, S.~Schafer-Nameki, and Y.-N. Wang, ``{5d and 4d
  SCFTs: Canonical Singularities, Trinions and S-Dualities},''
  \href{http://dx.doi.org/10.1007/JHEP05(2021)274}{{\em JHEP} {\bfseries 05}
  (2021) 274}, \href{http://arxiv.org/abs/2012.12827}{{\ttfamily
  arXiv:2012.12827 [hep-th]}}.

\bibitem{Giacomelli:2020ryy}
S.~Giacomelli, N.~Mekareeya, and M.~Sacchi, ``{New aspects of Argyres--Douglas
  theories and their dimensional reduction},''
  \href{http://dx.doi.org/10.1007/JHEP03(2021)242}{{\em JHEP} {\bfseries 03}
  (2021) 242}, \href{http://arxiv.org/abs/2012.12852}{{\ttfamily
  arXiv:2012.12852 [hep-th]}}.

\bibitem{future}
C.~Hwang and S.~Kim, ``{Future work},''.

\bibitem{Jafferis:2010un}
D.~L. Jafferis, ``{The Exact Superconformal R-Symmetry Extremizes Z},''
  \href{http://dx.doi.org/10.1007/JHEP05(2012)159}{{\em JHEP} {\bfseries 05}
  (2012) 159}, \href{http://arxiv.org/abs/1012.3210}{{\ttfamily arXiv:1012.3210
  [hep-th]}}.

\bibitem{Bhattacharya:2008zy}
J.~Bhattacharya, S.~Bhattacharyya, S.~Minwalla, and S.~Raju, ``{Indices for
  Superconformal Field Theories in 3,5 and 6 Dimensions},''
  \href{http://dx.doi.org/10.1088/1126-6708/2008/02/064}{{\em JHEP} {\bfseries
  02} (2008) 064}, \href{http://arxiv.org/abs/0801.1435}{{\ttfamily
  arXiv:0801.1435 [hep-th]}}.

\bibitem{Bhattacharya:2008bja}
J.~Bhattacharya and S.~Minwalla, ``{Superconformal Indices for N = 6 Chern
  Simons Theories},''
  \href{http://dx.doi.org/10.1088/1126-6708/2009/01/014}{{\em JHEP} {\bfseries
  01} (2009) 014}, \href{http://arxiv.org/abs/0806.3251}{{\ttfamily
  arXiv:0806.3251 [hep-th]}}.

\bibitem{Kapustin:2011vz}
A.~Kapustin, H.~Kim, and J.~Park, ``{Dualities for 3d Theories with Tensor
  Matter},'' \href{http://dx.doi.org/10.1007/JHEP12(2011)087}{{\em JHEP}
  {\bfseries 12} (2011) 087}, \href{http://arxiv.org/abs/1110.2547}{{\ttfamily
  arXiv:1110.2547 [hep-th]}}.

\bibitem{Hwang:2015wna}
C.~Hwang and J.~Park, ``{Factorization of the 3d superconformal index with an
  adjoint matter},'' \href{http://dx.doi.org/10.1007/JHEP11(2015)028}{{\em
  JHEP} {\bfseries 11} (2015) 028},
  \href{http://arxiv.org/abs/1506.03951}{{\ttfamily arXiv:1506.03951
  [hep-th]}}.

\bibitem{Hama:2011ea}
N.~Hama, K.~Hosomichi, and S.~Lee, ``{SUSY Gauge Theories on Squashed
  Three-Spheres},'' \href{http://dx.doi.org/10.1007/JHEP05(2011)014}{{\em JHEP}
  {\bfseries 05} (2011) 014}, \href{http://arxiv.org/abs/1102.4716}{{\ttfamily
  arXiv:1102.4716 [hep-th]}}.

\bibitem{Closset:2018ghr}
C.~Closset, H.~Kim, and B.~Willett, ``{Seifert fibering operators in 3d
  $\mathcal{N}=2$ theories},''
  \href{http://dx.doi.org/10.1007/JHEP11(2018)004}{{\em JHEP} {\bfseries 11}
  (2018) 004}, \href{http://arxiv.org/abs/1807.02328}{{\ttfamily
  arXiv:1807.02328 [hep-th]}}.

\bibitem{Closset:2019hyt}
C.~Closset and H.~Kim, ``{Three-dimensional $\mathcal{N}$ = 2 supersymmetric
  gauge theories and partition functions on Seifert manifolds: A review},''
  \href{http://dx.doi.org/10.1142/S0217751X19300114}{{\em Int. J. Mod. Phys. A}
  {\bfseries 34} no.~23, (2019) 1930011},
  \href{http://arxiv.org/abs/1908.08875}{{\ttfamily arXiv:1908.08875
  [hep-th]}}.

\bibitem{Closset:2023vos}
C.~Closset and O.~Khlaif, ``{Twisted indices, Bethe ideals and 3d $ \mathcal{N}
  $ = 2 infrared dualities},''
  \href{http://dx.doi.org/10.1007/JHEP05(2023)148}{{\em JHEP} {\bfseries 05}
  (2023) 148}, \href{http://arxiv.org/abs/2301.10753}{{\ttfamily
  arXiv:2301.10753 [hep-th]}}.

\bibitem{Closset:2023jiq}
C.~Closset and O.~Khlaif, ``{On the Witten index of 3d $\mathcal{N}=2$ unitary
  SQCD with general CS levels},''
  \href{http://dx.doi.org/10.21468/SciPostPhys.15.3.085}{{\em SciPost Phys.}
  {\bfseries 15} no.~3, (2023) 085},
  \href{http://arxiv.org/abs/2305.00534}{{\ttfamily arXiv:2305.00534
  [hep-th]}}.

\bibitem{Closset:2023bdr}
C.~Closset and O.~Khlaif, ``{Grothendieck lines in 3d $ \mathcal{N} $ = 2 SQCD
  and the quantum K-theory of the Grassmannian},''
  \href{http://dx.doi.org/10.1007/JHEP12(2023)082}{{\em JHEP} {\bfseries 12}
  (2023) 082}, \href{http://arxiv.org/abs/2309.06980}{{\ttfamily
  arXiv:2309.06980 [hep-th]}}.

\bibitem{Closset:2023izb}
C.~Closset and O.~Khlaif, ``{New results on 3d $\mathcal{N}$=2 SQCD and its 3d
  GLSM interpretation},''
  \href{http://dx.doi.org/10.1142/S0217751X24460114}{{\em Int. J. Mod. Phys. A}
  {\bfseries 39} no.~33, (2024) 2446011},
  \href{http://arxiv.org/abs/2312.05076}{{\ttfamily arXiv:2312.05076
  [hep-th]}}.

\bibitem{Closset:2024sle}
C.~Closset, E.~Furrer, and O.~Khlaif, ``{One-form symmetries and the 3d
  $\mathcal{N}=2$ $A$-model: Topologically twisted indices and CS theories},''
  \href{http://dx.doi.org/10.21468/SciPostPhys.18.2.066}{{\em SciPost Phys.}
  {\bfseries 18} no.~2, (2025) 066},
  \href{http://arxiv.org/abs/2405.18141}{{\ttfamily arXiv:2405.18141
  [hep-th]}}.

\bibitem{Closset:2025lqt}
C.~Closset, E.~Furrer, A.~Keyes, and O.~Khlaif, ``{The 3d $A$-model and
  generalised symmetries, Part I: bosonic Chern-Simons theories},''
  \href{http://arxiv.org/abs/2501.11665}{{\ttfamily arXiv:2501.11665
  [hep-th]}}.

\bibitem{Pufu:2016zxm}
S.~S. Pufu, ``{The F-Theorem and F-Maximization},''
  \href{http://dx.doi.org/10.1088/1751-8121/aa6765}{{\em J. Phys. A} {\bfseries
  50} no.~44, (2017) 443008}, \href{http://arxiv.org/abs/1608.02960}{{\ttfamily
  arXiv:1608.02960 [hep-th]}}.

\end{thebibliography}\endgroup
\end{document}